\documentclass{article}

\usepackage{arxiv}
\usepackage[utf8]{inputenc} 
\usepackage[T1]{fontenc}    
\usepackage{hyperref}       
\usepackage{url}            
\usepackage{booktabs}       
\usepackage{amsfonts}       
\usepackage{nicefrac}       
\usepackage{microtype}      
\usepackage{lipsum}         
\usepackage{graphicx}
\usepackage{doi}

\usepackage{amsfonts}
\usepackage{epstopdf}
\usepackage{algorithmic}
\ifpdf
  \DeclareGraphicsExtensions{.eps,.pdf,.png,.jpg}
\else
  \DeclareGraphicsExtensions{.eps}
\fi

\usepackage{float}
\usepackage{placeins}
\usepackage{cite}
\usepackage{color}
\usepackage[dvipsnames,svgnames,table]{xcolor}
\usepackage{colortbl}

\usepackage{stmaryrd} 
\SetSymbolFont{stmry}{bold}{U}{stmry}{m}{n}

\usepackage{bm}
\usepackage{bbm}
\usepackage{xspace}
\usepackage{amsmath}
\DeclareMathOperator*{\argmin}{argmin}
\usepackage{mathtools}
\usepackage{amssymb}
\usepackage{amsbsy}
\usepackage{eucal}
\usepackage{mathrsfs}
\usepackage{tcolorbox}
\usepackage{wrapfig}
\usepackage{pdfpages}
\usepackage{caption}
\usepackage{changepage}
\usepackage{multirow}
\usepackage{cancel}
\usepackage{subcaption}
\usepackage{grffile}
\usepackage{pgf}
\usepackage{pgfgantt}
\usepackage{verbatim}
\usepackage{siunitx}
\usepackage{listings}
\usepackage[normalem]{ulem}

\usepackage{longtable}

\definecolor{cadetblue}{RGB}{47,85,151}
\definecolor{lightblue}{RGB}{180,199,231}

\usepackage{cleveref}       

\title{Model Error Embedding with Orthogonal Gaussian Processes}

\newif\ifuniqueAffiliation
\uniqueAffiliationtrue

\ifuniqueAffiliation 
\author{ Mridula Kuppa\\
Department of Aerospace Engineering\\
University of Illinois Urbana Champaign\\
Champaign, IL 61801 USA\\
\texttt{vkuppa2@illinois.edu} \\
\And
Khachik Sargsyan \thanks{Corresponding author} \\
Sandia National Laboratories\\
Livermore, CA 94551 USA\\
\texttt{ksargsy@sandia.gov} \\
\AND
Marco Panesi \\
University of California Irvine\\
Mechanical and Aerospace Engineering\\
Irvine, CA 92697 USA\\
\texttt{mpanesi@uci.edu} \\
\And
Habib N. Najm\\
Sandia National Laboratories\\
Livermore, CA 94551 USA\\
\texttt{hnnajm@sandia.gov} \\
}
\else
\usepackage{authblk}

\newbox{\orcid}\sbox{\orcid}{\includegraphics[scale=0.06]{orcid.pdf}} 
\author[1]{%
	\href{https://orcid.org/0000-0000-0000-0000}{\usebox{\orcid}\hspace{1mm}David S.~Hippocampus\thanks{\texttt{hippo@cs.cranberry-lemon.edu}}}%
}
\author[1,2]{%
	\href{https://orcid.org/0000-0000-0000-0000}{\usebox{\orcid}\hspace{1mm}Elias D.~Striatum\thanks{\texttt{stariate@ee.mount-sheikh.edu}}}%
}
\affil[1]{Department of Computer Science, Cranberry-Lemon University, Pittsburgh, PA 15213}
\affil[2]{Department of Electrical Engineering, Mount-Sheikh University, Santa Narimana, Levand}
\fi

\begin{document}
\maketitle

\begin{abstract}
Computational models of complex physical systems often rely on simplifying assumptions which inevitably introduce model error, with consequent predictive errors. 
Given data on model observables, the estimation of parameterized model-error representations, along with other model parameters, would be ideally done while separating the contributions of each of the two sets of parameters, in order to ensure meaningful stand-alone model predictions.
This work builds an embedded model error framework using a weight-space representation of Gaussian processes (GPs) to flexibly capture model-error spatiotemporal correlations and enable inference with GP-embedding in non-linear models. To disambiguate model and model-error/bias parameters, we extend an existing orthogonal GP method to the embedded model-error setting and derive appropriate orthogonality constraints.
To address the increased dimensionality introduced by the GP representation, we employ the likelihood-informed subspace method. The construction is demonstrated on linear and non-linear examples, where it effectively corrects model predictions to match data trends. Extrapolation beyond the training data recovers the prior predictive distribution, and the orthogonality constraints lead to meaningful stand-alone model predictions and nearly uncorrelated posteriors between model and model-error parameters.
\end{abstract}


\section{Introduction}
\label{sec:Introduction}

Computational models are ubiquitous in science and engineering, enabling the prediction of complex physical phenomena arising in fluid and solid mechanics, chemical kinetics, biological systems, and many other domains. These models are typically derived from theoretical principles or constructed using empirical relations when theoretical understanding is incomplete. In many applications, the underlying physics is not fully understood or experimental data are limited due to cost, complexity, or safety considerations. As a result, simplifying assumptions are often introduced during model development, inevitably leading to a degree of model inadequacy (also referred to in this work as model error, model discrepancy, or bias).

Moreover, computational models depend on a set of independent variables, such as space or time, and a collection of input parameters that must be specified to evaluate the model. For example, permeability must be prescribed in groundwater simulations to predict hydraulic head, and Arrhenius rate coefficients must be specified in chemical kinetics models to simulate the evolution of species concentrations. These parameters are typically inferred through model calibration, wherein limited observational data, obtained either from experiments or from higher-fidelity simulations, are used to learn the appropriate parameter values. In this work, we focus on the Bayesian formulation of such inverse problems while explicitly accounting for the model error.

Bayesian approaches to model calibration have been an active area of research for several decades~\cite{romanowicz1994evaluation,kaipio2005statistical,raftery1995inference,poole2000inference}. In particular, Gaussian processes (GPs) have been widely used to define priors over unknown functions and also as cheap emulators for the expensive computer models~\cite{matheron1963principles,kimeldorf1970correspondence,sacks1989design,Cressie:1990,handcock1993bayesian,bernardo1998regression,neal2012bayesian}.
Often, early work on Bayesian model calibration either subsumed model error into the observational error term or neglected it altogether by assuming it to be negligible. A key advance in explicitly accounting for model inadequacy was introduced by Kennedy and O’Hagan~\cite{kennedy2001bayesian} (hereafter KOH). Their framework augments the computational model with a GP representation of the discrepancy between the model predictions and reality, and jointly infers both the model parameters and the discrepancy function using observational data.
While the flexibility of the additive GP allows to capture the spatial correlation structure of model error~\cite{Kennedy:2002,Higdon:2004,Bayarri:2009a}, several studies have highlighted key challenges with this formulation, most notably the conflation of the model parameters and the GP~\cite{arendt2012quantification, tuo2015efficient, gramacy2015calibrating, santner2003design, Arendt:2016, Brown:2018, Oreluk:2022}.
This makes it difficult to assess the overall predictive capability of the computational model. 
In addition, without careful prior-design~\cite{brynjarsdottir2014learning}, the GP-augmented predictions may not adhere to the underlying physical constraints and governing equations. 

Recent theoretical work has shown that under simplifying assumptions of an inexpensive computer code, deterministic data, and a known covariance function for the GP bias, KOH calibration produces maximum likelihood estimates of the model parameters that depend on the reproducing kernel Hilbert space (RKHS) norm based on the bias covariance~\cite{tuo2015efficient, tuo2016theoretical}. 
\textcolor{black}{This analysis has also been extended to noisy data settings where similar conclusions are drawn regarding the asymptotic properties of KOH calibration \cite{tuo2020improved}}. 
This implies that the estimated parameter values are tied to the prior specification of the bias, which results in the KOH approach being $L_2$-inconsistent, meaning that the parameter values minimizing the $L_2$ loss are not recovered even in the infinite-data regime~\cite{tuo2016theoretical}. Several modifications have been proposed to address this limitation, including modular approaches that separately fit the model and the GP~\cite{bayarri2009modularization,andres2025model}; prior tuning strategies that shape the discrepancy GP using physical knowledge about the parameters~\cite{brynjarsdottir2014learning}; orthogonalizing the discrepancy GP with respect to model gradients, leading to the orthogonal Gaussian process (OGP) framework~\cite{plumlee2017bayesian, plumlee2018orthogonal}; and scaled GPs that reconcile $L_2$ calibration of~\cite{tuo2016theoretical} with the KOH framework~\cite{gu2018scaled}. Among these, the OGP approach is particularly appealing because it provides a foundationally principled mechanism to disentangle the roles of the model and discrepancy within joint Bayesian calibration.

A more physically informed approach to model discrepancy is the embedded/internal model correction framework, in which stochastic correction terms are embedded within the computational model~\cite{oliver2015validating,sargsyan2015statistical,morrison2018representing,pernot2017critical,bandy2024stochastic} rather than being additive on model outputs. In the embedded model-error formulation of~\cite{sargsyan2015statistical}, a statistical model is embedded within specific physical model components where approximations are known to have been made. Estimation of parameters in this model error construction provides predictions with uncertainty accounting for model error, while also serving to provide diagnostic information on the importance of different approximations on resulting predictive uncertainty. The simplest version of this approach involves enriching existing model parameters with additive random variables. Further, when polynomial chaos expansions (PCEs) are used to construct a surrogate for the dependence of model quantities of interest (QoIs) on uncertain model parameters, additional stochastic dimensions (germs) representing model error are employed to augment select model parameters, and the associated PCE coefficients are inferred alongside the physical model parameters. Subsequent uncertainty propagation yields predictions that remain consistent with the underlying physics while enabling the attribution of discrepancy to specific submodels. This framework has been successfully applied in diverse contexts, including chemical ignition modeling~\cite{sargsyan2015statistical,Hakim:2018}, scramjet design~\cite{huan2017global}, materials modeling~\cite{Hegde:2024}, hypersonic flow modeling~\cite{kuppa2025stochastic}, and has been extended to account for noisy observations~\cite{sargsyan2019embedded}. 

Most of the existing work on embedded model error has focused on random variable embeddings, which typically ignore the spatial and/or temporal dependence of the model error. A recent exception is the work of~\cite{Fan:2025}, which employs a 2-degree of freedom GP embedding, relying on the Karhunen Lo\`eve expansion (KLE)~\cite{Karhunen:1946,Sakamoto:2002,Chowdhary:2016}, to represent model error, leveraging the flexibility of GPs to capture some spatial correlation in the context of nonlocal operator regression.
\textcolor{black}{Another approach, closely related to our proposed method, is functional calibration \cite{brown2018nonparametric,tuo2023reproducing} where the model parameters to be calibrated are expressed as functions of control inputs to account for model inadequacy.
Brown et al. \cite{brown2018nonparametric} rightfully point out that their method has the advantages of model assessment for missing physics and increased confidence in extrapolation to new conditions, when compared to additive model error formulations.
They also emphasize that the prior on functional calibration parameters should be decided carefully to avoid identifiability issues. 
Tuo et al. \cite{tuo2023reproducing} presented a frequentist version of functional calibration with theoretical analysis for prediction consistency and consistency of the calibration function.}  
While there is an extensive literature on regression and generative modeling for non-linear data -- such as warped GPs~\cite{snelson2003warped}, deep GPs~\cite{damianou2013deep}, and multifidelity GPs~\cite{perdikaris2017nonlinear} -- the problem considered in this work is fundamentally different. In our setting, the computational model is already known, and the objective is to calibrate this model using observational data while explicitly accounting for model error. Thus, rather than learning a purely data-driven surrogate, we aim to infer model parameters and model discrepancy in a Bayesian framework. Similarly, while there is significant work~\cite{Marzouk:2007,marzouk2009dimensionality,Martin:2012} on Bayesian learning of random fields, \emph{e.g.} spatially dependent model parameters/inputs, the disambiguation challenges that emerge with the combination of physical parameterization and embedded random field model error constructions distinguishes the present context.

One reason for the relatively limited literature on GP-based embedded model error is the difficulty of inferring function-space GPs when they are embedded within non-linear, often non-invertible, computational models. To address this challenge, we propose adopting the weight-space representation of Gaussian processes~\cite{williams2006gaussian}, in which the GP is expressed as a weighted sum of eigenfunctions of the underlying covariance kernel. This representation enables inference of a GP embedded within a non-linear model without requiring invertibility.
Nevertheless, Bayesian inference remains computationally challenging due to the increased dimensionality of the parameter space, as the GP weights must be inferred jointly with the computational model parameters. We argue that recent advances in dimensionality reduction of high-dimensional Bayesian inference problems~\cite{marzouk2009dimensionality,spantini2015optimal,cui2014likelihood} can be leveraged to mitigate these challenges. However, the issue of disambiguation between the computational model and the embedded model bias persists and must be carefully addressed in this embedded model error setting.

This work has three main contributions. 
First, we extend the general embedded model error framework of~\cite{sargsyan2015statistical} to increase the flexibility of model error representations by adopting the weight-space perspective of Gaussian processes. 
Second, to address the disambiguation issue and disentangle the contributions of the model from those of the embedded GP, we adapt the OGP methodology of~\cite{plumlee2017bayesian} to the embedded model error setting. This ensures that the model’s standalone predictions remain meaningful (with respect to a specified loss function), which is especially valuable when extrapolating beyond the data inputs.
And lastly, we show how the likelihood-informed subspace (LIS) approach~\cite{spantini2015optimal,cui2014likelihood}, which enables efficient dimensionality reduction for Bayesian inference problems, can be employed to retain a sufficiently large set of basis functions in the weight-space representation of GPs for model error embedding. 

The remainder of the paper is organized as follows. Section~\ref{sec:GPs} reviews Gaussian processes, highlighting the equivalence between the function-space and weight-space perspectives. Section~\ref{sec:Additive_error} presents the additive model error formulation along with the corresponding OGP construction. In Section~\ref{sec:EMB_OGP}, we introduce our embedded model error methodology using the weight-space GP formulation and describe two approaches for enforcing orthogonality in the embedded setting: linearized OGP (LOGP) and regularized OGP (ROGP). Section~\ref{sec:LIS} reviews highlights of the LIS approach, which enables efficient dimensionality reduction for Bayesian inference. In Section~\ref{sec:Appl}, three examples -- a linear model, a non-linear model with interacting submodels, and an advection-diffusion-reaction PDE, demonstrate the features of the proposed methodology. Finally, Section~\ref{sec:concl} provides concluding remarks and directions for future work.

\section{Gaussian Processes}\label{sec:GPs}

The weight-space perspective of Gaussian processes is particularly convenient for GP-based model error embedding within non-linear computational models, as it enables working with a finite set of weights associated with a fixed set of basis functions, in contrast to the non-parametric function space view. While the two perspectives are equivalent in principle, it is useful to first examine how they compare in practice, both in terms of predictive capability and inference using data. We study this comparison in the regression setting with a training dataset $\mathcal{D} = \{(\mathbf{x}_i, y_i) \mid i = 1,\dots, N\}$, 
with inputs $\mathbf{x}_i \in \mathcal{X} \subset \mathbb{R}^D$ and outputs $y_i \in \mathbb{R}$.  
The objective is to learn the relationship between $\mathbf{x}$ and $y$ in order to make predictions at new, unseen inputs.  
Let $X \in \mathbb{R}^{N \times D}$ denote the matrix of all training inputs and  
$\mathbf{y} \in \mathbb{R}^N$ the vector of corresponding training outputs.  
We next summarize how the function-space and weight-space views achieve this goal. 

\subsection{Function-Space View}\label{GP_FV}

In the function-space (FS) formulation, a prior distribution over \emph{functions} is specified by a mean function $m(\mathbf{x})$ and a covariance function $k(\mathbf{x}, \mathbf{x}')$:
\begin{align}
    m(\mathbf{x}) &= \mathbb{E}[f(\mathbf{x})], \label{eq:gp_fv_mean}\\[4pt]
    k(\mathbf{x}, \mathbf{x}') &= \mathbb{E}\!\left[(f(\mathbf{x}) - m(\mathbf{x}))(f(\mathbf{x}') - m(\mathbf{x}'))\right]. \label{eq:gp_fv_cov}
\end{align}
We write
\begin{align*}
    f(\mathbf{x}) \sim GP(m(\mathbf{x}), k(\mathbf{x}, \mathbf{x}')),
\end{align*}
to indicate that for any finite collection of inputs $\mathbf{x}_1,\dots,\mathbf{x}_N \in \mathcal{X}$, the vector  
$[f(\mathbf{x}_1),\dots,f(\mathbf{x}_N)]$ follows a multivariate normal distribution with the given mean and covariance functions.
The posterior predictive distribution at test inputs is obtained by conditioning the joint prior distribution, over both training and test inputs, on the observed data.
\textcolor{black}{Since regression for FS GP is well established, we omit the detailed equations here and  refer the reader to \cite{williams2006gaussian}}. 
 
\subsection{Weight-Space View}\label{sec:GP_WV}

The weight-space (WS) GP interpretation begins with Mercer's theorem~\cite{konig2013eigenvalue}, which states that any positive definite kernel function can be written as an (in general infinite) series expansion in the kernel eigenfunctions and eigenvalues:
\begin{align*}
    k(\mathbf{x}, \mathbf{x}') = \sum_{i=0}^{\infty} \lambda_i \, \phi_i(\mathbf{x}) \phi_i(\mathbf{x}'). 
    \label{eq:gp_mercer}
\end{align*}
The eigenfunctions, $\phi_i$, and eigenvalues, $\lambda_i$, are obtained from the integral equation
\begin{align*}
    \int_{\mathcal{X}} k(\mathbf{x}, \mathbf{x}') \, \phi(\mathbf{x}) \, d\mu(\mathbf{x})
    = \lambda \, \phi(\mathbf{x}'),
\end{align*}
where $\mu$ denotes a measure over $\mathcal{X}$. 
These eigenfunctions form an orthogonal basis and may be normalized so that, 
\begin{align*}
    \int_{\mathcal{X}} \phi_i(\mathbf{x}) \phi_j(\mathbf{x}) \, d\mu(\mathbf{x})
    = \delta_{ij}.
\end{align*}
\textcolor{black}{If $\mathbf{x}$ is sampled from a density $p(\mathbf{x})$, the associated measure can be written as $d\mu(\mathbf{x}) = p(\mathbf{x}) d\mathbf{x}$. In the absence of a known sampling density, however, and given an arbitrary set of data points, $\mathbf{x}_p, p=1,\ldots,N$, with no information about their underlying distribution, we assume a uniform density over the domain. Under this assumption, the integral is approximated using the trapezoidal rule, leading to a matrix eigenvalue problem from which the basis functions are computed numerically using eigenvalue decomposition.}

\begin{figure}[H]
    \centering
    \subcaptionbox{$\ell = 0.2$}{\includegraphics[width=0.35\linewidth]{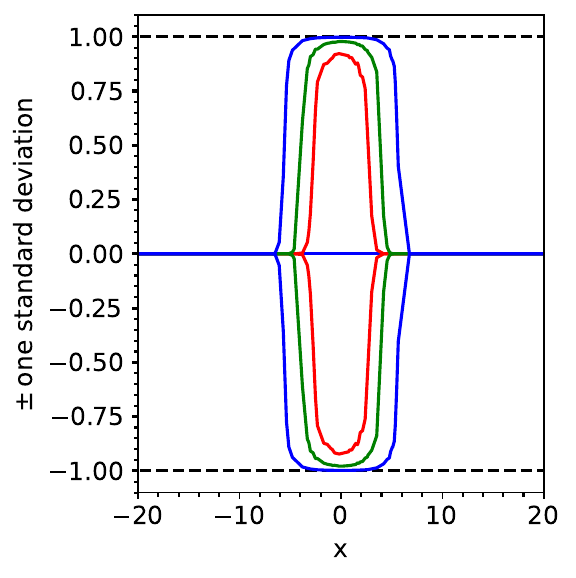}}
    \subcaptionbox{$\ell = 1$}{\includegraphics[width=0.35\linewidth]{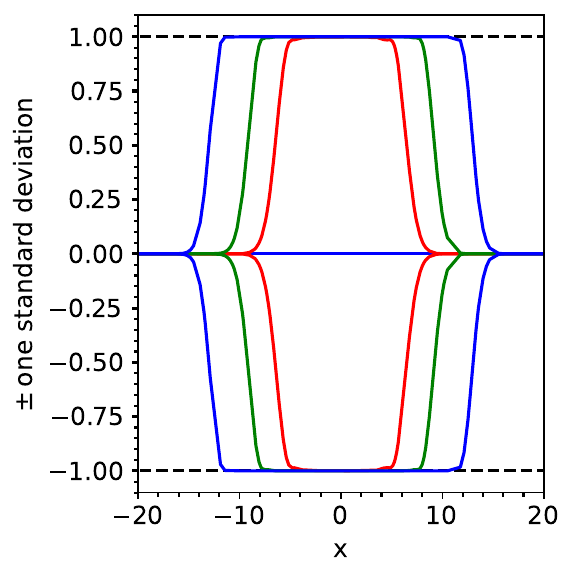}}
    \caption{Panels (a) and (b) show the comparison of prior-predictive plus/minus one standard deviation obtained using the FS view (dashed black) and the WS view with $m = 10$ (red), $20$ (green), and $40$ (blue) for correlation lengths 0.2 and 1 respectively.}
    \label{fig:std_fv_vs_wv}
\end{figure}

Any $L^2$ function that lies in the span of $\{\phi_i\}$ can therefore be expressed as a weighted summation \cite{murphy2012machine}:
\begin{align}
    f(\mathbf{x}) = \sum_{i=0}^{\infty} \phi_i(\mathbf{x}) \, w_i,
\end{align}
where $w_i$ are the weights associated with the basis functions.  
In practice, the infinite series is truncated to the first $m$ terms:
\begin{align}\label{eq:GP_WV}
    f_m(\mathbf{x})
    = \boldsymbol{\phi}(\mathbf{x})^{\top} \boldsymbol{w},
\end{align}
where  
$\boldsymbol{\phi}(\mathbf{x}) = [\phi_0(\mathbf{x}), \dots, \phi_{m-1}(\mathbf{x})]^{\top}$  
and $\boldsymbol{w} \in \mathbb{R}^m$, and a Gaussian prior is placed on the weights,
\begin{align*}
    \boldsymbol{w} \sim \mathcal{N}(\mathbf{0}, \Sigma_p),
    \qquad \Sigma_p = \mathrm{diag}(\lambda_0,\dots,\lambda_{m-1}). 
\end{align*}
For comparison with the FS view, the corresponding mean and covariance functions can be derived follows~\cite{williams2006gaussian}:
\begin{align}
    m(\mathbf{x})
        &= \boldsymbol{\phi}(\mathbf{x})^{\top} \mathbb{E}[\boldsymbol{w}]
         = \mathbf{0},\\[4pt]
    k(\mathbf{x}, \mathbf{x}')
        &= \boldsymbol{\phi}(\mathbf{x})^{\top}
           \mathbb{E}[\boldsymbol{w}\boldsymbol{w}^{\top}]
           \boldsymbol{\phi}(\mathbf{x}')
         = \boldsymbol{\phi}(\mathbf{x})^{\top} \Sigma_p \boldsymbol{\phi}(\mathbf{x}')
         = \sum_{i=0}^{m-1} \lambda_i \phi_i(\mathbf{x}) \phi_i(\mathbf{x}').
         \label{eq:gp_wv_cov}
\end{align}
Therefore, we see that truncation introduces a local prior-predictive variance deficit given by
\begin{equation}
    \Delta \mathrm{Var}(\mathbf{x})
    = \mathrm{Var}[f(\mathbf{x})] - \mathrm{Var}[f_m(\mathbf{x})]
    = \sum_{i=m}^{\infty} \lambda_i \phi_i(\mathbf{x})^2.
\end{equation}
\textcolor{black}{To illustrate this deficit, consider a one-dimensional input, $x \sim \mathcal{N}(0,2)$, and the squared exponential (SQE) covariance function, $k(x,x') = \sigma^2 \exp\!\left(-(x-x')^2/(2\ell^2)\right)$, with $\sigma=1$.  
For this setting, analytical expressions for the eigenfunctions and eigenvalues are available~\cite{williams2006gaussian}, allowing us to compare the FS and WS views without any numerical errors arising from eigensystem computation.  
Figure~\ref{fig:std_fv_vs_wv} compares prior-predictive plus/minus one standard deviation obtained from the FS view alongside its truncated WS representation for $m = 10, 20,$ and $40$.  
The variance deficit is minimal near $x = 0$ and grows as $|x|$ increases. 
While its spatial profile depends on both the measure of $x$ and the kernel, the reduction in prior-predictive variance is fundamentally a consequence of truncating the WS representation, and thus persists regardless of the choice of input measure. 
For shorter correlation lengths, the eigenvalues decay more slowly, leading to a more pronounced loss of variance for a fixed truncation level. Consequently, a larger number of basis functions is required to recover the FS behaviour, whereas kernels with longer correlation lengths can be well-approximated with fewer terms.}

In the WS representation, the noisy observations are used to compute the posterior distribution of the weights via Bayes' theorem:
\begin{align*}
    \pi(\boldsymbol{w} \mid \mathbf{y})
    \propto \pi(\mathbf{y} \mid \boldsymbol{w}) \, \pi(\boldsymbol{w}).
\end{align*}
For standard, independent and identically distributed (iid) Gaussian noise of known variance, $\sigma^2_d$, on the observations, the likelihood is written as
\begin{align*}
    \pi(\mathbf{y} \mid \boldsymbol{w})
    = \frac{1}{(2\pi\sigma_d^2)^{N/2}}
      \exp\!\left(-\frac{1}{2\sigma_d^2}
      \lVert \mathbf{y} - \Phi^{\top}\boldsymbol{w} \rVert^2 \right),
\end{align*}
where $\Phi$ is the $m\times N$ matrix formed by stacking the basis evaluations $\boldsymbol{\phi}(\mathbf{x}_i)$ for all training points.  
Since both the likelihood and prior are Gaussian, the posterior is also Gaussian, given by~\cite{williams2006gaussian}:
\begin{align*}
    \pi(\boldsymbol{w} \mid \mathbf{y})
    &\sim \mathcal{N}(\bar{\boldsymbol{w}}, A^{-1}),\\
    A &= \sigma_d^{-2} \Phi \Phi^{\top} + \Sigma_p^{-1},\\
    \bar{\boldsymbol{w}}
      &= \sigma_d^{-2} A^{-1} \Phi \mathbf{y}.
\end{align*}
Finally, pushing forward the posterior distribution of the weights through the linear model in Eq.\eqref{eq:GP_WV} yields the push forward posterior (PFP) distribution at a new input $\mathbf{x}^*$~\cite{williams2006gaussian}:
\begin{equation}
    f^* \mid \mathbf{x}^*, X, \mathbf{y}
    \sim \mathcal{N}\!\left(
        \frac{1}{\sigma_d^2}\,
        \boldsymbol{\phi}(\mathbf{x}^*)^{\top} A^{-1} \Phi \mathbf{y},\;
        \boldsymbol{\phi}(\mathbf{x}^*)^{\top} A^{-1} \boldsymbol{\phi}(\mathbf{x}^*)
    \right). \label{eq:gp_pred_dist} 
\end{equation}

\begin{figure}[ht]
\centering
\subcaptionbox{Function space}{\includegraphics[width=0.33\linewidth]{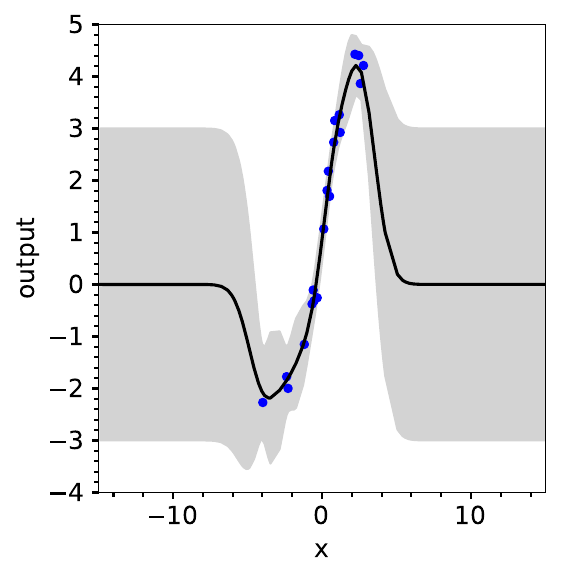}}
\subcaptionbox{Weight space}{\includegraphics[width=0.33\linewidth]{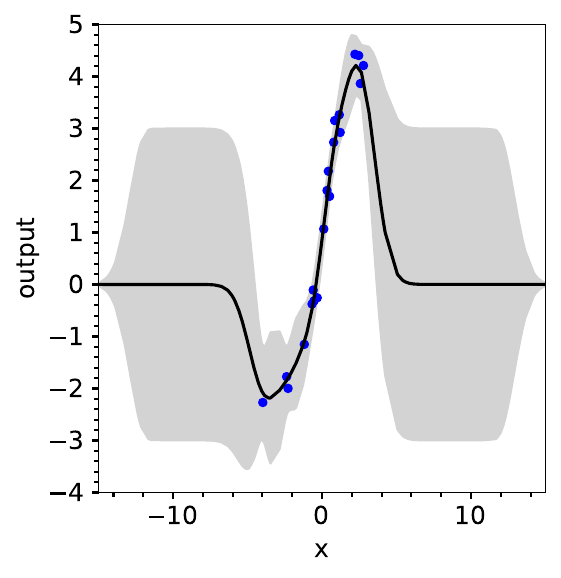}}
\caption{Comparison of PFP mean and plus/minus three standard deviations using FS and WS formulations. GP covariance function hyperparameters are $\sigma=1$ and $l=1$. In both panels, the noise variance is equal to 0.1.}
\label{fig:pfp_fv_vs_wv}
\end{figure}

Truncation at a finite number of basis functions that leads to the collapse of prior-predictive standard deviation in the WS perspective, as seen in Figure~\ref{fig:std_fv_vs_wv}, also translates to the PFP in Eq~\eqref{eq:gp_pred_dist} when the test input $\mathbf{x}^*$ is far away from the training data.
Figure~\ref{fig:pfp_fv_vs_wv} shows the PFP distributions with three standard deviations obtained using the FS and WS formulations. $N=20$ data points were generated using the function $f(x) = 1 + x+\sin{x}$ with $x 
\sim \mathcal{N}(0, 2)$ and corrupted with iid noise of $\sigma^2_d = 0.1$. 
The number of basis functions is fixed to $m=40$ in the WS view.
An exact correspondence is observed between FS and WS in the vicinity of the training data with a collapse in prediction uncertainty using WS perspective as we move far away from the observations. 

To summarize, the WS view allows us to deal with a finite set of weights while maintaining a close correspondence with the nonparametric FS view as long as we do not extrapolate too far away from the observations. 
With this precaution in mind, the WS view can be used in practice as a flexible GP-based embedded model error representation for both linear and non-linear computational models.

\section{Additive Model Error}\label{sec:Additive_error}

In this section, we first recall the general Bayesian framework for handling model error using additive GP representations. \textbf{}We then introduce the OGP formulation, which addresses the well-known issue of confounding between model parameters and the bias term.

\subsection{General Formulation}

In a typical model calibration problem, we are given $N$ observations, collected in a vector $\mathbf{y} \in \mathbb{R}^N$, corresponding to a matrix of $N$ inputs, $X \in \mathbb{R}^{N \times D}$. The data are assumed to be contaminated by iid Gaussian noise with known standard deviation $\sigma_d$, such that 
\begin{equation}
    y(\mathbf{x}) = f_t(\mathbf{x}) + \textcolor{black}{\varepsilon},
\end{equation}
where $f_t(\mathbf{x})$ denotes the unknown true function and $\varepsilon \sim \mathcal{N}(0, \sigma_d^2)$. 
The goal is to infer the model parameters $\boldsymbol{\lambda} \in \Lambda \subset \mathbb{R}^p$ of a computational model $f(\mathbf{x}; \boldsymbol{\lambda})$, hereafter referred to as the fit model, and to subsequently make predictions at new input locations.

Kennedy and O'Hagan~\cite{kennedy2001bayesian} proposed an additive GP approach to explicitly model the discrepancy between the truth and the predictions from the fit model. Their data model is given by
\begin{align}
    y(\mathbf{x}) 
    = \underbrace{f(\mathbf{x}; \boldsymbol{\lambda}) + \delta(\mathbf{x})}_{g(\mathbf{x}; \boldsymbol{\lambda})}
    + \varepsilon, \label{eq:KOH_data_model}
\end{align}
where $\delta(\mathbf{x}) \sim GP(0, k(\mathbf{x}, \mathbf{x}'))$ represents the additive model error (a non-zero mean could be used, but we adopt the zero-mean function for simplicity) and $g(\mathbf{x};\boldsymbol{\lambda})$ denotes the \textit{augmented} fit model. Let $\mathbf{f}_{\boldsymbol{\lambda}}$ denote the vector of model evaluations at locations $X$.
Using Bayes’ rule, the posterior distribution of the model parameters is
\begin{align*}
\pi(\boldsymbol{\lambda} \mid \mathbf{y}) 
\propto \pi(\mathbf{y} \mid \boldsymbol{\lambda}) \, \pi(\boldsymbol{\lambda}),
\end{align*}
where $\pi(\boldsymbol{\lambda})$ is the parameter prior and $\pi(\mathbf{y} \mid \boldsymbol{\lambda})$ is the likelihood.  
From~\eqref{eq:KOH_data_model}, the data satisfy
\[
    y \sim GP\bigl(f(\mathbf{x}; \boldsymbol{\lambda}),\; k(\mathbf{x},\mathbf{x}') + \sigma_d^2 \delta_{\mathbf{x}\mathbf{x}'}\bigr),
\]
where $\delta_{\mathbf{x}\mathbf{x}'}$ is the Kronecker delta function. 
To avoid clutter in the notation, we do not show the GP hyperparameters and assume that they are given, or estimated, \emph{e.g.,} by maximizing the marginal likelihood. 
The parameter posterior may thus be written as  
\begin{align}
    \pi(\boldsymbol{\lambda} \mid \mathbf{y})
    \propto 
    \frac{\pi(\boldsymbol{\lambda})}{|K + \sigma_d^2 I|^{1/2}}
    \exp\!\left(
        -\frac{1}{2}
        (\mathbf{y} - \mathbf{f}_{\boldsymbol{\lambda}})^{\!\top}
        (K + \sigma_d^2 I)^{-1}
        (\mathbf{y} - \mathbf{f}_{\boldsymbol{\lambda}})
    \right) \label{eq:add_fun_param_post}
\end{align}
where $K$ is the $N \times N$ matrix with elements $k(\mathbf{x}_i, \mathbf{x}_j)$. 
Having obtained the parameter posteriors, the posterior distribution of the bias function at some arbitrary location, $\mathbf{x}'$, can now be obtained with~\cite{plumlee2017bayesian}
\begin{align}
    \pi(\delta(\mathbf{x}') \mid \mathbf{y})
    = \int 
        \pi(\delta(\mathbf{x}') \mid \mathbf{y}, \boldsymbol{\lambda})
        \pi(\boldsymbol{\lambda} \mid \mathbf{y})
    \, d\boldsymbol{\lambda}, \label{eq:add_fun_gp_post}
\end{align}
where the distribution $\pi(\delta(\mathbf{x}') \mid \mathbf{y}, \boldsymbol{\lambda})$ follows from the standard GP conditional distribution \cite{williams2006gaussian}. 

\textcolor{black}{The KOH formulation conflates the bias prior and parameter posterior, a challenge widely recognized in the literature as the structural identifiability issue \cite{bayarri2007,han2009simultaneous,gramacy2015calibrating,tuo2016theoretical}.} 
This makes it difficult to compare the contributions of the various parameters to the predictive performance of the model and more importantly assess the overall predictive capability of the fit model.
In the next section, we describe the OGP formulation of~\cite{plumlee2017bayesian}, which provides a principled mechanism to address this issue within a joint Bayesian framework.

\subsection{Formulation with Orthogonal Gaussian Processes}\label{sec:OGP_plumlee}

We note from Eq.~\eqref{eq:add_fun_param_post} that the parameter posterior depends on the bias covariance matrix $K$. 
The method introduced in~\cite{plumlee2017bayesian} modifies this covariance based on a “best” parameter value, $\boldsymbol{\lambda}^*$, defined according to a suitable loss function, so that the parameter posterior lies in the vicinity of $\boldsymbol{\lambda}^*$. 
Specifically, the bias covariance is modified to enforce orthogonality between the discrepancy and the gradient of the computer model evaluated at $\boldsymbol{\lambda}^*$. 

Assuming both the true function and the computer model to be deterministic, the approach begins by defining a loss functional that can uniquely identify the best parameter value. Using the $L^2$ loss weighted by a measure $\mu$ over $\mathcal{X}$~\cite{plumlee2017bayesian},
\begin{align}\label{eq:LL2}
    L_{L^2(\mu)}\{f_t(\cdot) - f(\cdot,\boldsymbol{\lambda})\}
    = \int_{\mathcal{X}} \left(f_t(\mathbf{x}) - f(\mathbf{x};\boldsymbol{\lambda})\right)^2 \, d\mu(\mathbf{x}),
\end{align}
the optimal parameter $\boldsymbol{\lambda}^*$ is defined as 
\begin{align*}
    \boldsymbol{\lambda}^* = \argmin_{\boldsymbol{\lambda} \in \Lambda} L_{L^2(\mu)}\{f_t(\cdot) - f(\cdot,\boldsymbol{\lambda})\}
\end{align*}
where $\Lambda$ denotes an admissible parameter space. 
The true function may then be approximated as
\begin{align*}
    f_t(\mathbf{x}) \approx f(\mathbf{x};\boldsymbol{\lambda}^*) + \delta(\mathbf{x}),
\end{align*}
where $\delta(\mathbf{x})$ represents the model discrepancy. Substituting the above approximation into Eq. \eqref{eq:LL2} yields
\begin{align*}
L_{L^2(\mu)}\{f_t(\cdot) - f(\cdot,\boldsymbol{\lambda})\}
= \int_{\mathcal{X}} 
\left(f(\mathbf{x};\boldsymbol{\lambda}^*) + \delta(\mathbf{x}) 
- f(\mathbf{x};\boldsymbol{\lambda})\right)^2 d\mu(\mathbf{x}).
\end{align*}
Because this loss function is minimized at $\boldsymbol{\lambda}^*$, its gradient with respect to $\boldsymbol{\lambda}$ must vanish at $\boldsymbol{\lambda}^*$, leading to the orthogonality condition~\cite{plumlee2017bayesian}
\begin{align}
\int_{\mathcal{X}} 
\nabla_{\boldsymbol{\lambda}} f(\mathbf{x};\boldsymbol{\lambda})
\big|_{\boldsymbol{\lambda}^*} 
\, \delta(\mathbf{x}) 
\, d\mu(\mathbf{x})
= \mathbf{0}. 
\label{eq:additive_model_ortho_const}
\end{align}
This condition results in $p$ linear constraints on the discrepancy process:
\begin{align}
    \mathcal{L}_k(\delta)
    :=
    \int_{\mathcal{X}}
    \frac{\partial f}{\partial \lambda_k}(\mathbf{x};\boldsymbol{\lambda})
    \big|_{\boldsymbol{\lambda}^*}
    \delta(\mathbf{x}) 
    \, d\mu(\mathbf{x})
    = 0,
    \qquad k=1,\dots,p.
\end{align}

Starting with the prior $\delta(\mathbf{x}) \sim GP(0, k(\mathbf{x},\mathbf{x}'))$, the modified covariance obtained by conditioning the prior GP on these linear constraints is given by~\cite{plumlee2017bayesian,plumlee2018orthogonal}
\begin{align}
k_{\boldsymbol{\lambda}^*}(\mathbf{x},\mathbf{x}')
= k(\mathbf{x},\mathbf{x}')
- h_{\boldsymbol{\lambda}^*}(\mathbf{x})^\top 
  H_{\boldsymbol{\lambda}^*}^{-1}
  h_{\boldsymbol{\lambda}^*}(\mathbf{x}'), 
  \label{eq:add_OGP_cov}
\end{align}
where
\begin{align}
h_{\boldsymbol{\lambda}^*}(\mathbf{x})
&= \int_{\mathcal{X}}
   \nabla_{\boldsymbol{\lambda}}f(\mathbf{x}';\boldsymbol{\lambda})\big|_{\boldsymbol{\lambda}^*}
   \, k(\mathbf{x},\mathbf{x}')
   \, d\mu(\mathbf{x}')
   \in \mathbb{R}^{p}, 
   \label{eq:add_OGP_inth}\\[6pt]
H_{\boldsymbol{\lambda}^*}
&= \int_{\mathcal{X}}\!\!\int_{\mathcal{X}}
   \nabla_{\boldsymbol{\lambda}}f(\mathbf{x}'; \boldsymbol{\lambda})\big|_{\boldsymbol{\lambda}^*}
   \nabla_{\boldsymbol{\lambda}}f(\mathbf{x}; \boldsymbol{\lambda})\big|_{\boldsymbol{\lambda}^*}^\top
   k(\mathbf{x}',\mathbf{x})
   \, d\mu(\mathbf{x}') \, d\mu(\mathbf{x})
   \in \mathbb{R}^{p \times p}. 
   \label{eq:add_OGP_intH}
\end{align}

In practice, $\boldsymbol{\lambda}^*$ is obtained by solving the least squares (LS) problem using the observed data $\mathbf{y}$ and the computer model. The integrals in Eqs.~\eqref{eq:add_OGP_inth}--\eqref{eq:add_OGP_intH} are typically computed numerically, except in certain special cases where closed-form expressions are available (see~\cite{plumlee2018orthogonal}). Because $k_{\boldsymbol{\lambda}^*}$ incorporates the orthogonality condition implied by Eq.~\eqref{eq:additive_model_ortho_const}, adopting the prior
\[
    \delta(\mathbf{x}) \sim GP(0, k_{\boldsymbol{\lambda}^*}(\mathbf{x},\mathbf{x}'))
\]
prevents the discrepancy from being confounded with the best-fit parameter value $\boldsymbol{\lambda}^*$. 
This results in a posterior distribution $\pi(\boldsymbol{\lambda} \mid \mathbf{y})$ that lies in the vicinity of $\boldsymbol{\lambda}^*$. 
Also, the posterior distribution in the FS view using the OGP framework has the same form as Eq.~\eqref{eq:add_fun_param_post}, with $K$ replaced by $K_{\boldsymbol{\lambda}^*}$. 

The WS view may also be used by expanding the discrepancy in the eigenfunctions of $k_{\boldsymbol{\lambda}^*}$. 
To see how orthogonality modifies the basis functions is WS GP representation, $\delta_{\boldsymbol{w}}(\mathbf{x}) = \boldsymbol{\phi}(\mathbf{x})^\top \boldsymbol{w}, \, \boldsymbol{w} \in \mathbb{R}^m$, consider a linear model $f(\mathbf{x}; \boldsymbol{\lambda}) = a(\mathbf{x})^\top \boldsymbol{\lambda}$, where $a(\mathbf{x}) = [a_1(\mathbf{x}), \cdots, a_p(\mathbf{x})]^\top$. 
Substituting the GP basis expansion and model gradients in Eq.~\eqref{eq:additive_model_ortho_const} results in 
\begin{align*}
    \int_{\mathcal{X}} \phi_k(\mathbf{x}) a_i(\mathbf{x}) d\mu(\mathbf{x}) = 0, \quad k=0,\cdots,m-1 \, , i=1,\cdots,p  
\end{align*}
Thus, each basis function is orthogonalized with respect to the $a_i(\mathbf{x})$ functions.  
In the next section, we build upon this framework for orthogonality in model error embedding with WS GPs.

\section{Model Error Embedding with Orthogonal Gaussian Processes}\label{sec:EMB_OGP}

As proposed in~\cite{sargsyan2015statistical}, the idea of internal model correction is to embed model error terms within the computer model so that propagation of model error satisfies the constraints of the model leading to physically consistent predictions.
Along the same lines, we extend the methodology of~\cite{sargsyan2015statistical} and embed our model $f$ with a WS GP to make the model error characterization more flexible. 
The data model can be written as 
\begin{align}\label{eq:emgp_datamod}
    y = \underbrace{\tilde{f}(\mathbf{x}; \boldsymbol{\lambda}, \delta_{\boldsymbol{w}}(\mathbf{x}))}_{g(\mathbf{x}; \boldsymbol{\lambda}, \boldsymbol{w})} + \varepsilon,
\end{align}
where $\delta_{\boldsymbol{w}}(\mathbf{x}) = \boldsymbol{\phi}(\mathbf{x})^\top \boldsymbol{w}$ is the WS GP whose basis functions correspond to a covariance function $k(\mathbf{x}, \mathbf{x}')$. 
In the case of $\tilde{f}$ being a linear model, we focus on additive embedding in order to retain linearity in $(\bm{\lambda},\bm{w})$. As a result, this case reduces to the KOH framework. Similarly, our framework for additive model error embedding with OGP when applied to linear models simply recovers the OGP framework of~\cite{plumlee2017bayesian} described in the previous section. 
Therefore, for the subsequent development of our methodology we focus only on non-linear models.

\textcolor{black}{The specific selection of sub-models for embedding with model error GP is a problem dependent task and a choice of the modeller.
For instance, one may be interested in the attribution of error introduced by model components where simplifications, approximations or phenomenological modeling have been employed \cite{sargsyan2019embedded}.
Such embeddings, using random variables, have been demonstrated in the context of chemical kinetics in combustion to calibrate the Arrhenius rate parameters of a simplified reaction mechanism using data from a detailed kinetics model \cite{sargsyan2015statistical,sargsyan2019embedded}, as well as in large eddy simulations for scramjet design, where different treatments of turbulence characterization and domain geometries have been assessed through model error embedding \cite{huan2017global}.}  

\textcolor{black}{
As an example, consider a general PDE operator, $\mathcal{G}$, acting on a state variable, $u$, such that
\begin{align}
    \mathcal{G}(u(\mathbf{x}))=s(u(\mathbf{x}),\mathbf{x};\boldsymbol{\lambda})
\end{align}
where $s$ represents a non-linear source term with parameters $\boldsymbol{\lambda}$. 
Data can be some noisy observations of a functional, $f$, of the state variables: 
\begin{align}
y = f(u(\mathbf{x});\boldsymbol{\lambda}) + \varepsilon
\end{align}
One can think of augmenting $s$ with model error terms to account for uncertainty introduced by simplifying assumptions, so that
\begin{align}
    \mathcal{G}(u(\mathbf{x})) =  \tilde{s}(u(\mathbf{x}), \mathbf{x};\boldsymbol{\lambda}, \delta_w(\mathbf{x}))
\end{align}
which also modifies the QoI functional as 
\begin{align}
y = \tilde{f}(u(\mathbf{x});\boldsymbol{\lambda},\delta_w(\mathbf{x})) + \varepsilon.
\end{align}
Now, by inferring $\{\boldsymbol{\lambda},\boldsymbol{w}\}$ jointly, one can assess the importance of structural modifications to $s$. 
A concrete example of this with advection-diffusion-reaction (ADR) PDE is provided in Sec.~\ref{sec:appl_ADR}. 
One can also envision an additive embedding in the parameters of $s$, 
\begin{align}
    \mathcal{G}(u(\mathbf{x})) = s(u(\mathbf{x}), \mathbf{x};\boldsymbol{\lambda}+\mathbf{\delta}_w(\mathbf{x}))
\end{align}
in which case, the approach closely resembles functional calibration \cite{brown2018nonparametric,tuo2023reproducing}. 
When multiple such embeddings can be made one could use Bayesian model selection based on evidence computation to determine the best probabilistic embedding \cite{sargsyan2019embedded}.}

Following Bayes rule, the joint posterior of $(\boldsymbol{\lambda}, \boldsymbol{w})$, according to the data model in Eq.~\eqref{eq:emgp_datamod}, is given by 
\begin{align}
    \pi(\boldsymbol{\lambda}, \boldsymbol{w} | \mathbf{y}) \propto \frac{\pi(\boldsymbol{\lambda}, \boldsymbol{w})}{(2\pi\sigma_d^2)^{N/2}} \times \exp{\left(-\frac{\sum_{i=1}^{N}\left(y_i - \tilde{f}\left(\mathbf{x}_i; \boldsymbol{\lambda}, \boldsymbol{\phi}(\mathbf{x}_i)^\top\boldsymbol{w}\right)\right)^2}{2 \sigma_d^2}\right)}.
    \label{eq:em_GP_bayes}
\end{align}
Sampling of the above posterior will again not avoid confounding of model parameters and bias weights. 
To achieve the required disambiguation, we propose two approaches -- one based on model linearization (LOGP) and the other based on regularization (ROGP) -- that are built on the same OGP foundations introduced in Section \ref{sec:OGP_plumlee}. 

\subsection{Linearized OGP}

Similar to OGP for additive model error, we start by defining a best parameter value, $\boldsymbol{\lambda}^*$, according to the $L^2$ loss weighted by a measure $\mu$ over $\mathcal{X}$ 
\begin{align*}
    \boldsymbol{\lambda}^* = \argmin_{\boldsymbol{\lambda} \in \Lambda} L_{L^2(\mu)}\{f_t(\cdot) - f(\cdot,\boldsymbol{\lambda})\}.
\end{align*}
To approximate the true function using the predictive function, $\tilde{f}$, we first linearize $\tilde{f}$ using a first-order Taylor series expansion about $\boldsymbol{w}^*$ (yet to be defined) and $\boldsymbol{\lambda}^*$:
\begin{align*}
    g(\mathbf{x}; \boldsymbol{\lambda}, \boldsymbol{w}) &\approx \tilde{f}(\mathbf{x}; \boldsymbol{\lambda}^*, \delta_{\boldsymbol{w}^*}(\mathbf{x})) + \nabla_{\boldsymbol{\lambda}} \tilde{f}\big|_{(\boldsymbol{\lambda}^*, \boldsymbol{w}^*)}^\top \Delta \boldsymbol{\lambda} + \nabla_{\boldsymbol{w}} \tilde{f}\big|_{(\boldsymbol{\lambda}^*, \boldsymbol{w}^*)}^\top \Delta \boldsymbol{w} \nonumber \\
    &= \tilde{f}(\mathbf{x}; \boldsymbol{\lambda}^*, \delta_{\boldsymbol{w}^*}(\mathbf{x})) + \nabla_{\boldsymbol{\lambda}} \tilde{f}\big|_{(\boldsymbol{\lambda}^*, \boldsymbol{w}^*)}^\top \Delta \boldsymbol{\lambda} + \partial_{\delta_{\boldsymbol{w}}(\mathbf{x})} \tilde{f}\big|_{(\boldsymbol{\lambda}^*, \boldsymbol{w}^*)} \nabla_{\boldsymbol{w}} \delta_{\boldsymbol{w}}(\mathbf{x})\big|_{\boldsymbol{w}^*}^\top \Delta \boldsymbol{w} \nonumber \\
    &= \tilde{f}(\mathbf{x}; \boldsymbol{\lambda}^*, \delta_{\boldsymbol{w}^*}(\mathbf{x})) + \nabla_{\boldsymbol{\lambda}} \tilde{f}\big|_{(\boldsymbol{\lambda}^*, \boldsymbol{w}^*)}^\top \Delta \boldsymbol{\lambda} + \partial_{\delta_{\boldsymbol{w}}(\mathbf{x})} \tilde{f}\big|_{(\boldsymbol{\lambda}^*, \boldsymbol{w}^*)}\sum_{j=0}^{m-1} \phi_j(\mathbf{x}) (w_j - w_j^*),
\end{align*}
where $\Delta \boldsymbol{\lambda} = \boldsymbol{\lambda} - \boldsymbol{\lambda}^*$, $\Delta \boldsymbol{w} = \boldsymbol{w} - \boldsymbol{w}^*$, $\nabla_{\boldsymbol{\lambda}} \tilde{f}:= \nabla_{\boldsymbol{\lambda}}\tilde{f}(\mathbf{x}; \boldsymbol{\lambda}, \delta_{\boldsymbol{w}}(\mathbf{x}))$ and $\partial_{\delta_{\boldsymbol{w}}(\mathbf{x})} \tilde{f}:=\partial_{\delta_{\boldsymbol{w}}(\mathbf{x})}\tilde{f}(\mathbf{x}; \boldsymbol{\lambda}, \delta_{\boldsymbol{w}}(\mathbf{x}))$. 
Now, setting $\boldsymbol{w}^* = \boldsymbol{0}$ and evaluating the above linearized function at $\boldsymbol{\lambda} = \boldsymbol{\lambda}^*$, we get
\begin{align}
    f_t(\mathbf{x}) &\approx f(\mathbf{x}; \boldsymbol{\lambda}^*) + \partial_{\delta_{\boldsymbol{w}}(\mathbf{x})} \tilde{f}\big|_{(\boldsymbol{\lambda}^*, \boldsymbol{0})}\sum_{j=0}^{m-1} \phi_j(\mathbf{x})w_j \nonumber \\
    &= f(\mathbf{x}; \boldsymbol{\lambda}^*) + \partial_{\delta_{\boldsymbol{w}}(\mathbf{x})} \tilde{f}\big|_{(\boldsymbol{\lambda}^*, \boldsymbol{0})} \boldsymbol{\phi}(\mathbf{x})^\top \boldsymbol{w}. \label{eq:em_LOGP_lin}
\end{align}
\textcolor{black}{This means that the true function is approximated by a first-order perturbation of the augmented fit model around the fit model evaluated at $\boldsymbol{\lambda}^*$.}

With this separation of the computer model and model error contributions, the orthogonality constraint on $\delta_{\boldsymbol{w}}(x)$ can be derived by substituting Eq.~\eqref{eq:em_LOGP_lin} into the loss function:
\begin{align*}
    L_{L^2(\mu)}\{f_t(.) - f(.; \boldsymbol{\lambda})\} &= \int_{\mathcal{X}} \left(f(\mathbf{x}; \boldsymbol{\lambda}^*) + \partial_{\delta_{\boldsymbol{w}}(\mathbf{x})}\tilde{f} \big|_{(\boldsymbol{\lambda^*}, \mathbf{0})} \delta_{\boldsymbol{w}}(\mathbf{x}) - f(\mathbf{x}; \boldsymbol{\lambda})\right)^2 d\mu(\mathbf{x}).
\end{align*}
Evaluating the gradient of the loss with respect to $\boldsymbol{\lambda}$ at $\boldsymbol{\lambda}^*$ results in the required constraints, 
\begin{align}
    \int_{\mathcal{X}} \partial_{\delta_{\boldsymbol{w}}(\mathbf{x})}\tilde{f} \big|_{(\boldsymbol{\lambda}^*, \mathbf{0})}\nabla_{\boldsymbol{\lambda}} f\big|_{\boldsymbol{\lambda}^*} \delta_{\boldsymbol{w}}(\mathbf{x}) d\mu(\mathbf{x}) = \mathbf{0}.
\end{align}
We see that, as in the additive GP case, we have $p$ linear constraints involving the embedded GP:
\begin{align}
\mathcal{L}_k(\delta_{\boldsymbol{w}}(\mathbf{x})) := \int_{\mathcal{X}} \partial_{\delta_{\boldsymbol{w}}(\mathbf{x})}\tilde{f} \big|_{(\boldsymbol{\lambda}^*, \mathbf{0})} \partial_{\lambda_k} f\big|_{\boldsymbol{\lambda}^*} \delta_{\boldsymbol{w}}(\mathbf{x}) d\mu(\mathbf{x}) = 0, \qquad k=1, \cdots, p \: . \label{eq:LOGP_constraints}
\end{align}
The modified covariance for LOGP, $k_{\boldsymbol{\lambda}^*}(\mathbf{x}, \mathbf{x}')$, obtained by conditioning the GP prior on the linear constraints in Eq.~\eqref{eq:LOGP_constraints}, retains the structure of Eq.~\eqref{eq:add_OGP_cov}. 
However, the integrals in Eqs.~\eqref{eq:add_OGP_inth}--\eqref{eq:add_OGP_intH} now also incorporate the gradient, $\partial_{\delta_{\boldsymbol{w}}(\mathbf{x})}\tilde{f} \big|_{(\boldsymbol{\lambda}^*, \mathbf{0})}$, which arises from model linearization:
\begin{align}
h_{\boldsymbol{\lambda}^*}(\mathbf{x})
&= \int_{\mathcal{X}} \partial_{\delta_{\boldsymbol{w}}(\mathbf{x})} \tilde{f}\big|_{(\boldsymbol{\lambda}^*, \mathbf{0})}(\mathbf{x}')
\nabla_{\boldsymbol{\lambda}}f(\mathbf{x}';\boldsymbol{\lambda})\big|_{\boldsymbol{\lambda}^*}
\, k(\mathbf{x},\mathbf{x}')
\, d\mu(\mathbf{x}')
\in \mathbb{R}^{p}, 
\label{eq:em_LOGP_inth}\\[6pt]
H_{\boldsymbol{\lambda}^*}
&= \int_{\mathcal{X}}\!\!\int_{\mathcal{X}}
\partial_{\delta_{\boldsymbol{w}}(\mathbf{x})} \tilde{f}\big|_{(\boldsymbol{\lambda}^*, \mathbf{0})}(\mathbf{x}')
\nabla_{\boldsymbol{\lambda}}f(\mathbf{x}';\boldsymbol{\lambda})\big|_{\boldsymbol{\lambda}^*}
 \nonumber\\
&\quad \quad \times \partial_{\delta_{\boldsymbol{w}}(\mathbf{x})} \tilde{f}\big|_{(\boldsymbol{\lambda}^*, \mathbf{0})}(\mathbf{x})
\nabla_{\boldsymbol{\lambda}}f(\mathbf{x};\boldsymbol{\lambda})\big|_{\boldsymbol{\lambda}^*}^\top
\,
k(\mathbf{x}',\mathbf{x})
\, d\mu(\mathbf{x}') \, d\mu(\mathbf{x})
\in \mathbb{R}^{p \times p}. 
\label{eq:em_LOGP_intH}
\end{align}
We can now obtain the WS GP for $k_{\boldsymbol{\lambda}^*}(\mathbf{x}, \mathbf{x}')$, numerically, which can be used for inference in Eq.~\eqref{eq:em_GP_bayes}. 

\subsection{Regularized OGP}
A different approach to enforce the orthogonality constraints is to use regularization terms with the prior. 
The advantage of this approach is that the constraints in the embedded GP can be applied without any linearization of the model. 
In addition, unlike the LOGP construction, this approach leaves the eigenfunctions of the original covariance kernel unchanged. Consequently, one may exploit analytical eigenfunctions, such as those available for the SQE kernel under a Gaussian measure on the inputs, thereby avoiding the numerical errors that often arise in estimating eigenfunctions. To construct the necessary regularization terms, we begin again by approximating the unknown true function but without any model linearization, so that
\begin{align*}
    f_t(\mathbf{x}) \approx \tilde{f}(\mathbf{x}; \boldsymbol{\lambda}^*, \boldsymbol{\phi}(\mathbf{x})^\top \boldsymbol{w}).
\end{align*}
Now, we expand the loss as 
\begin{align*}
    L_{L^2(\mu)}\{f_{t}(.) - f(., \boldsymbol{\lambda})\} = \int_{\mathcal{X}} \left(\tilde{f}(\mathbf{x}; \boldsymbol{\lambda}^*, \boldsymbol{\phi}(\mathbf{x})^\top \boldsymbol{w}) - f(\mathbf{x}; \boldsymbol{\lambda}) \right)^2 d\mu(\mathbf{x}).
\end{align*}
Evaluating the gradient of this loss with respect to $\boldsymbol{\lambda}$ at $\boldsymbol{\lambda}^*$ results in the required constraints
\begin{align*}
    \int_{\mathcal{X}} \left[\tilde{f}(\mathbf{x}; \boldsymbol{\lambda}^*, \boldsymbol{\phi}(\mathbf{x})^\top \boldsymbol{w}) - f(\mathbf{x}; \boldsymbol{\lambda}^*)\right] \nabla_{\boldsymbol{\lambda}} f(\mathbf{x}; \boldsymbol{\lambda})\big|_{\boldsymbol{\lambda}^*} d\mu(\mathbf{x}) = \mathbf{0}.
\end{align*}
Again, we have $p$ constraints but they are \textit{non-linear} in $\delta_{\boldsymbol{w}}(\mathbf{x})$:
\begin{align}
\mathcal{R}_k(\boldsymbol{\lambda}^*, \delta_{\boldsymbol{w}}(\mathbf{x})) := \int_{\mathcal{X}}  \left[\tilde{f}(\mathbf{x}; \boldsymbol{\lambda}^*, \boldsymbol{\phi}(\mathbf{x})^\top \boldsymbol{w}) - f(\mathbf{x}; \boldsymbol{\lambda}^*)\right] \partial_{\lambda_k} f(\mathbf{x}; \boldsymbol{\lambda})\big|_{\boldsymbol{\lambda}^*} d\mu(\mathbf{x}) = 0, \quad k=1, \cdots, p.
\end{align}
To enforce these constraints, regularization terms are used along with the prior, resulting in a joint log-posterior written as  
\begin{align}
\log{\pi(\boldsymbol{\lambda}, \boldsymbol{w} \mid \mathbf{y})} \propto &\ \log \pi(\boldsymbol{\lambda}, \boldsymbol{w}) - \sum_{i=1}^{N}\left[\frac{1}{2}\log(2 \pi \sigma_d^2) + \frac{1}{2\sigma_d^2}\left\{y_i - \tilde{f}(\mathbf{x}_i; \boldsymbol{\lambda}, \boldsymbol{\phi}(\mathbf{x}_i)^\top\boldsymbol{w})\right\}^2\right]\nonumber \\
&\ - \frac{1}{2} \sum_{k=1}^{p} \alpha_k \mathcal{R}_k(\boldsymbol{\lambda}^*, \delta_{\boldsymbol{w}}(\mathbf{x}))^2
\label{eq:rogp_log_post}
\end{align}
where $\{\alpha_k\}$ denote the penalty parameters. 
The values of $\boldsymbol{\alpha} = \{\alpha_k\}$ must be decided according to how strongly one wishes to enforce the orthogonality constraint. The larger the value of $\alpha_k$, the closer will be the parameter posterior to ${\lambda}_k^*$. 
However, our numerical experiments also demonstrate that large values of $\boldsymbol{\alpha}$ introduce strong correlations among the GP weights, leading to larger compute times with Markov chain Monte Carlo (MCMC) sampling. 
Therefore, we recommend choosing $\boldsymbol{\alpha}$ so that the parameter posteriors are ``close enough" to $\boldsymbol{\lambda}^*$ as deemed by the user, while avoiding an overly expensive MCMC sampling.  

\section{Likelihood Informed Subspace}\label{sec:LIS}

As demonstrated in Section \ref{sec:GP_WV}, achieving a close correspondence between the WS and FS formulations requires retaining a sufficiently large number of basis functions and associated weights, especially when making predictions far from the training data. However, retaining many weights substantially increases the computational cost of MCMC sampling, since the resulting inverse problem becomes high-dimensional. In practice, given data size limitations, only a subset of the model parameters and GP weights are likely to be strongly informed by the data, while the remaining weights, typically those corresponding to higher-order basis functions, would have posteriors that remain close to their priors. 
This observation motivates the use of the Likelihood-Informed Subspace (LIS) method~\cite{spantini2015optimal,cui2014likelihood}, which allows the retention of a large basis set while reducing the effective dimensionality of the Bayesian inverse problem. Beyond dimensionality reduction, LIS provides a principled mechanism for combining the posterior of the data-informed parameters, obtained via projection onto the LIS, with the prior of the uninformed parameters, obtained via projection onto the complementary subspace (CS).

LIS has been developed for both linear and nonlinear model constructions~\cite{spantini2015optimal,cui2014likelihood}.
\textcolor{black}{In the case of linear forward models with Gaussian prior and likelihood, it was observed that, consistent with the ill-posed nature of inverse problems, the posterior covariance is a low-rank update of the prior covariance. 
As a result, projecting the likelihood function onto a low-dimensional likelihood-informed subspace provided useful means for approximating the Bayesian update, particularly in high-dimensional problems \cite{spantini2015optimal}.
The optimal projector, which spans the LIS, is found using the eigendecomposition of the prior-preconditioned Hessian of the data misfit function.} 

In the extension to non-linear setting, the method identifies a low-dimensional subspace in the parameter space, referred to as the \emph{global} LIS, along which the likelihood most strongly alters the prior. 
This global LIS is estimated as the Monte Carlo mean of \emph{local} LISs from regions of high posterior probability
\textcolor{black}{These local LISs are constructed at uncorrelated parameter samples identified using an adaptive procedure initialized at the LIS-projected posterior mode, employing a sequence of subspace MCMC chains.}
The posterior can then be approximated as the product of a reduced posterior supported on the global LIS and the prior marginalized onto its CS. 
With calibration parameters $\boldsymbol{\theta} = (\boldsymbol{\lambda}, \boldsymbol{w}) \in \mathbb{R}^{p+m}$, the posterior admits the approximation
\begin{align}
\pi(\boldsymbol{\theta} | \mathbf{y}) \approx \hat{\pi}(\boldsymbol{\theta} | \mathbf{y}) 
&\propto \pi(\mathbf{y}|P_r\boldsymbol{\theta}) \pi(\boldsymbol{\theta}) \nonumber \\
&= \pi(\mathbf{y} | U_r \boldsymbol{\theta}_r) \pi_r(\boldsymbol{\theta}_r) \pi_{\perp}(\boldsymbol{\theta}_{\perp}) \nonumber \\
&= \pi(\boldsymbol{\theta}_r | \mathbf{y})\pi_{\perp}(\boldsymbol{\theta}_{\perp}) \label{eq:LIS_redpost}
\end{align}
where $P_r$ denotes the rank-$r$ projector onto the global LIS; $U_r$ denotes the global LIS basis with $\boldsymbol{\theta}_r$ representing the corresponding coordinates and $\boldsymbol{\theta}_{\perp}$
denotes the global CS coordinates.
Thus, the effective dimension for MCMC sampling is reduced to $r \le p+m$. The value of $r$ is chosen to retain all meaningful/significant likelihood-informed directions.

We provide a summary of the method in Appendix~\ref{sec:LIS_sum} and refer the reader to~\cite{spantini2015optimal, cui2014likelihood} for further details. 

\section{Applications}\label{sec:Appl}

Next we demonstrate the key features of our methodology through three examples -- a linear model, a non-linear model with interacting sub-models, and an ADR PDE. 
In each case, the goal is to calibrate a fit model with respect to a truth model while accounting for the model error using the embedded GP (additive GP in the linear case). 
\textcolor{black}{A uniform measure is placed on $\mathbf{x}$ and the basis functions are estimated numerically for the SQE covariance function.}  
In all the applications, we compare the push forward posterior through the fit model on its own, denoted by $\text{PFP-}f$, the push forward posterior through the GP-augmented fit model, denoted by $\text{PFP-}g$, and the posterior predictive, denoted by PP, which includes both the GP-augmented fit model and the measurement noise.
This comparison will allow us to demonstrate the key strengths of the proposed method which are: (i) the orthogonality constraints on the embedded GP lead to predictive distributions obtained with the fit model on its own that are meaningful, in the sense that they closely follow predictions obtained with the least-squares estimates of the model parameters, (ii) to the extent allowed by the fit model structure, the embedded GP provides flexibility in capturing the spatial correlations of the model error and hence addressing the structural deficiencies of the fit model. 
The No-U-Turn-Sampler (NUTS) algorithm~\cite{hoffman2014no} implemented in the python package \texttt{PyMC}~\cite{pymc2023} is used for all the MCMC results obtained with ROGP.
For LOGP with LIS, we use a combination of \texttt{emcee}~\cite{foreman2013emcee} and \texttt{PyMC}. 

\subsection{Linear Model}
Let the truth model,
\begin{equation*}
    f_{t}(x) = 2 + 2x + 3x^2 - 5x^3,
\end{equation*}
and the fit model, 
\begin{align*}
    f(x; \boldsymbol{\lambda}) = \lambda_0 + \lambda_1 x,
\end{align*}
be defined on $x \in [-3, 3]$ with $\boldsymbol{\lambda} = (\lambda_0, \lambda_1)$ denoting the calibration parameters. 
The measurement data are generated by evaluating the truth model and adding iid Gaussian noise with standard deviation $\sigma_d = 0.2$.  

\begin{figure}[ht]
\centering
\subcaptionbox{Marginals of $\lambda_0$ \label{fig:lin_KOH_marg_l0}}{\includegraphics[width=0.32\linewidth]{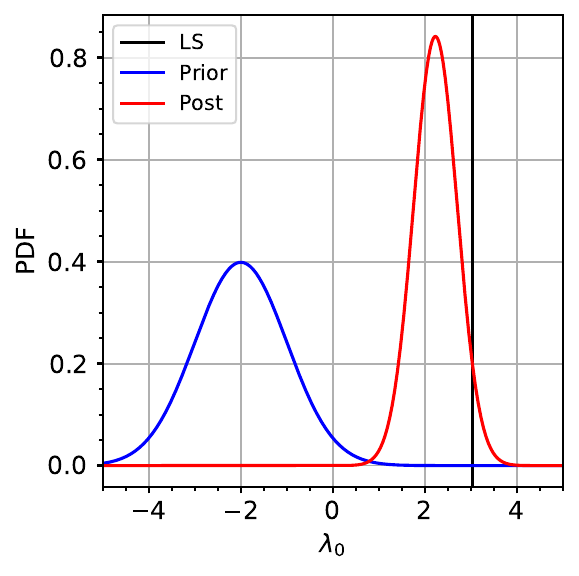}}
\subcaptionbox{Marginals of $\lambda_1$ \label{fig:lin_KOH_marg_l1}}{\includegraphics[width=0.32\linewidth]{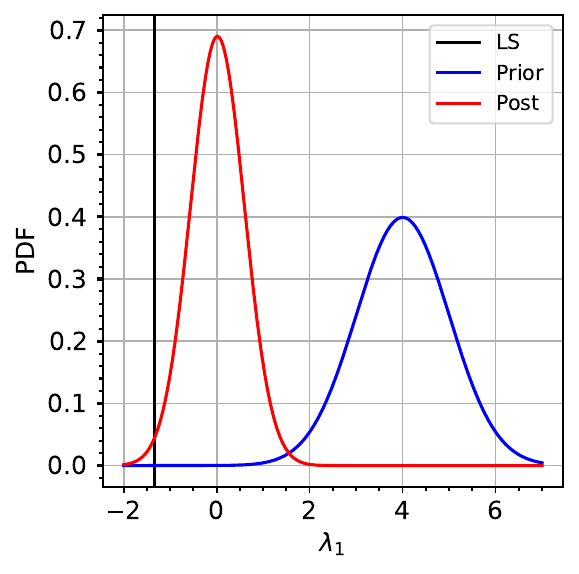}}
\subcaptionbox{Push forward predictions \label{fig:lin_KOH_pf}}{\includegraphics[width=0.32\linewidth]{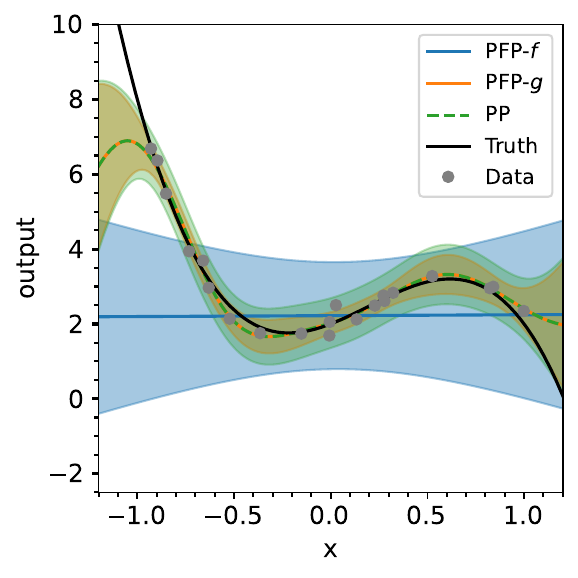}}
\caption{Frames (a) and (b) show marginal priors, posteriors and least-squares estimates for $\lambda_0$ and $\lambda_1$ obtained with KOH. Panel (c) shows the truth function, data points and push forward posterior predictions. 
PFP lines show the mean prediction and bands show $\pm$ three standard deviations.}
\label{fig:lin_KOH_marg_pf}
\end{figure}

\begin{figure}[ht]
\centering
\subcaptionbox{Marginals of $\lambda_0$ \label{fig:lin_OGP_marg_l0}}{\includegraphics[width=0.32\linewidth]{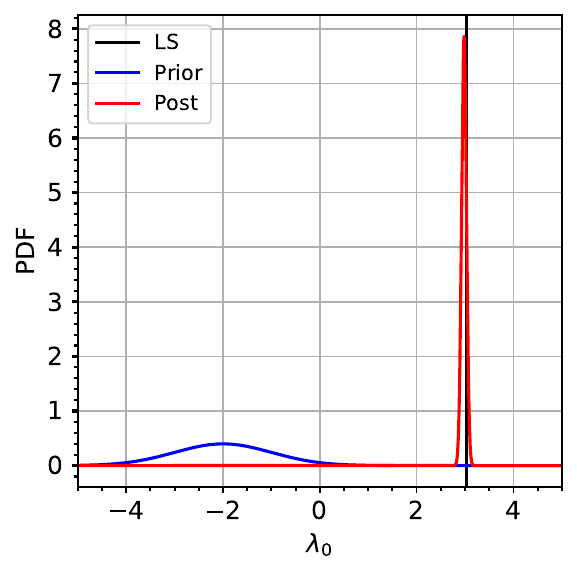}}
\subcaptionbox{Marginals of $\lambda_1$ \label{fig:lin_OGP_marg_l1}}{\includegraphics[width=0.32\linewidth]{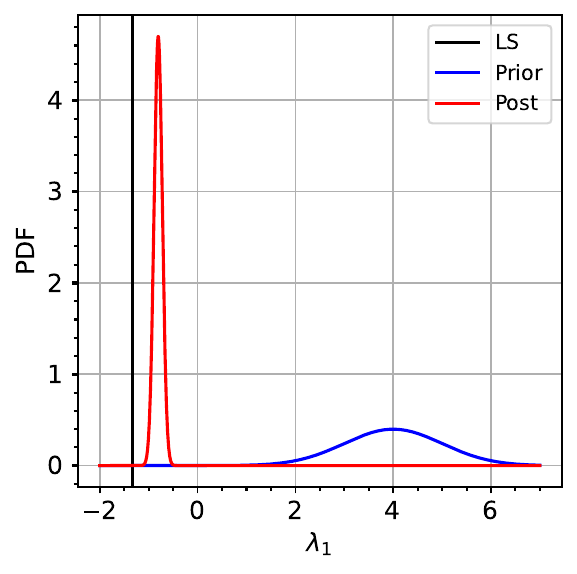}}
\subcaptionbox{Push forward predictions \label{fig:lin_OGP_pf}}{\includegraphics[width=0.32\linewidth]{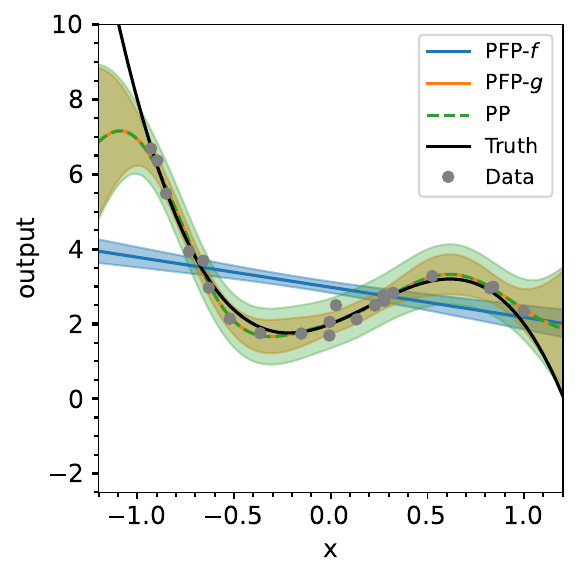}}
\caption{Frames (a) and (b) show marginal priors, posteriors and least-squares estimates for $\lambda_0$ and $\lambda_1$ obtained with OGP. Panel (c) shows the truth function, data points and push forward posterior predictions. 
PFP lines show the mean prediction and bands show $\pm$ three standard deviations.}
\label{fig:lin_OGP_marg_pf}
\end{figure}

We first demonstrate the model calibration using the KOH method to highlight its inability to disambiguate between model and bias predictions. 
The GP augmented fit model is given by 
\begin{align*}
g(x; \boldsymbol{\lambda}, \boldsymbol{w}) = \lambda_0 + \lambda_1 x + \boldsymbol{\phi}(x)^\top\boldsymbol{w}.
\end{align*}
We start with $N = 20$ data points across the $x$-axis, drawn randomly from a uniform distribution on $[-1, 1]$, and use a SQE kernel with $l=0.3$ and $\sigma_f = 1$. 
We restrict the data set to this domain in order to show the behaviour of predictive distributions when extrapolating away from the data.  
The GP expansion uses $m=20$ weights (this choice is based on a convergence study that is detailed later in this section). 
A Gaussian prior is placed on the model parameters,
\[
    \pi(\boldsymbol{\lambda}) \sim \mathcal{N}\big( [-2,\, 4],\, \mathrm{diag}(1,\, 1) \big).
\]
As described previously in Section \ref{sec:GP_WV}, the prior on the GP weights is Gaussian with a diagonal covariance matrix consisting of the kernel eigenvalues in order to maintain equivalence with the corresponding function view.  
We also note that the posteriors in this linear model case are obtained analytically, for both KOH and OGP,  since both the prior and the likelihood are Gaussian. 
Figures \ref{fig:lin_KOH_marg_l0} and \ref{fig:lin_KOH_marg_l1} present the marginal posterior distributions of the parameters $\boldsymbol{\lambda}$.  
The solid black lines denote the optimal parameter values obtained by solving the corresponding least squares problem.  
While the marginal posteriors indicate a degree of information gain relative to the priors, becoming narrower, they do still indicate a significant degree of uncertainty despite the highly-informative data set, and are further away from the LS estimates in comparison to the OGP results in Figure~\ref{fig:lin_OGP_marg_pf} discussed later below.  
Figure~\ref{fig:lin_KOH_pf} shows the PFP and PP distributions. 
Because the marginal parameter posteriors are weakly informed, the resulting predictions from the fit model are not reliable.  
As expected, the GP-augmented fit model, shown in orange, accurately captures the trend of the data given the flexibility of the GP correction and lastly, the posterior predictive plotted in green fully ``covers" all the data points. 
\vspace{1em}
\vspace{-1em}

Figure \ref{fig:lin_OGP_marg_pf} shows the results for marginal posteriors, PFP, and PP distributions obtained with the OGP method. 
The parameter posteriors are now better informed by the data and lie in the vicinity of the LS parameter values. 
Accordingly, the push forward posterior through the fit model on its own provides meaningful predictions, as the model now has full latitude in learning from the data, while letting the additive GP capture the residual discrepancy. 
We note that, given the infinite support of our prior, the LS estimate is expected to be on top of the MAP estimate without model error consideration in the infinite data limit, making it an ideal reference. 

\begin{figure}[ht]
    \centering
    \subcaptionbox{KOH \label{fig:lin_KOH_joint}}{\includegraphics[width=0.37\linewidth]{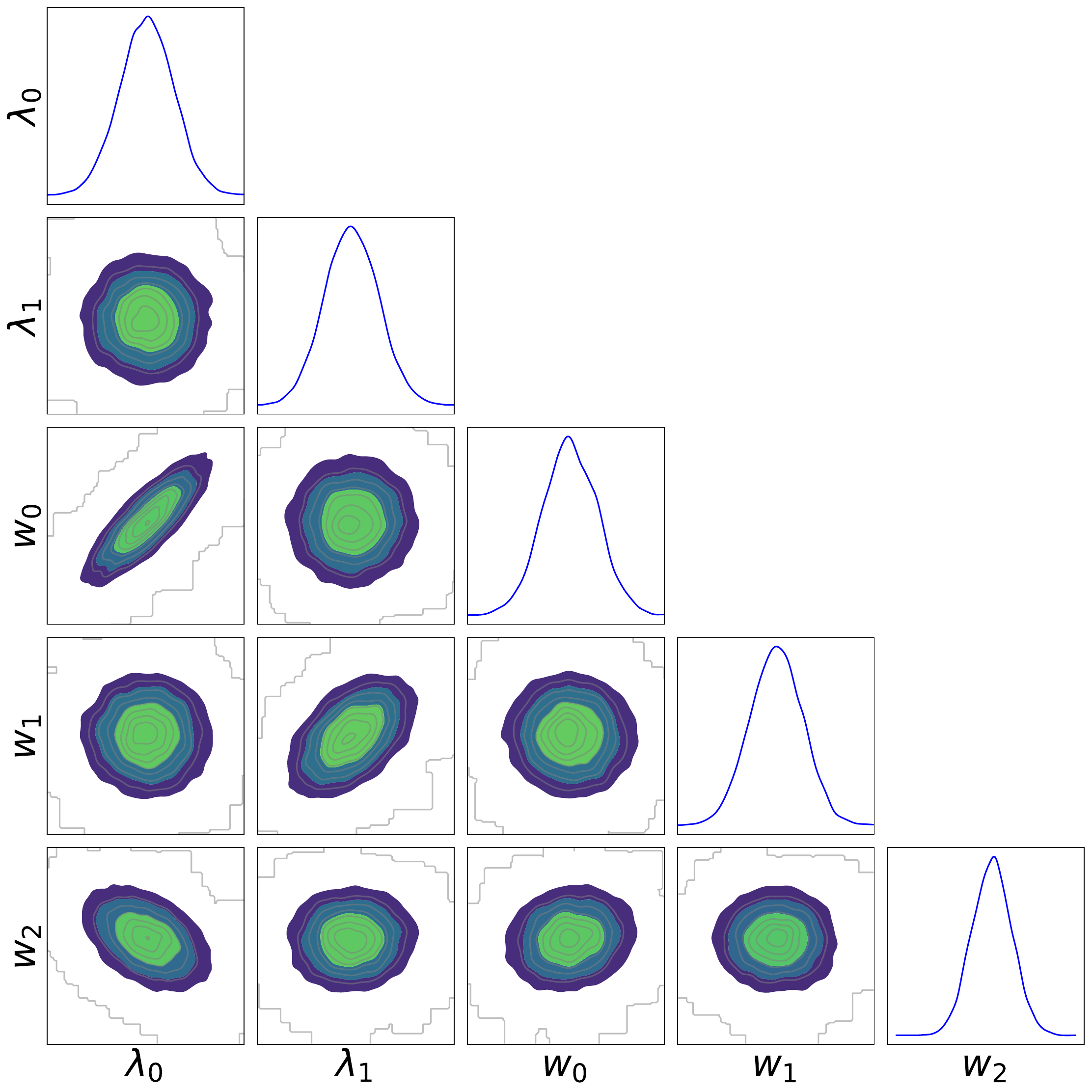}}
    \subcaptionbox{ OGP\label{fig:lin_OGP_joint}}{\includegraphics[width=0.37\linewidth]{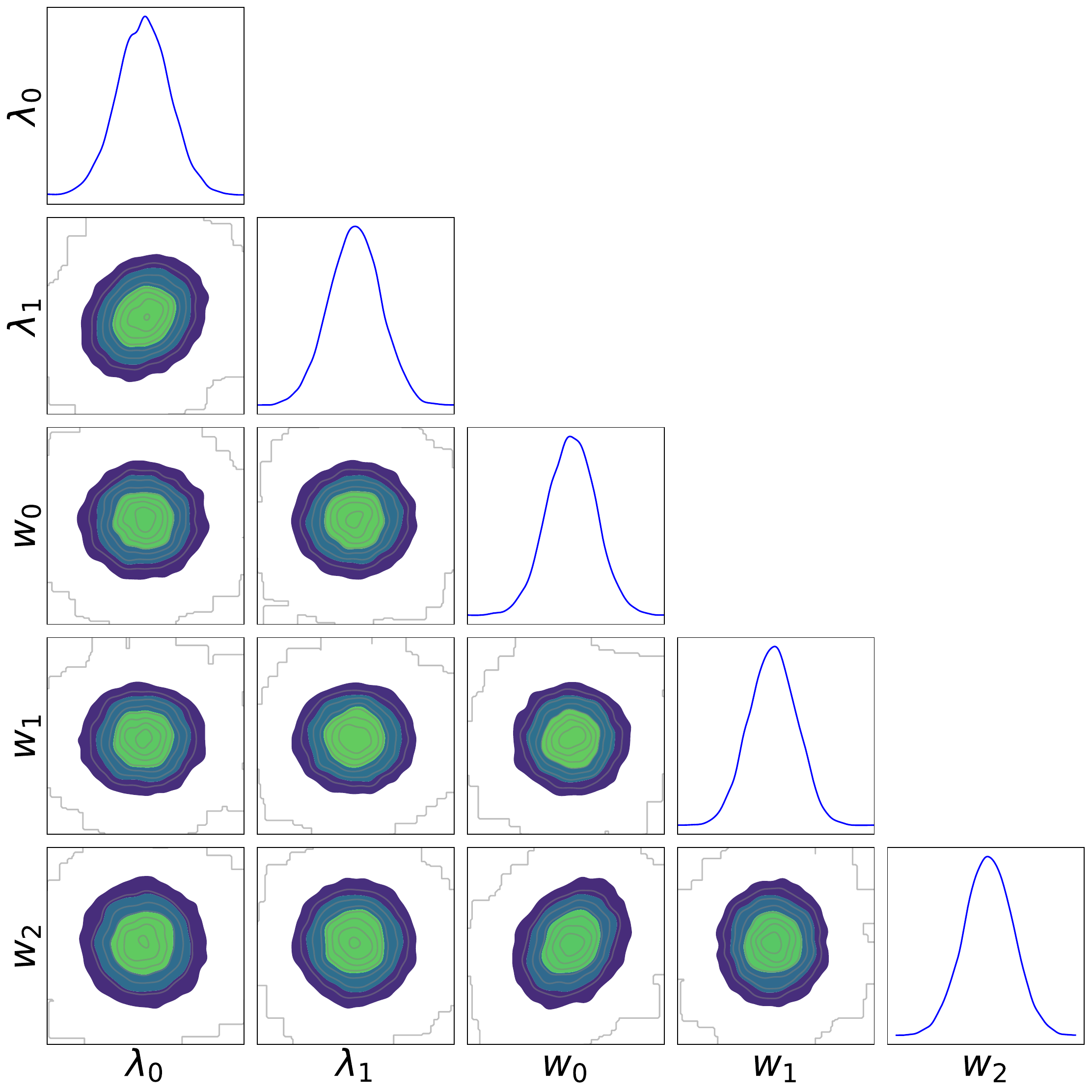}}
    \caption{Frames (a) and (b) show the joint posterior distributions of model parameters and GP weights for $N=20$ and $m=20$ obtained with KOH and OGP respectively.}
    \label{fig:lin_joint}
\end{figure}

It is also instructive to compare the joint distributions of the model parameters and GP weights obtained using the two methods, as shown in Figure \ref{fig:lin_joint}. 
In the KOH formulation, a clear correlation emerges between the parameter $\lambda_0$ and the weight $w_0$, illustrating the conflation between the fit model and the GP component. 
In contrast, the OGP-based formulation yields joint distributions in which the model parameters and GP weights are effectively uncorrelated. 
This decoupling not only reflects the intended separation between model and bias contributions but also leads to more efficient MCMC sampling as will be described in the next example. 

The emergence of only mild correlations under OGP is easy to see in this linear case. 
The linear model can be rewritten as, 
\vspace{-0.5em}
\begin{equation*}
\mathbf{y} = [G^{\top} \quad \Phi^{\top}]
\begin{bmatrix}
 \boldsymbol{\lambda}\\
 \boldsymbol{w}
\end{bmatrix} + \boldsymbol{\varepsilon}, \quad \boldsymbol{\varepsilon} \sim \mathcal{N}(\mathbf{0}, \Sigma_{\text{obs}})
\end{equation*}
where $G^\top = [\mathbf{1} \quad X] \in \mathbb{R}^{N \times 2}$ and $\Phi^{\top} \in \mathbb{R}^{N \times m}$, obtained by stacking the basis function evaluations at all data points.
With a Gaussian prior
\[
\pi(\boldsymbol{\lambda}, \boldsymbol{w}) = \mathcal{N}\left(\begin{bmatrix}
\boldsymbol{\mu}_{\lambda}\\
\boldsymbol{\mu}_{w}
\end{bmatrix}, \begin{bmatrix}
\Sigma_{\lambda} & \mathbf{0}\\
\mathbf{0} & \Sigma_w
\end{bmatrix}\right)
\]
and a Gaussian likelihood, the posterior is also Gaussian with the precision matrix given by 
\begin{align*}
\Sigma_{\mathrm{pos}}^{-1} &= \begin{bmatrix}
G \\
\Phi
\end{bmatrix} \Sigma_{\mathrm{obs}}^{-1}\begin{bmatrix}
G^\top & \Phi^\top
\end{bmatrix} + \begin{bmatrix}
\Sigma_{\lambda}^{-1} & \mathbf{0} \\
\mathbf{0} & \Sigma_{w}^{-1} 
\end{bmatrix}\\
&= 
\renewcommand{\arraystretch}{1.5}
\begin{bmatrix}
G\Sigma_{\mathrm{obs}}^{-1}G^\top + \Sigma_{\lambda}^{-1} & G\Sigma_{\mathrm{obs}}^{-1}\Phi^\top \\ 
\Phi\Sigma_{\mathrm{obs}}^{-1}G^\top & \Phi\Sigma_{\mathrm{obs}}^{-1}\Phi^\top + \Sigma_{w}^{-1}
\end{bmatrix}
\end{align*}
The applied orthogonality constraint 
$\int_{\mathcal{X}} 
\nabla_{\boldsymbol{\lambda}} f(x;\boldsymbol{\lambda})
\big|_{\boldsymbol{\lambda}^*} 
\, \delta(x) 
\, d\mu(x)
= \mathbf{0}$ is equivalent to $G \Sigma_{\mathrm{obs}}^{-1}\Phi^\top = \mathbf{0}$ (since $\bm{\varepsilon}$ is \emph{iid} and $\Sigma_{\text{obs}}\propto I$)  given that $\nabla_{\boldsymbol{\lambda}} f(x;\boldsymbol{\lambda}) = [1 \quad x]^\top$. 
Therefore, the off-diagonal terms in the posterior precision are zero and hence the cross-correlations between the model parameters and GP weights in the posterior distribution are also zero. 
As will be demonstrated in the next example, orthogonalization also leads to reduced correlations between model parameters and the GP weights in the non-linear model error embedding case.
When $\Sigma_{\text{obs}}$ is not diagonal, the cross-correlations are expected to be non-zero. 

\begin{figure}[ht]
\centering
\subcaptionbox{PDF of $\lambda_0$\label{fig:l0_conv_wm_nd20}}{\includegraphics[width=0.35\linewidth]{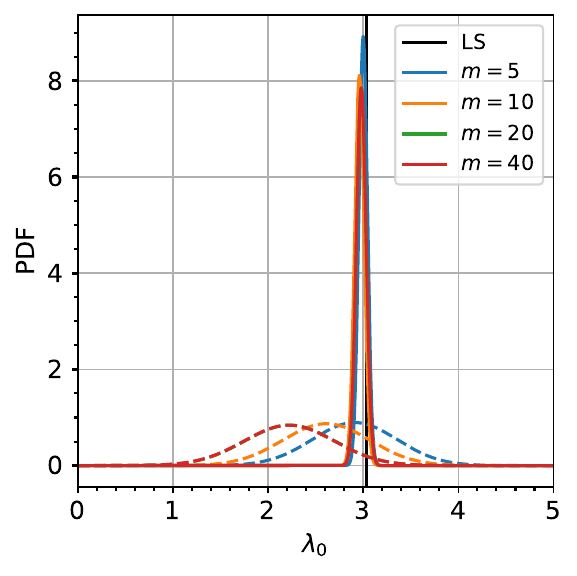}}
\subcaptionbox{PDF of $\lambda_1$\label{fig:l1_conv_wm_nd20}}{\includegraphics[width=0.35\linewidth]{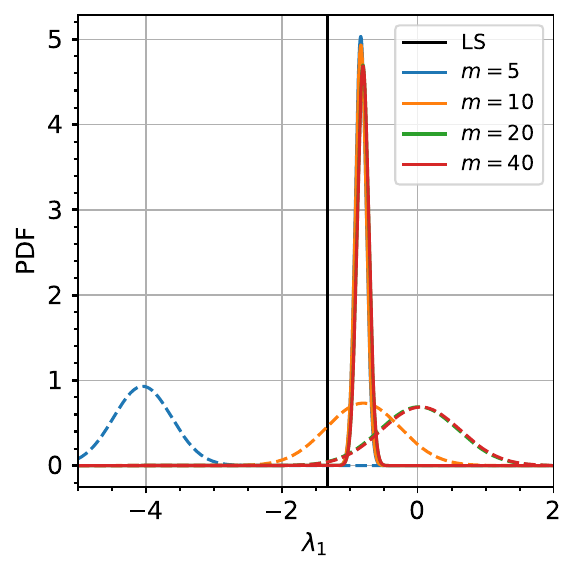}}
\vspace{-1em}
\caption{Panels (a) and (b) show the convergence in marginal posteriors of $\lambda_0$ and $\lambda_1$, respectively, with increasing number of basis functions. In both plots, the dashed lines show results with KOH and solid lines show results with OGP.}
\label{fig:lin_conv_wm}
\end{figure}

We now examine convergence of the marginal posterior distributions of model parameters for both KOH and OGP as the number of GP basis functions increases. Results are shown in Figure~\ref{fig:lin_conv_wm}.
For KOH, shown with dashed lines, we observe that approximately 20 basis functions are sufficient to achieve converged marginal posteriors for $\boldsymbol{\lambda}$. However, the posterior distributions change substantially when increasing the number of basis functions from 5 to $\ge$20. This behaviour is a direct consequence of the confounding between the model parameters and the GP weights inherent in this approach.
In contrast, the posterior distributions obtained using OGP, shown with solid lines, converge faster, remaining largely unchanged across different numbers of basis functions, and staying close to the LS solution. 
This is again attributed to the orthogonality constraint in OGP.

\begin{figure}[ht]
\centering
\subcaptionbox{Interpolation region\label{fig:ppstd_int_wm_nd20}}{\includegraphics[width=0.35\linewidth]{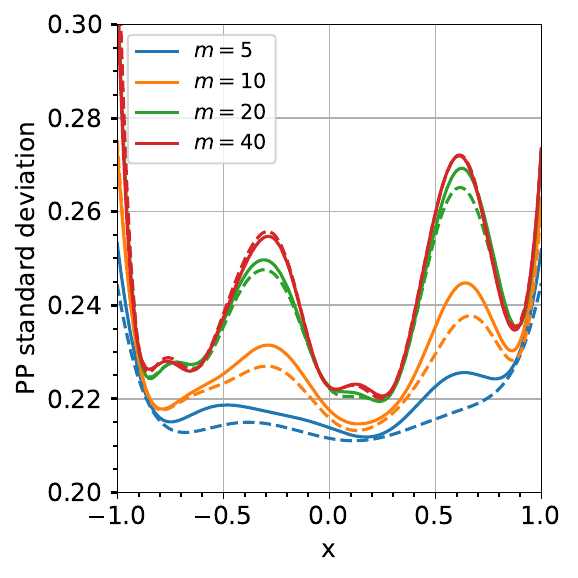}}
\subcaptionbox{Extrapolation region\label{fig:ppstd_ext_wm_nd20}}{\includegraphics[width=0.35\linewidth]{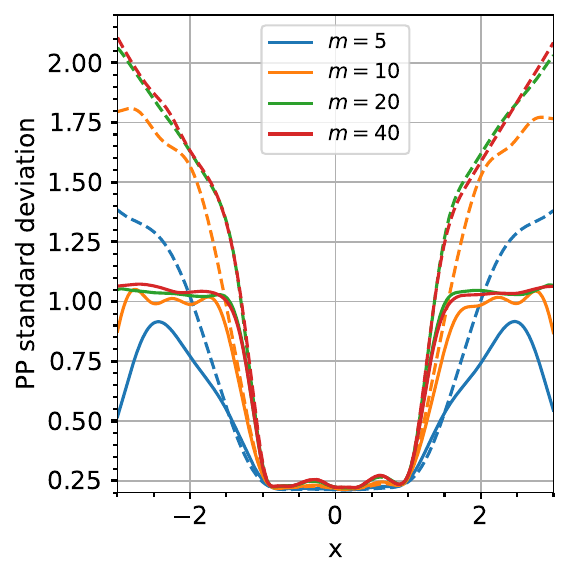}}
\vspace{-1em}
\caption{Panels (a) and (b) show convergence in posterior predictive standard deviation with increasing number of basis functions in interpolation and extrapolation regions respectively. In both plots, the dashed lines show results with KOH and solid lines show results with OGP.}
\label{fig:ppstd_conv_wm}
\end{figure}

The number of requisite GP basis functions can also be selected by analyzing the PP standard deviation, which is a reliable convergence indicator since it accounts for combined effects of model parameters and GP weights. Figure~\ref{fig:ppstd_conv_wm} shows the convergence of the PP standard deviation with increasing number of basis functions, for both interpolation  and extrapolation.
In the interpolation region, the PP standard deviation for both KOH and OGP is effectively converged with $m=20$. 
However, in the extrapolation region, the standard deviation is significantly larger for KOH. 
This increase arises from the high uncertainty in the model parameters, as shown in Figure~\ref{fig:lin_conv_wm}. 
Because this uncertainty scales with $x$, it leads to a growing standard deviation as predictions move farther away from $x = 0$.
In contrast, for OGP, the uncertainty in the extrapolation region is driven solely by the additive GP. 
As the number of basis functions increases, this uncertainty approaches the prior standard deviation, $\sigma_f = 1$, thereby recovering equivalence with the FS view of GP.

\begin{figure}[ht]
\centering
\subcaptionbox{\label{fig:l0_conv_wN_nb5}}{\includegraphics[width=0.4\linewidth]{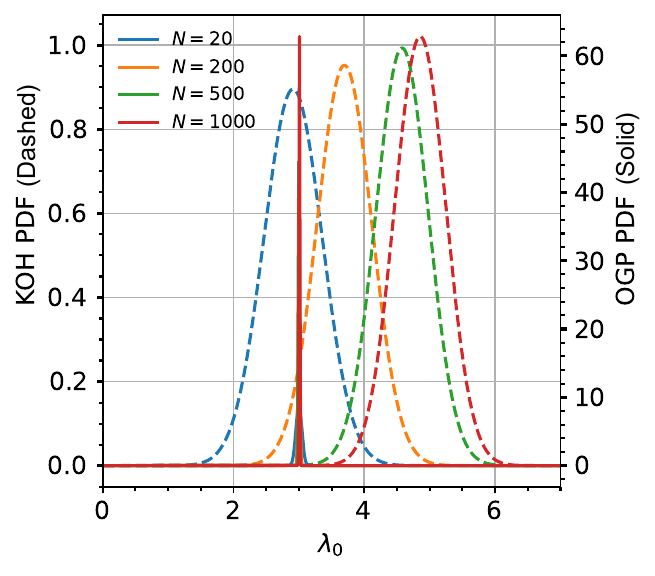}}
\subcaptionbox{\label{fig:l1_conv_wN_nb5}}{\includegraphics[width=0.4\linewidth]{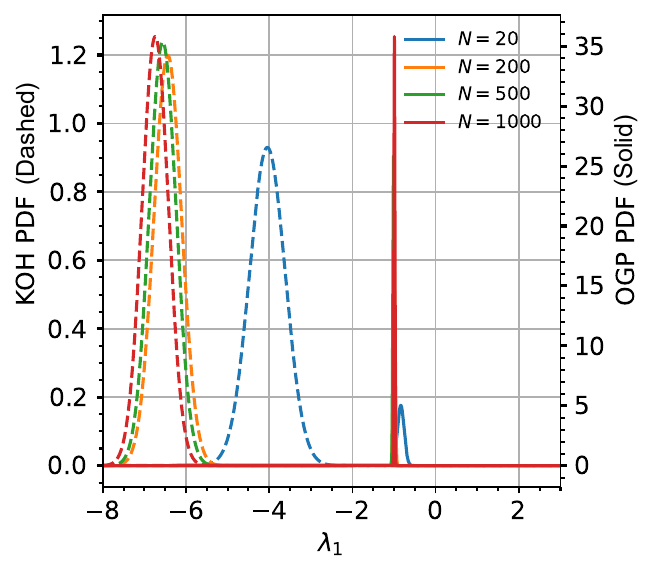}}
\caption{Panels (a) and (b) show the convergence in marginal posteriors of $\lambda_0$ and $\lambda_1$, respectively, with increasing number of data points for $m=5$. In both plots, the dashed lines show results with KOH and solid lines show results with OGP.}
\label{fig:lin_conv_wN}
\end{figure}

\begin{figure}[ht]
\centering
\subcaptionbox{\label{fig:l0_conv_wN_nb5_zoom}}{\includegraphics[width=0.35\linewidth]{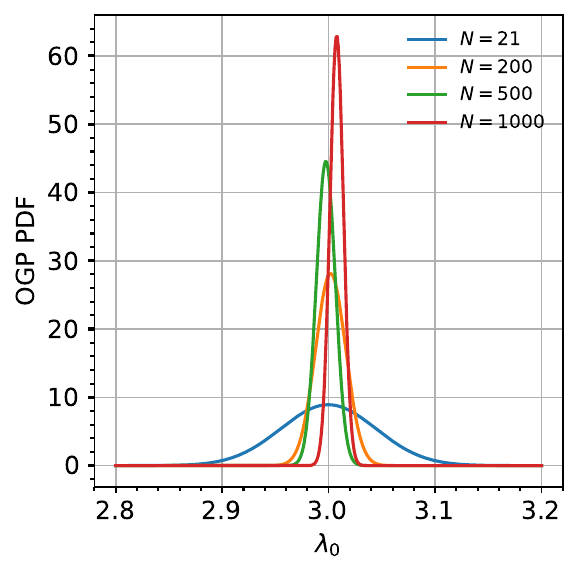}}
\subcaptionbox{\label{fig:l1_conv_wN_nb5_zoom}}{\includegraphics[width=0.35\linewidth]{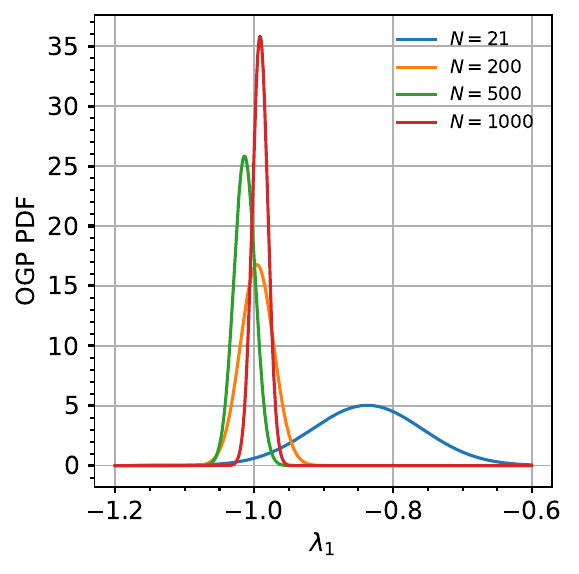}}
\caption{Zoomed-in view of the OGP marginal posteriors from Fig.~\ref{fig:lin_conv_wN}.}
\label{fig:lin_conv_wN_zoom}
\end{figure}

To illustrate the asymptotic behavior of KOH and OGP, Figure~\ref{fig:lin_conv_wN} shows the convergence of the marginal posterior distributions of $\boldsymbol{\lambda}$ as the number of data points increases. 
For this study, we use $m = 5$ basis functions, which allows the posterior distributions to converge more rapidly with increasing data because the calibration problem involves only seven parameters. 
The posterior modes obtained using KOH, shown with dashed lines, do not approach the LS solution even when the number of data points is increased to 1000. 
This observation is consistent with the theoretical results of~\cite{tuo2016theoretical}, where they show that the maximum likelihood estimate in the KOH framework converges asymptotically to a value that can differ substantially from the LS solution. Moreover, the posterior distributions remain relatively broad even for $N = 1000$.
In contrast, the posterior distributions obtained with OGP, shown with solid lines, remain tightly concentrated around the LS solution for each value of $N$ (the LS solutions themselves are omitted from the figure to avoid visual clutter). 
Moreover, the zoomed-in OGP posteriors shown in Fig.~\ref{fig:lin_conv_wN_zoom} indicate that the support does get narrower with increasing data, as would be expected given the added information content. 

\begin{figure}[ht]
\centering
\subcaptionbox{KOH \label{fig:KOH_joint_nd1000_nb5}}{\includegraphics[width=0.37\linewidth]{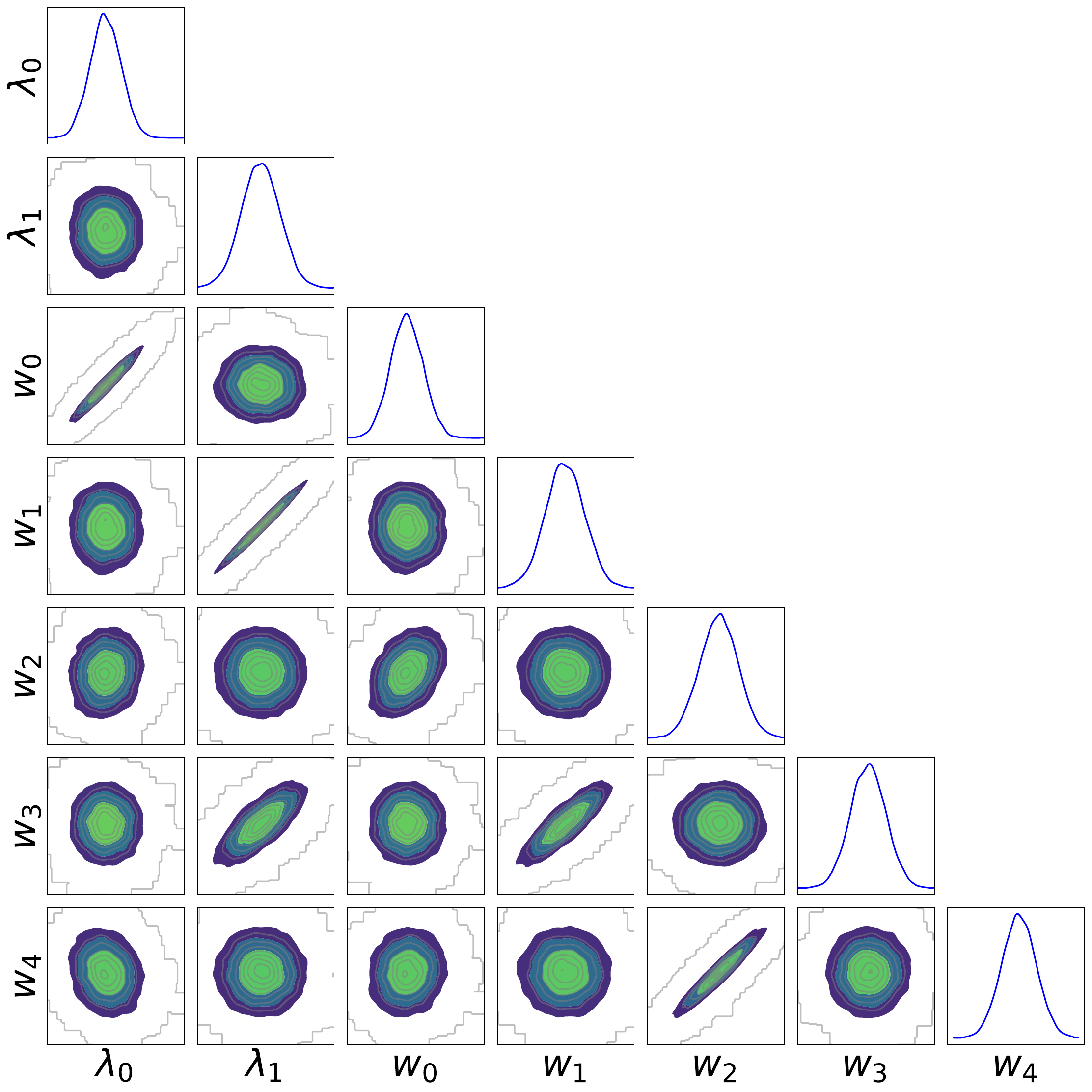}}
\subcaptionbox{ OGP\label{fig:OGP_joint_nd1000_nb5}}{\includegraphics[width=0.37\linewidth]{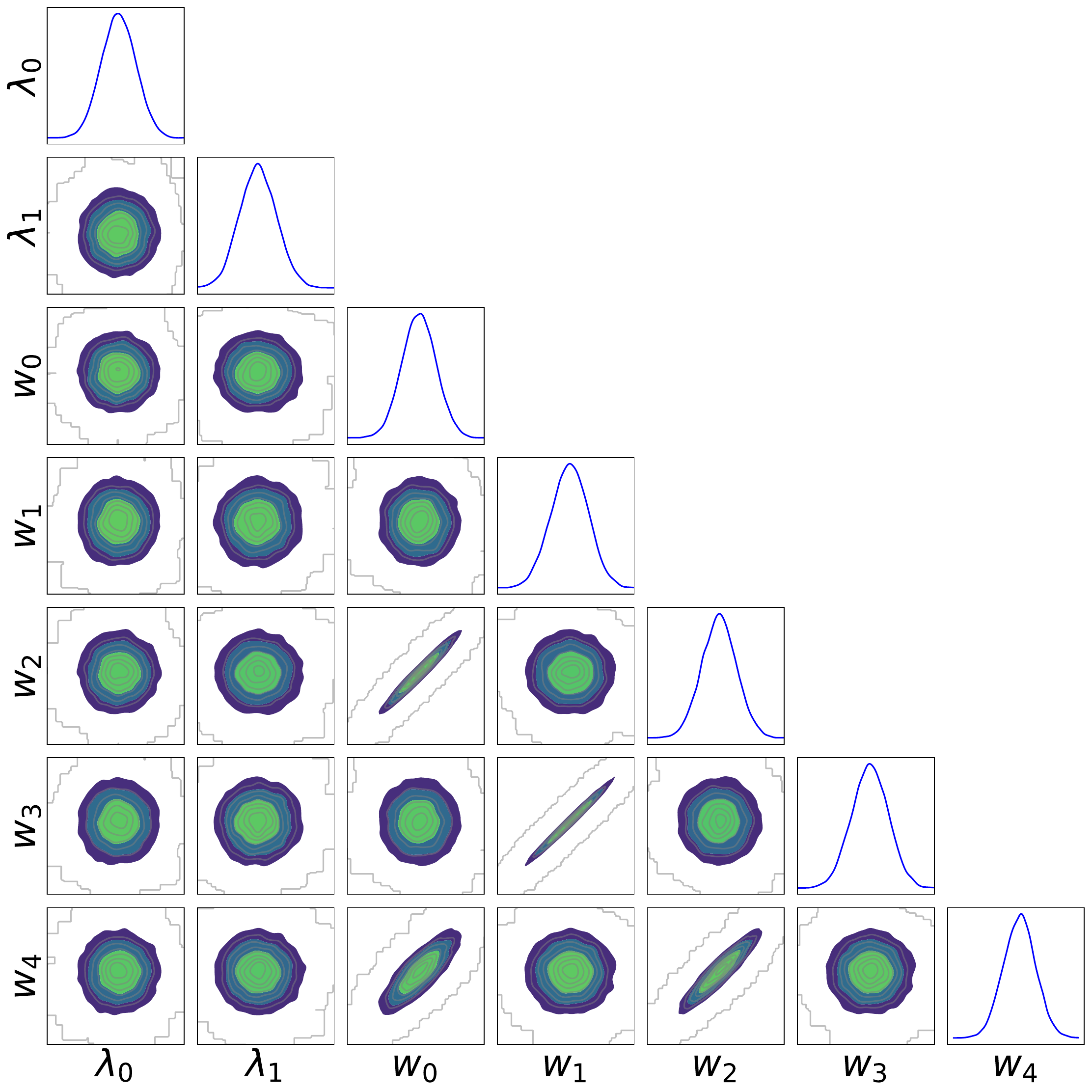}}
\caption{Panels (a) and (b) show the joint posterior distributions of model parameters and GP weights for $N=1000$ and $m=5$ obtained with KOH and OGP respectively.}
\label{fig:joint_nd1000_nb5}
\end{figure}

Further, the KOH diffuse posteriors are accompanied by strong correlations between the model parameters and the GP weights, specifically between $\lambda_0$ and $w_0$, and between $\lambda_1$ and $w_1$, as shown in Figure~\ref{fig:KOH_joint_nd1000_nb5}. By contrast, as shown in Figure~\ref{fig:OGP_joint_nd1000_nb5}, the OGP posteriors show no correlations between model parameters and GP weights. On the other hand, it is noteworthy that particularly strong correlations emerge among OGP weights whose indices differ by two, which is to some extent, also the case for some of the KOH GP weights.

\begin{figure}[ht]
\centering
\subcaptionbox{With LIS, $r = 10$ and $m=400$\label{fig:LIS_lin_nb400_nd20}}{\includegraphics[width=0.35\linewidth]{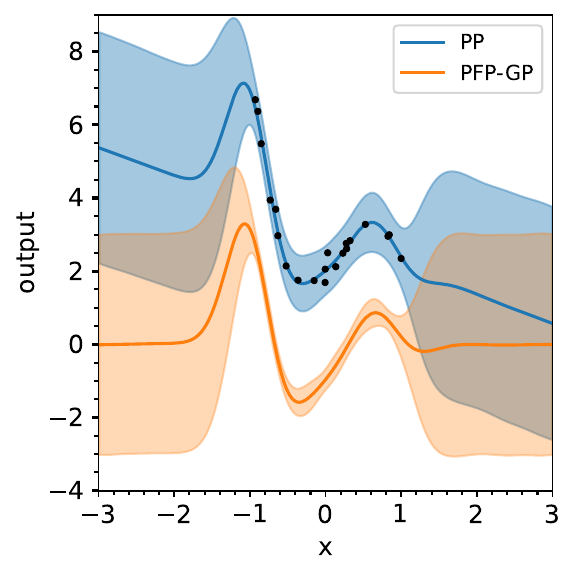}}
\subcaptionbox{Without LIS, $m = 10$\label{fig:LIS_lin_nb10_nd20}}{\includegraphics[width=0.35\linewidth]{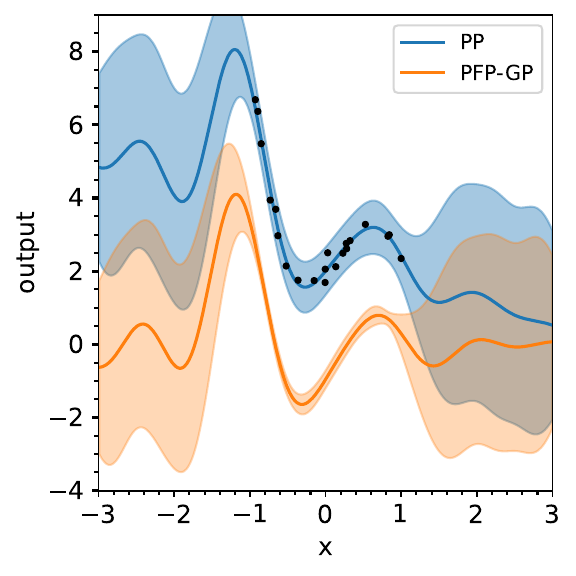}}
\caption{For $N=20$, panel (a) shows the posterior predictive and push-forward posterior through the GP when LIS with $r=10$ and $m=400$ is used; panel (b) shows the predictions without LIS and $m=10$. Solid lines show mean predictions and bands show $\pm$ three standard deviations.}
\label{fig:LIS_lin_nd20}
\end{figure}

Next, we demonstrate how the LIS approach enables us to use an arbitrarily large number of basis functions in the WS GP.
We start with $m=400$ basis functions in the GP, giving us a total of 402 calibration parameters.
With linear models, the local LIS from one point in the parameter space to another does not change because the Hessian of the log likelihood function is independent of the parameter value. 
Hence the global LIS and local LIS are equivalent. 
For $N = 20$ data points, the rank obtained for LIS, with an eigenvalue cut-off of 0.1, is $r=10$. 
This indicates that the data effectively inform only 10 directions in the high dimensional parameter space. 
Thus, it is sufficient to perform a reduced inference of dimension 10 in the LIS and sample from the prior projected onto the CS for the remaining 392 directions. 
Finally, the posterior samples from the LIS and the prior samples from the CS can be combined via Eq~\eqref{eq:LIS_redpost} to approximate the full high-dimensional posterior. 
The PP distribution, and the push-forward posterior through the GP component, denoted by PFP-GP, obtained with LIS, are shown in Figure~\ref{fig:LIS_lin_nb400_nd20}. 
For comparison, we also plot in Figure~\ref{fig:LIS_lin_nb10_nd20} the results obtained with $m=10$ basis functions but without employing LIS. 
We observe that the inclusion of a larger set of basis functions removes the oscillatory behaviour of the predictions seen in Figure~\ref{fig:LIS_lin_nb10_nd20}, which is an artifact of truncating the GP with very few weights as evidenced in PFP-GP. 
These oscillations are also visible in Figure~\ref{fig:ppstd_ext_wm_nd20}. 
The PFP-GP distribution of LIS cleanly attains the prior standard deviation of $\sigma_f=1$ as we extrapolate outside the data regime without any additional cost for inference. 

\begin{figure}[ht]
\centering
\subcaptionbox{With LIS, $r = 12$ and $m=400$\label{fig:LIS_lin_nb400_nd500}}{\includegraphics[width=0.35\linewidth]{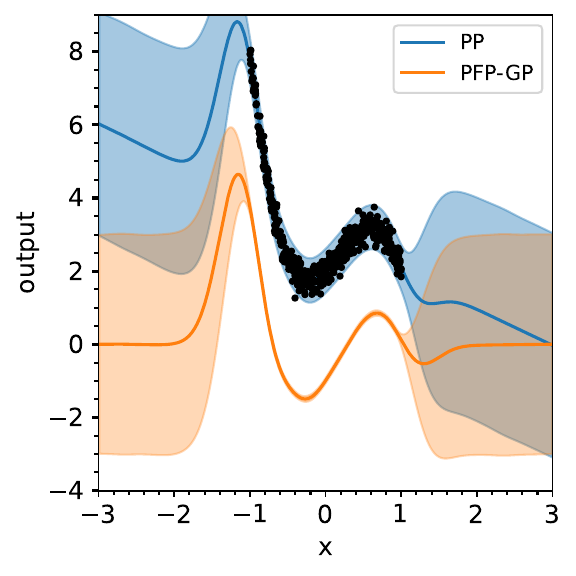}}
\subcaptionbox{Without LIS, $m = 12$\label{fig:LIS_lin_nb12_nd500}}{\includegraphics[width=0.35\linewidth]{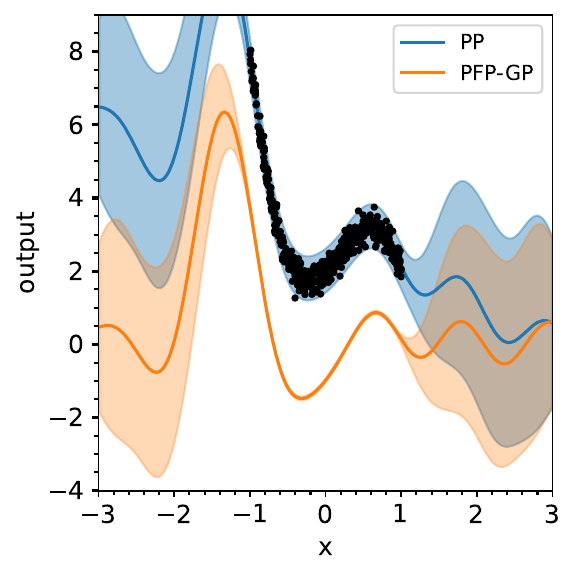}}
\caption{For $N=500$, panel (a) shows the posterior predictive and push-forward posterior through the GP when using LIS with $r=12$ and $m=400$. Panel (b) shows the predictions without LIS, and for $m=12$. Solid lines show mean predictions and bands show $\pm$ three standard deviations.}
\label{fig:LIS_lin_nd500}
\end{figure}

It is also informative to examine how increasing the number of data points affects the prediction intervals. 
This is illustrated in Figure~\ref{fig:LIS_lin_nd500} for $N = 500$.
The LIS rank with 0.1 eigenvalue cut-off increases to $r=12$ as the availability of more data can inform a larger set of directions in the parameter space. 
With a denser data set, the $\pm$ three standard deviation bands of PFP-GP contract to nearly zero within the interpolation region because of the \emph{iid} noise model. 
However, due to the inclusion of a large set of GP basis functions, the GP mean fully accounts for the structural deficiencies of the fit model.
Recall that random variable embedding (see~\cite{sargsyan2015statistical}), on the other hand, retains finite prediction intervals as a compromise for model error flexibility even as the data is increased. 
Lastly, even though the prediction intervals shrink in the interpolation region, the prior distributions are faithfully recovered in the extrapolation regime with LIS. 

\subsection{Non-linear Model}

Consider the truth model to be 
\begin{align*}
    f_{t}(x) = \exp\left(1-0.5x+x^2+x^3\right)
\end{align*}
and the fit model to be 
\begin{align*}
f(x;\boldsymbol{\lambda}) = \underbrace{\sin\left(\lambda_0x\right)}_{\mathrm{S}_1} + \underbrace{\exp\left(\lambda_1 x\right)}_{\mathrm{S}_2}
\end{align*}
defined on $x\in [-2, 2]$ with $\boldsymbol{\lambda} = (\lambda_0, \lambda_1)$. 
The fit model can be thought of as a combination of two sub-models, $\mathrm{S}_1$ and $\mathrm{S}_2$, whose predictive capability can be compared with GP embedded corrections. 
First, consider embedding in $\mathrm{S}_1$, \emph{e.g.} as
\begin{align*}
\tilde{f}(x;\boldsymbol{\lambda}, \delta_{\boldsymbol{w}}(x)) = \sin(\lambda_0x + \delta_{\boldsymbol{w}}(x)) + \exp(\lambda_1 x).
\end{align*}
It is evident that embedding model error within this particular submodel will not significantly improve the predictions, since the true underlying function is exponential and therefore structurally mismatched with the submodel. In practice, however, the form of the data-generating process is rarely known. In such cases, indeed, introducing embedded model error into individual submodels that comprise the fit model provides a systematic way to assess the relative contribution and importance of each submodel to overall model error. The fact that model error embedding in S$_1$ does not much improve predictions provides diagnostic information as to the dominant role of this misspecified term in subsequent predictive errors. In fact, its detrimental effects will still be evident in the present case when embedding in S$_2$, which we focus on now.

\begin{figure}[ht]
\centering
\subcaptionbox{Marginal of $\lambda_0$ \label{fig:PGP_l0_wm}}{\includegraphics[width=0.32\linewidth]{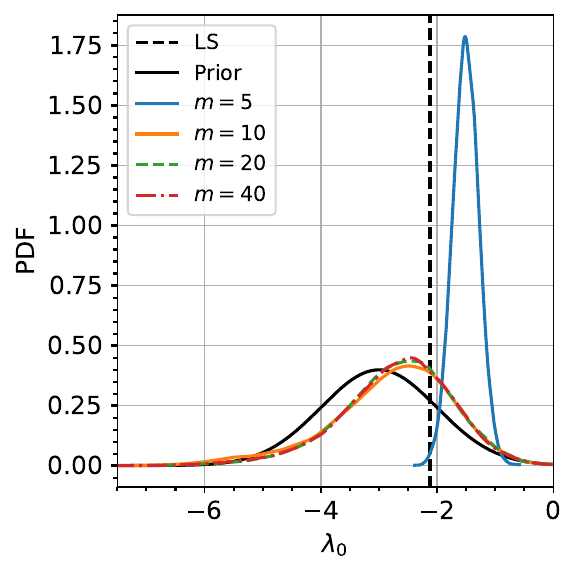}}
\subcaptionbox{Marginal of $\lambda_1$ \label{fig:PGP_l1_wm}}{\includegraphics[width=0.32\linewidth]{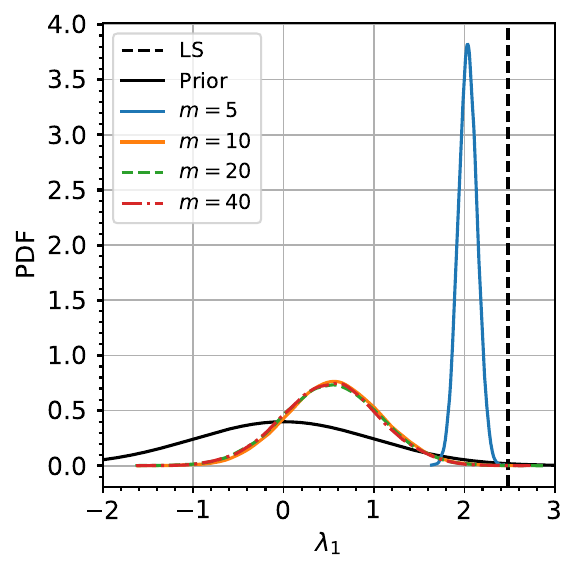}}
\subcaptionbox{Push forward predictions with $m=40$\label{fig:PGP_pf_nb40}}{\includegraphics[width=0.32\linewidth]{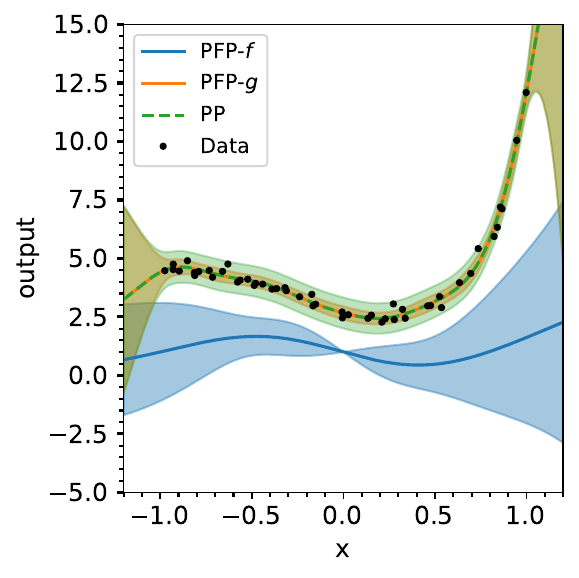}}
\caption{For $N=50$, frames (a) and (b) show marginal priors, posteriors and least-squares estimates for $(\lambda_0,\lambda_1)$ with conventional inference. Panel (c) shows the data and posterior push forward predictions. 
PFP lines show the mean prediction and bands show $\pm$ three standard deviations.}
\label{fig:PGP_nd50}
\end{figure}

\begin{figure}[ht]
    \centering
    \includegraphics[width=0.5\linewidth]{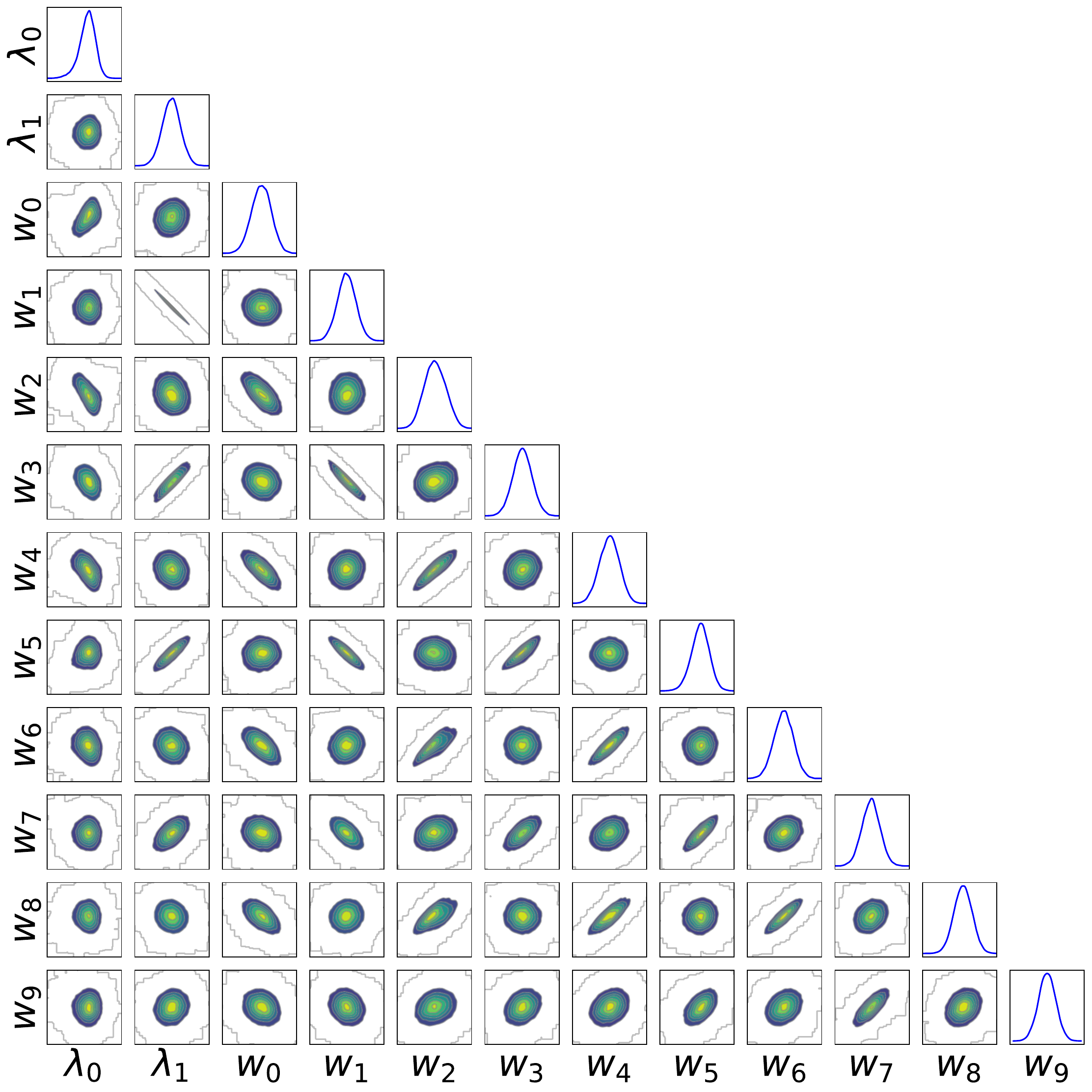}
    \caption{Joint posterior distribution obtained with conventional inference.}
    \label{fig:PGP_joint}
\end{figure}

Let us employ GP Embedding in $\mathrm{S_2}$ as follows:
\begin{align*}
\tilde{f}(x;\boldsymbol{\lambda}, \delta_{\boldsymbol{w}}(x)) = \sin(\lambda_0x) + \exp(\lambda_1 x+ \delta_{\boldsymbol{w}}(x))
\end{align*}
with a Gaussian prior $\pi(\boldsymbol{\lambda}) \sim \mathcal{N}\big( [-3,\, 0],\, \mathrm{diag}(1,\, 1) \big)$ on the model parameters, and an SQE kernel with $\sigma_f = 1$ and $l=0.3$ for the GP. 
First we analyse the performance of conventional inference without the application of any orthogonality constraint. 
The convergence of model parameter posteriors with increasing number of basis functions and PFP distributions for $N = 50$ data points are shown in Fig.~\ref{fig:PGP_nd50}. 
Similar to KOH with the linear model, the $\bm{\lambda}$ posteriors are quite broad because of strong correlations with the GP weights, as evident in Fig.~\ref{fig:PGP_joint}.
Also, the minimum effective sample size (ESS) for a total MCMC chain length of 40000 with a burn-in period of 10000 was 12500 for $\boldsymbol{\lambda}$ and 11000 for $\boldsymbol{w}$.  
These values are lower in comparison to both LOGP and ROGP discussed below. 
Lastly, the resulting PFP-$f$ distribution does not capture the data well, as the resulting posterior of $\lambda_1$, which controls the data trend, does not include the LS solution in its support.  

As shown in the linear model example, to decide the appropriate number of basis functions in the GP embedding, one can either use LIS or perform a convergence study of the PP standard deviation. 
We note that the LIS formulation of~\cite{cui2014likelihood,spantini2015optimal}, while in principle applicable for nonlinear models, is in fact feasible as-is only for LOGP in the present context, as it relies on the assumption of Gaussian priors. 
Extending this formulation to the ROGP setting, which incorporates additional regularization terms in the prior, would require non-trivial modifications to the existing LIS framework. Consequently, we employ LIS to retain a large number of basis functions for LOGP, while for ROGP we instead perform a convergence study to select an appropriate basis size. 

\begin{figure}[ht]
\centering
\subcaptionbox{Marginal of $\lambda_0$ \label{fig:LOGP_l0}}{\includegraphics[width=0.33\linewidth]{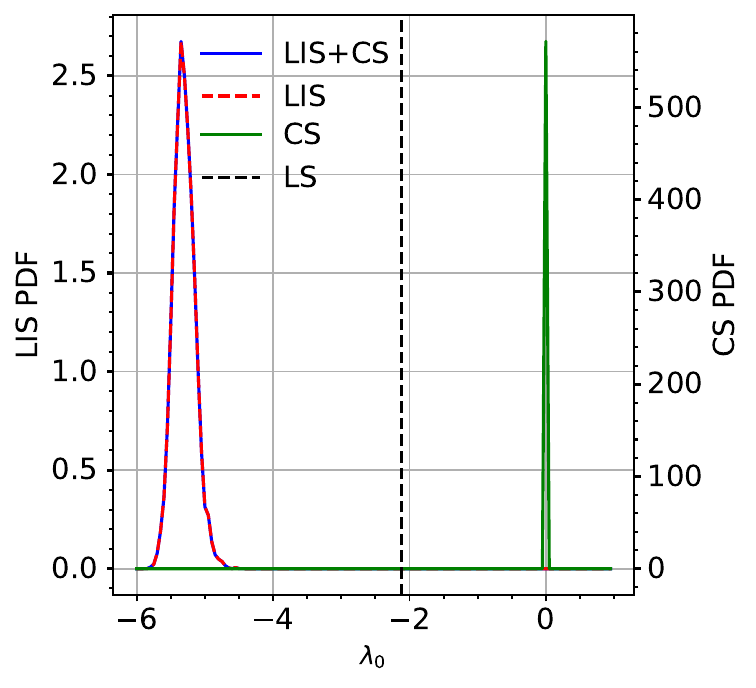}}
\subcaptionbox{Marginal of $\lambda_1$ \label{fig:LOGP_l1}}{\includegraphics[width=0.33\linewidth]{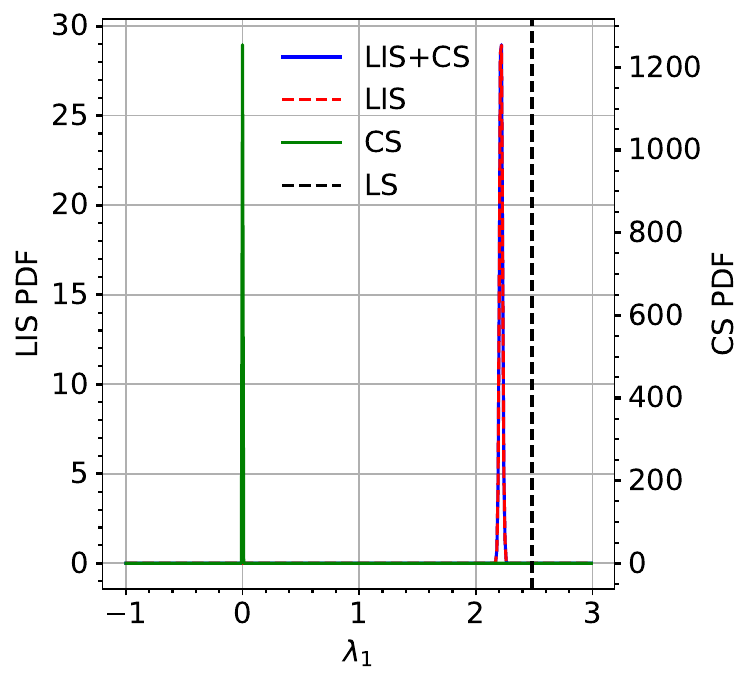}}
\subcaptionbox{Push forward predictions \label{fig:LOGP_pf_nb400}}{\includegraphics[width=0.32\linewidth]{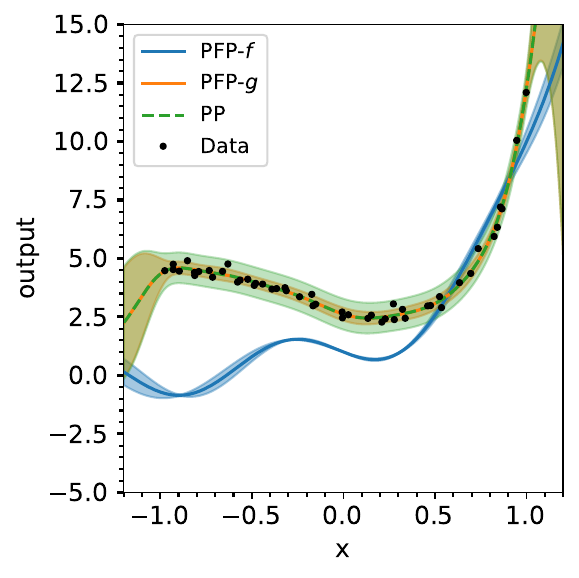}}
\caption{For $N=50$, $m=400$ and $r=13$, frames (a) and (b) show the marginal LIS posterior (red), CS prior (green) and full marginal posterior approximation (LIS+CS shown in blue) for $\lambda_0$ and $\lambda_1$ respectively. Note that LIS+CS marginal is on top of LIS marginal. Panel (c) shows the data and push forward posterior predictions.}
\label{fig:LOGP_nd50}
\end{figure}

\begin{figure}[ht]
\centering
\subcaptionbox{LIS parameters \label{fig:joint_LIS_LOGP}}{\includegraphics[width=0.43\linewidth]{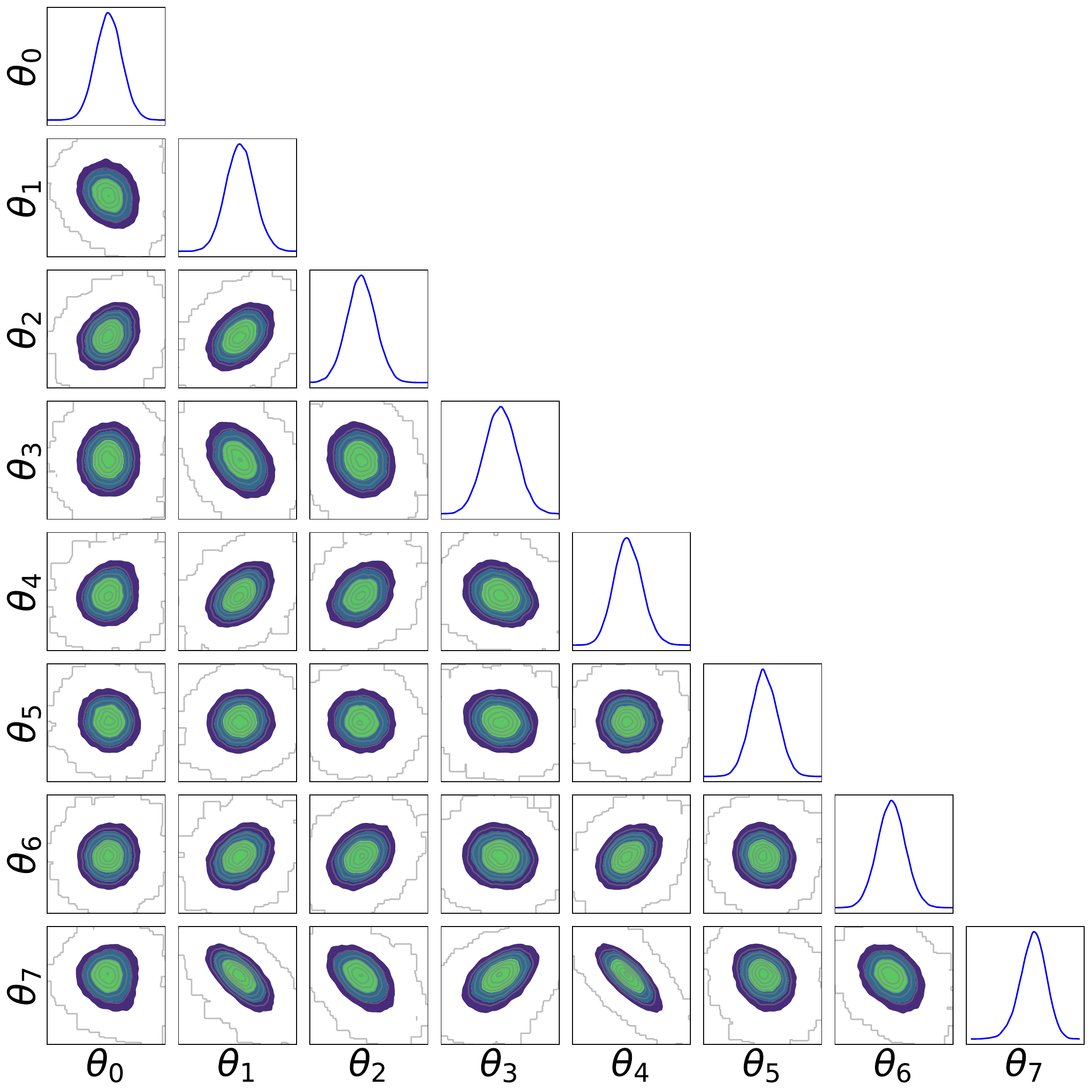}}
\subcaptionbox{Model parameters and first ten GP weights\label{fig:joint_param_LOGP}}{\includegraphics[width=0.43\linewidth]{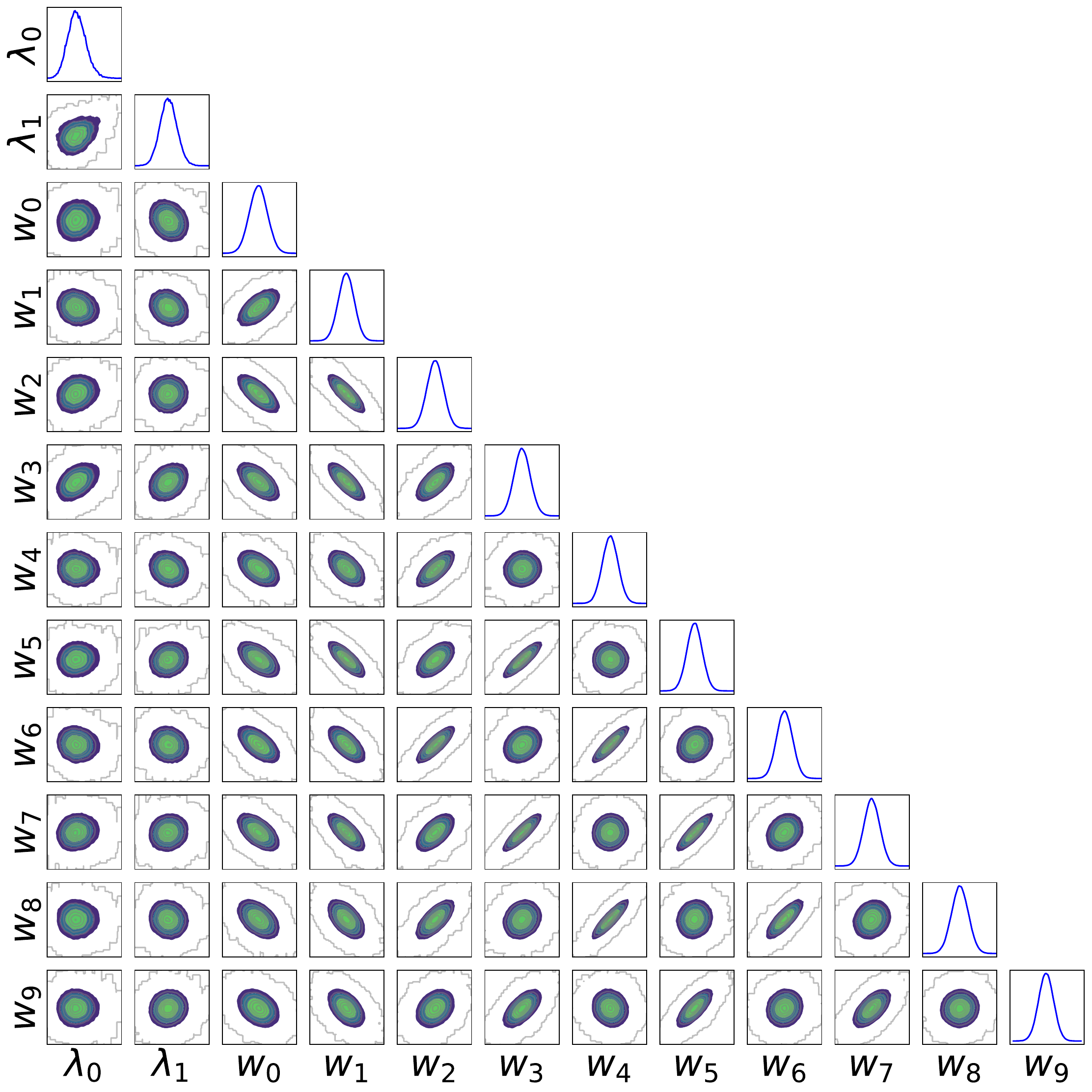}}
\caption{Frame (a) shows the joint posterior distributions of LIS parameters and frame (b) shows the joint posterior distributions of model parameters and GP weights for N = 50.}
\label{fig:joint_LOGP}
\end{figure}

Given the non-linear model, we need to first discover the global LIS, which is approximated as the Monte Carlo mean of a set of local LISs evaluated at a set of parameter values from the posterior distribution.
To this end, we use the adaptive procedure of~\cite{cui2014likelihood} (see Appendix \ref{sec:LIS_sum} for further details on this) and use the \texttt{emcee} package~\cite{foreman2013emcee} for subchain simulations. 
This package allows us to set the initial state of the chain which is a necessary requirement of the adaptive procedure, however, any other similarly-equipped MCMC algorithm can also be employed. 
After the global LIS is discovered, we use the NUTS algorithm~\cite{hoffman2014no} implemented in \texttt{PyMC}~\cite{pymc2023} for sampling in this global LIS. 
The posterior distributions for LOGP with $m=400$ are shown in Fig.~\ref{fig:LOGP_nd50}.
The parameter posterior distributions were approximated using LIS posteriors of rank $r = 13$ for an eigenvalue cut-off of 0.1. 
First, note that the CS distributions are quite narrow and centered at 0 indicating that the parameter posteriors are decided entirely by the LIS.
The posterior of $\lambda_0$ is not in the vicinity of its LS solution, which may be attributed to the linearization of the model for orthogonality constraints and also the small dataset size. 
However this is not an issue since the posterior of $\lambda_1$, which decides the data trend through the exponential term, is indeed closer to its LS solution resulting in a meaningful PFP-$f$ distribution, as shown in Fig.~\ref{fig:LOGP_pf_nb400}.
The ESS values for the LIS parameters (18000) are much higher than those obtained above using conventional inference, without imposition of orthogonality constraints. This improvement is also evident in the joint posterior distribution plots shown in Fig.~\ref{fig:joint_LIS_LOGP}. The weak correlations observed among the LIS parameters arise because the LIS basis diagonalizes the Hessian of the log-likelihood~\cite{cui2014likelihood}, leading to highly efficient MCMC sampling in the LIS space.
The joint posterior distributions between the model parameters and the first ten GP weights, estimated using samples from the LIS and CS, are shown in Fig.~\ref{fig:joint_param_LOGP}. Weak correlations between the model parameters and the GP weights are observed, which can be attributed to the enforced orthogonality. However, strong correlations are present among the GP weights themselves.

\begin{figure}[ht]
\centering
\subcaptionbox{Marginal of $\lambda_0$}{\includegraphics[width=0.33\linewidth]{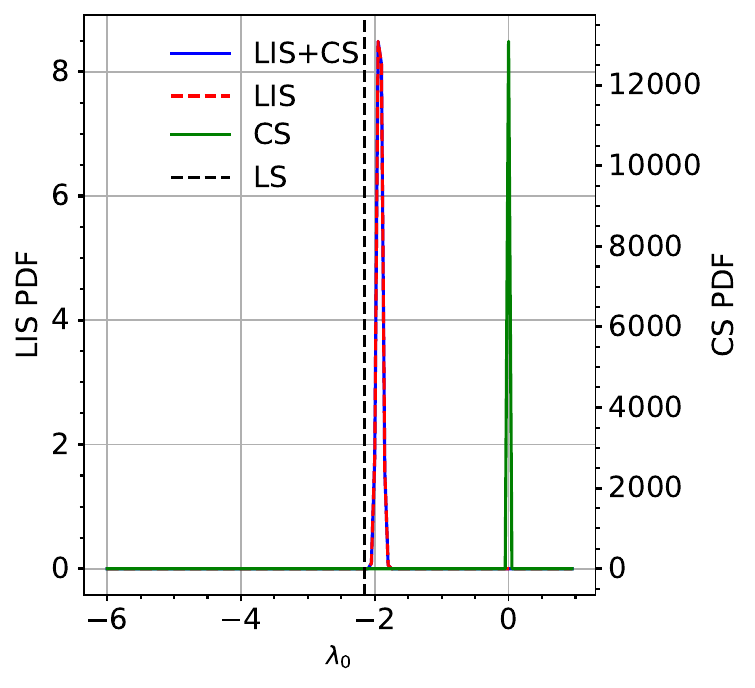}}
\subcaptionbox{Marginal of $\lambda_1$}{\includegraphics[width=0.33\linewidth]{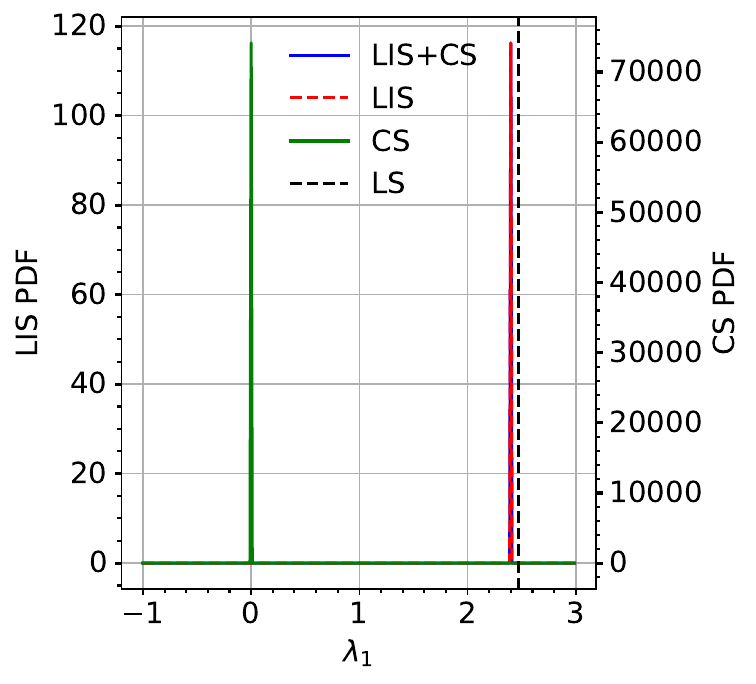}}
\caption{For $N=1000$, $m=400$ and $r=15$, frames (a) and (b) show the marginal LIS posterior (red), CS prior (green) and full marginal posterior approximation (LIS+CS shown in blue) for $\lambda_0$ and $\lambda_1$ respectively. Note that LIS+CS marginal is on top of LIS marginal.}
\label{fig:LOGP_nd1000}
\end{figure}

\begin{figure}[ht]
\centering
\subcaptionbox{LIS parameters \label{fig:joint_LIS_LOGP_nd1000}}{\includegraphics[width=0.43\linewidth]{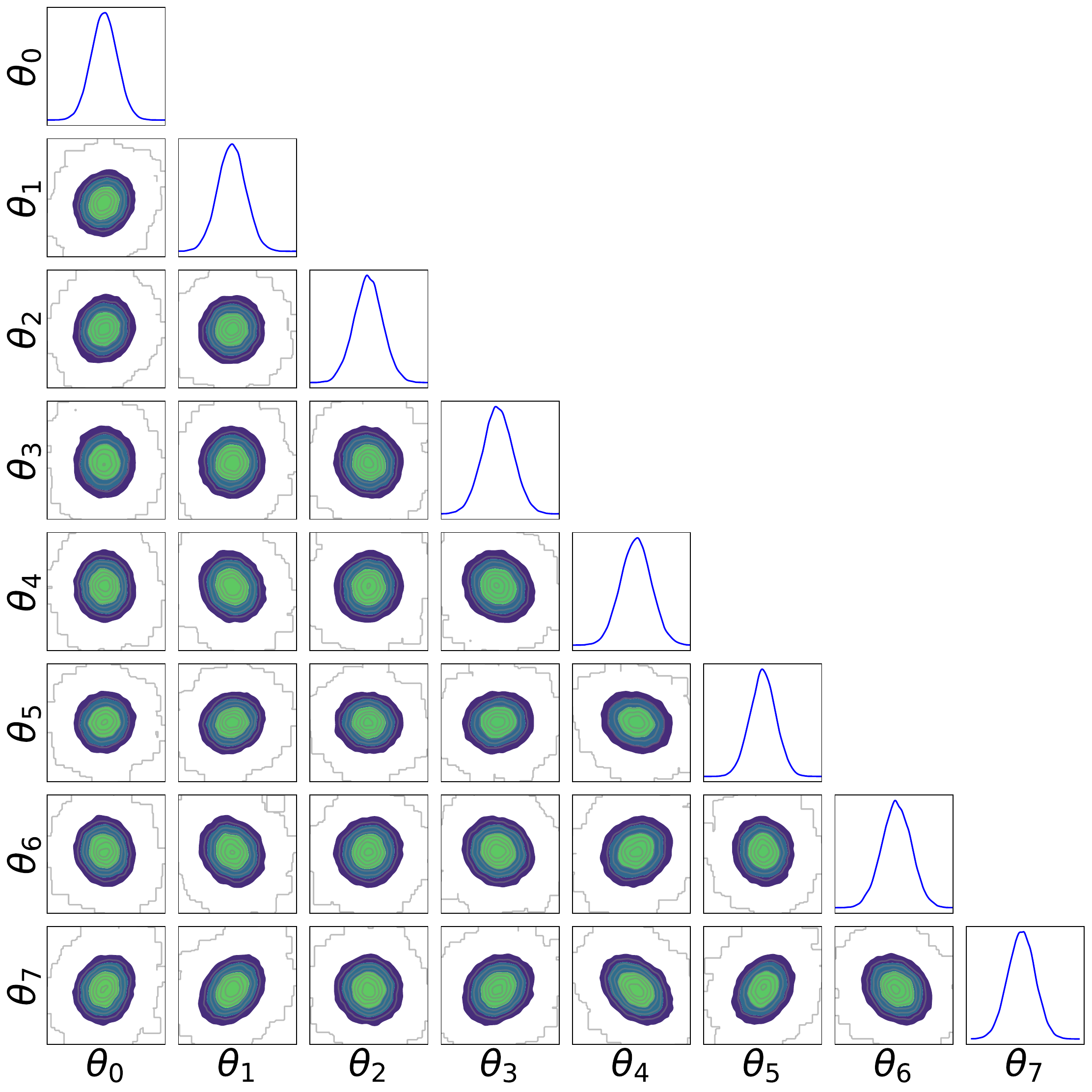}}
\subcaptionbox{Model parameters and five GP weights \label{fig:joint_param_LOGP_nd1000}}{\includegraphics[width=0.43\linewidth]{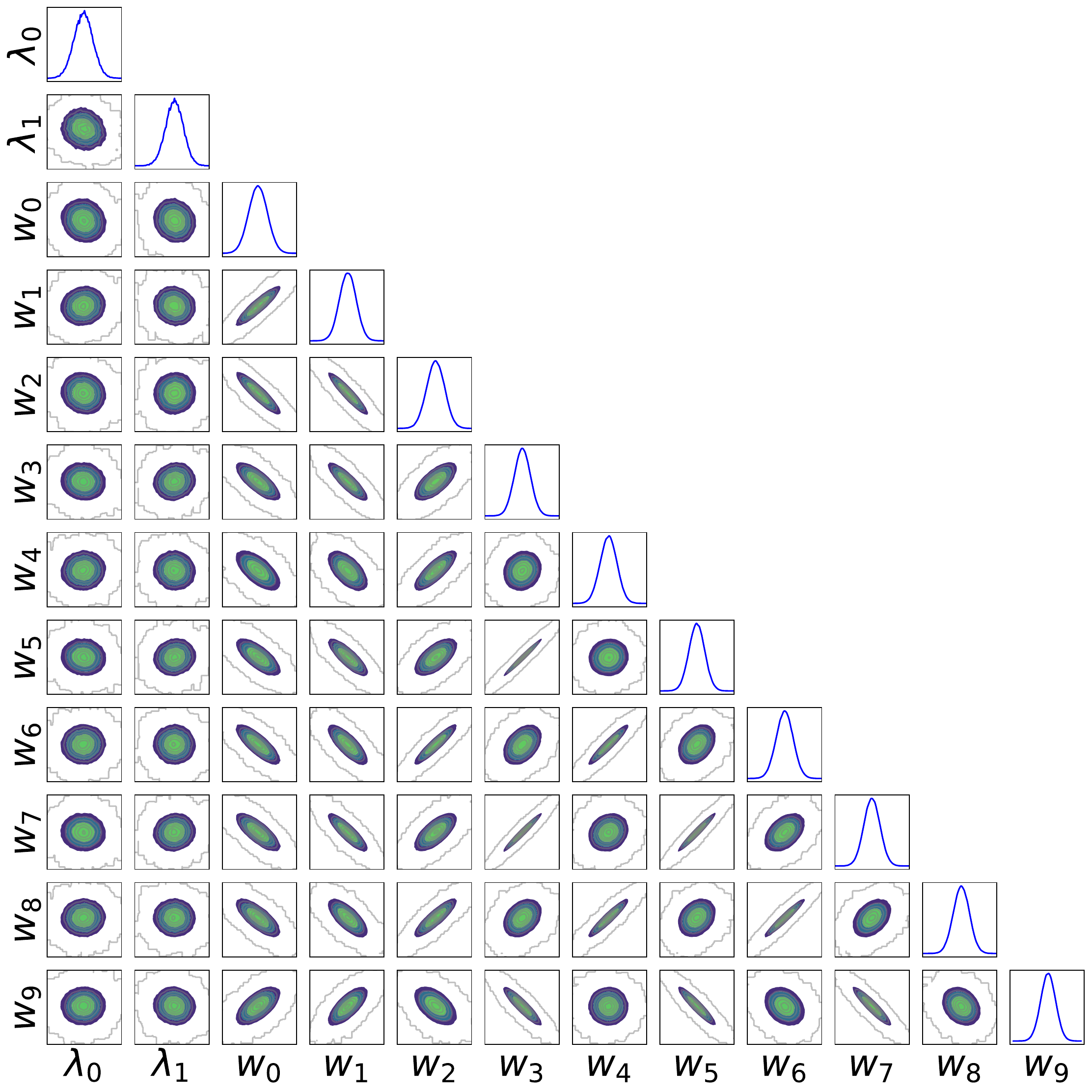}}
\caption{Frame (a) shows the joint posterior distributions of LIS parameters and frame (b) shows the joint posterior distributions of model parameters and GP weights for N = 1000.}
\label{fig:joint_LOGP_nd1000}
\end{figure}

With a larger data set of size $N = 1000$, the resulting parameter posteriors and joint distributions are shown in Figs.~\ref{fig:LOGP_nd1000} and \ref{fig:joint_LOGP_nd1000} respectively.
Both model parameters now have posterior distributions more concentrated around the LS solution when compared to the $N=50$ case.
We also observe much stronger correlations between the GP weights according to Fig.~\ref{fig:joint_param_LOGP_nd1000}. 
However, they can still be captured accurately as MCMC sampling happens in the LIS where the parameters have uncorrelated posteriors, as seen in Fig.~\ref{fig:joint_LIS_LOGP_nd1000}.
Overall, LOGP in conjunction with LIS serves as a very efficient strategy for GP error embedding in non-linear models. 

\begin{figure}[ht]
\centering
\subcaptionbox{Marginal of $\lambda_0$ \label{fig:ROGP_l0_wm}}{\includegraphics[width=0.32\linewidth]{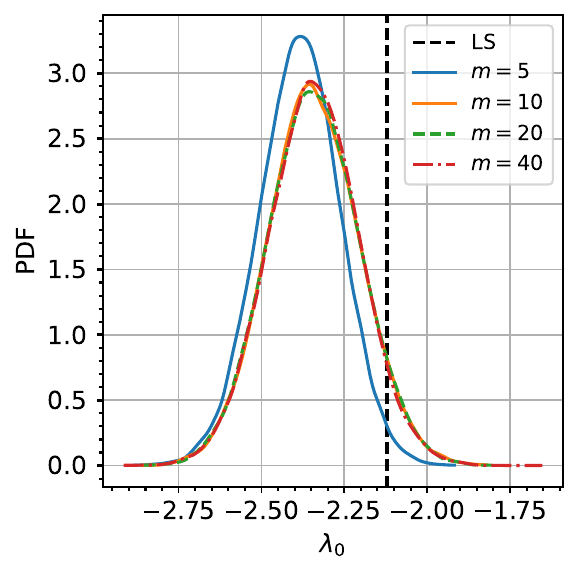}}
\subcaptionbox{Marginal of $\lambda_1$ \label{fig:ROGP_l1_wm}}{\includegraphics[width=0.32\linewidth]{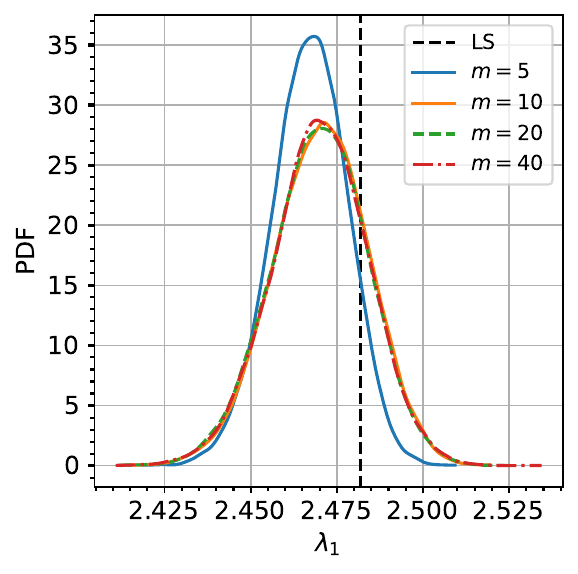}}
\subcaptionbox{Push forward predictions with $m=40$\label{fig:ROGP_pf_nb40}}{\includegraphics[width=0.32\linewidth]{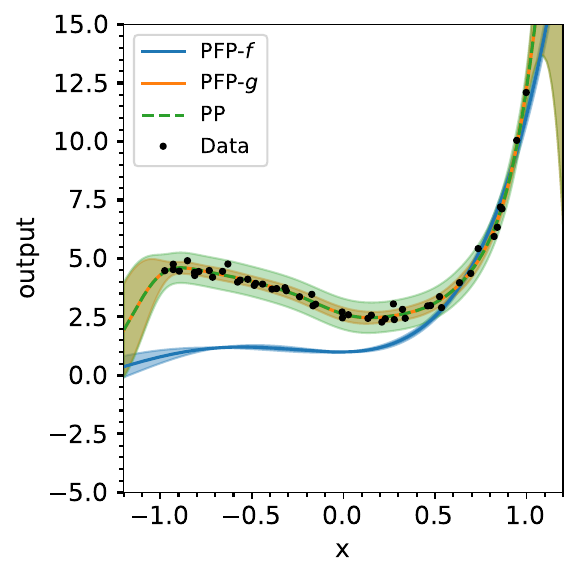}}
\caption{For $N=50$, frames (a) and (b) show the marginal priors, posteriors and least-squares estimates for $\lambda_0$ and $\lambda_1$ obtained with ROGP ($\alpha_1 = 10^8, \alpha_2=50^2$). Panel (c) shows the data and posterior push forward predictions. 
Lines in PFP show the mean prediction and bands show $\pm$ three standard deviations.}
\label{fig:ROGP_nd50}
\end{figure}

\begin{figure}[ht]
\centering
\subcaptionbox{Weak constraints, $\alpha_1 = 10^7, \alpha_2 = 50$ \label{fig:joint_ROGP_weak}}{\includegraphics[width=0.43\linewidth]{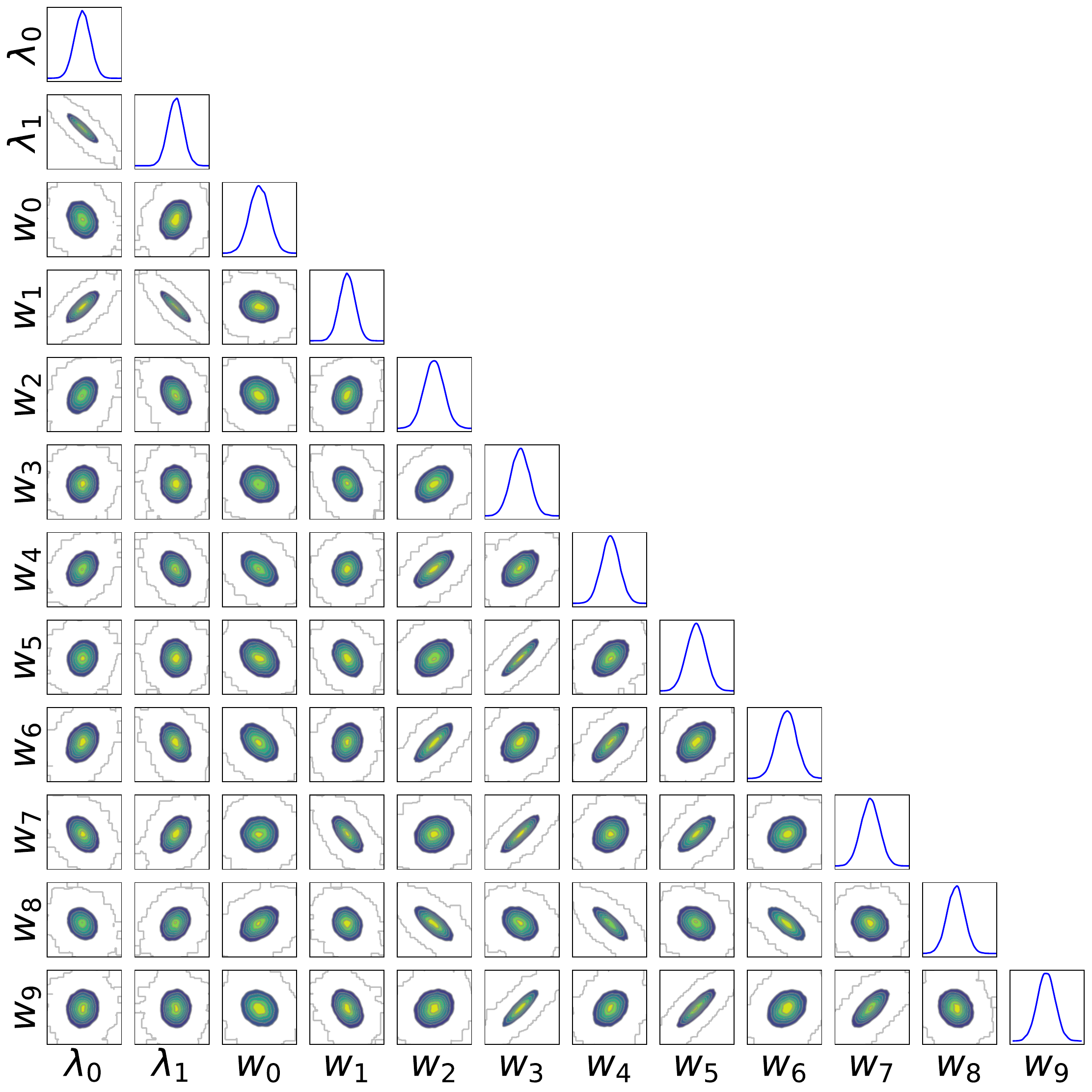}}
\subcaptionbox{Strong constraints, $\alpha_1 = 10^8, \alpha_2 = 50^2$ \label{fig:joint_ROGP_strong}}{\includegraphics[width=0.43\linewidth]{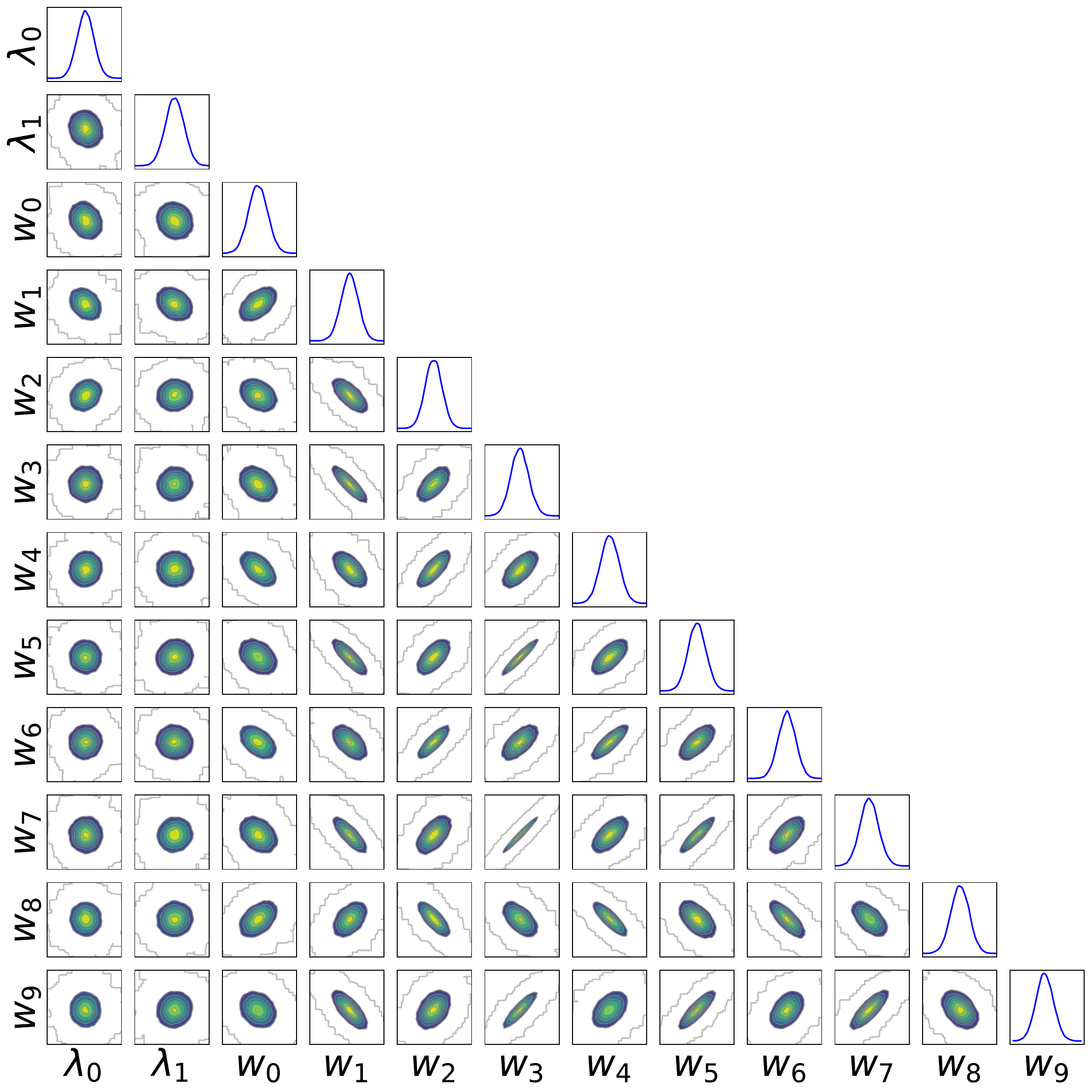}}
\caption{Panels (a) and (b) show the joint posterior distributions obtained using ROGP with weak and strong constraints respectively for $N=50$.}
\label{fig:ROGP_joint}
\end{figure}

The posterior distributions with ROGP are shown in Figs.~\ref{fig:ROGP_l0_wm} and \ref{fig:ROGP_l1_wm}. 
We observe convergence to the LS solution with just $m=10$ basis functions; however convergence study of PP standard deviation showed that at least $m=20$ basis functions are required for predictions in both interpolation and extrapolation regions.
A meaningful PFP-$f$ distribution is recovered as shown in Fig.~\ref{fig:ROGP_pf_nb40} corresponding to the LS prediction, in the high $x>0$ region, as the problematic sine function S$_1$ term becomes less significant with respect to the exponential S$_2$ term in this region, and its detrimental effects are ameliorated.  
The joint posterior distributions for the $m=40$ case, for both weak ($\alpha_1 = 10^7, \alpha_2=50$) and strong ($\alpha_1 = 10^8, \alpha_2=50^2$) $\boldsymbol{\alpha}$-enforcement of the orthogonality constraints,  are shown in Fig.~\ref{fig:ROGP_joint}. 
For the weaker constraints case, we observe correlations between the parameters and the weight $w_1$. 
This also results in a negative correlation between $\lambda_0$ and $\lambda_1$. 
These correlations mostly disappear when stronger constraints are used, however, strong correlations emerge between the GP weights. 
Further, with weaker constraints we get an ESS of 31000 for $\boldsymbol{\lambda}$ and 15000 for $\boldsymbol{w}$, both significantly higher than the numbers we reported above for the conventional unconstrained inference case. However, with stronger constraints, the strong $\bm{w}$ correlations led to lower ESS values for it compared to conventional inference without orthogonality, thereby increasing the computational cost of sampling. 

\begin{figure}[ht]
\centering
\subcaptionbox{Marginal of $\lambda_0$ \label{fig:ROGP_l0_nd1000}}{\includegraphics[width=0.32\linewidth]{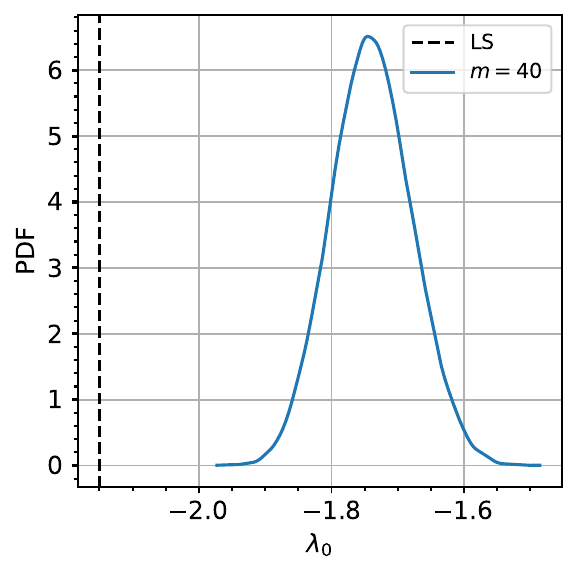}}
\subcaptionbox{Marginal of $\lambda_1$ \label{fig:ROGP_l1_nd1000}}{\includegraphics[width=0.32\linewidth]{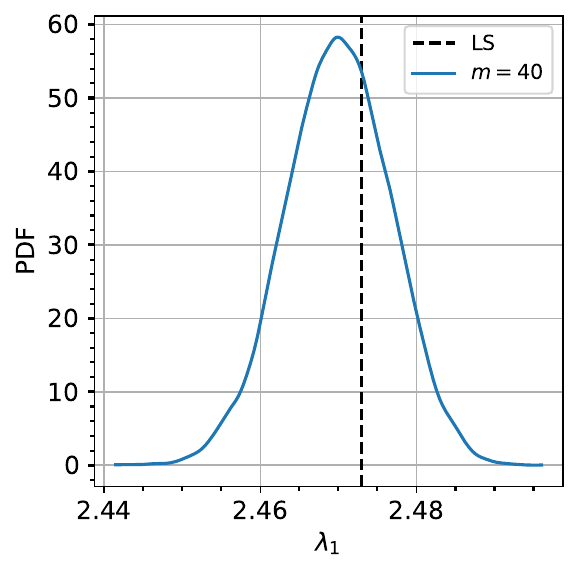}}
\caption{Parameter posteriors obtained with ROGP ($\alpha_1 = 10^8, \alpha_2=50^2$) for $N = 1000$ and $m=40$.}
\label{fig:ROGP_nd1000}
\end{figure}

\begin{figure}[ht]
\centering
\includegraphics[width=0.5\linewidth]{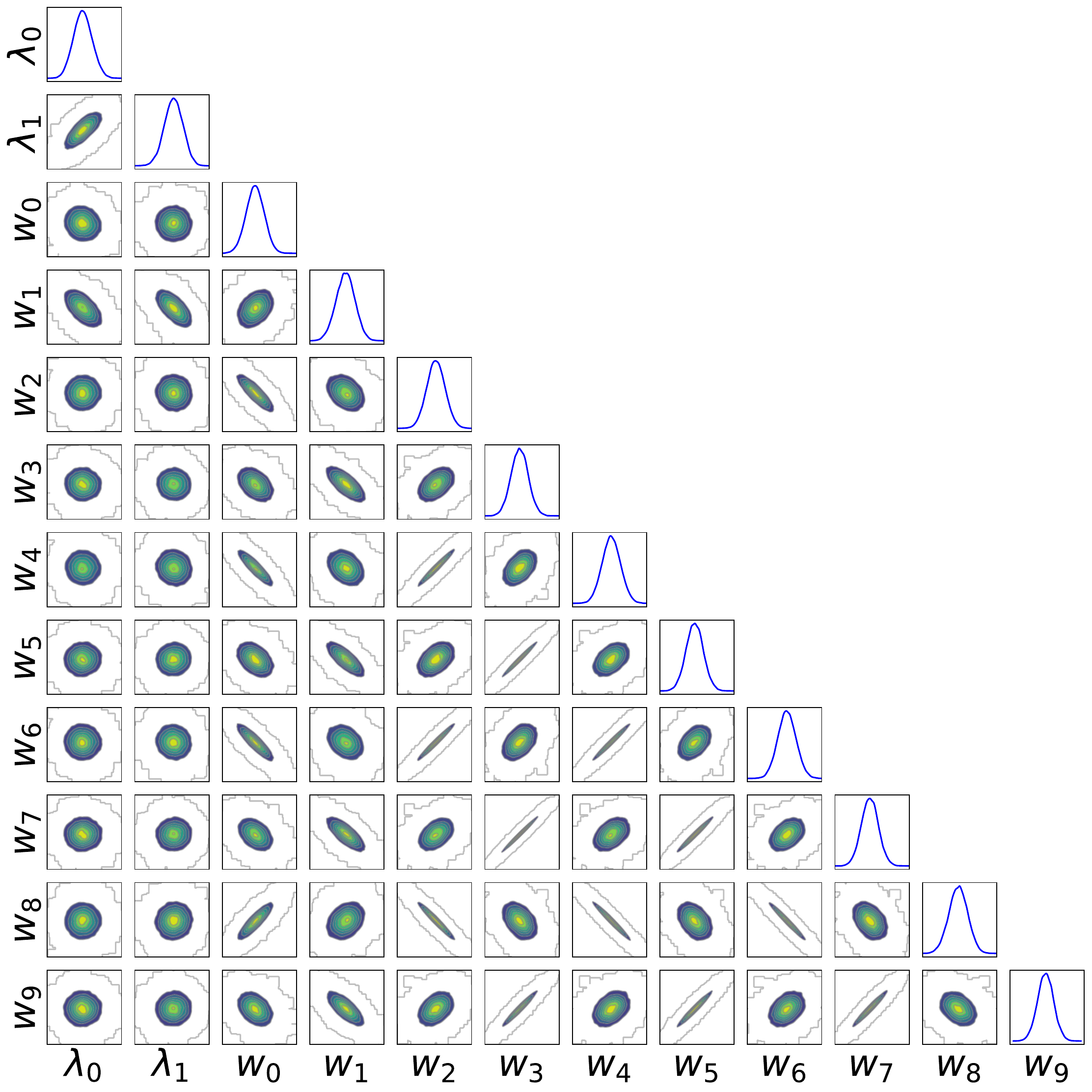}
\caption{Joint posterior distributions of model parameters and first ten weights obtained with ROGP ($\alpha_1 = 10^8, \alpha_2=50^2$) for $N=1000$ and $m=40$.}
    \label{fig:ROGP_joint_nd1000}
\end{figure}

The resulting parameter posteriors and joint distributions for $N=1000$ data points are shown in Figs.~\ref{fig:ROGP_nd1000} and \ref{fig:ROGP_joint_nd1000} respectively.
The support of $\lambda_1$ is narrower than the $N=50$ case, and the posterior mode of $\lambda_0$ has shifted to the right side of the LS estimate. 
Note that, here, the same penalty parameter values of strong $\boldsymbol{\alpha}$-enforcement were used as in the $N=50$ case and hence, the orthogonality constraint on $\lambda_0$ turned out to be slightly weaker.
This also results in a positive correlation between $\lambda_0$ and $\lambda_1$ as seen in Fig.~\ref{fig:ROGP_joint_nd1000}. 
We observe much stronger correlations between the GP weights as was the case also with LOGP. 
However, with ROGP we directly sample these GP weights which is much more challenging than operating with LIS. 
Consequently, the penalty parameter must be tuned on a case-by-case basis to enforce sufficient orthogonality while also avoiding the introduction of excessive correlations among the GP weights.
We emphasize that the primary objectives are (a) achieving a good fit between the stand-alone model and the data, and (b) ensuring that the GP-embedded model also fits the data well, providing sufficient correction of the residual discrepancy.
In our current setup, orthogonality helps us achieve goal (a). 
Therefore, to decide the penalty (hyper)-parameter, one can set up an outer optimization loop, where for each proposed value of $\boldsymbol{\alpha}$ in Eq.~\eqref{eq:rogp_log_post}, one first solves the Bayesian problem and then evaluates some statistic of the joint parameter posterior, say the mean and/or the magnitudes of the errors in (a) and (b). One can then decide how to choose the penalty according to some user-defined combined error measure which may also account for some measure of the problem complexity to avoid solving an unnecessarily hard problem without much to gain in terms of predictive accuracy.


\begin{figure}[ht]
\centering
\subcaptionbox{LOGP prior predictive \label{fig:LOGP_prior_ext}}{\includegraphics[width=0.33\linewidth]{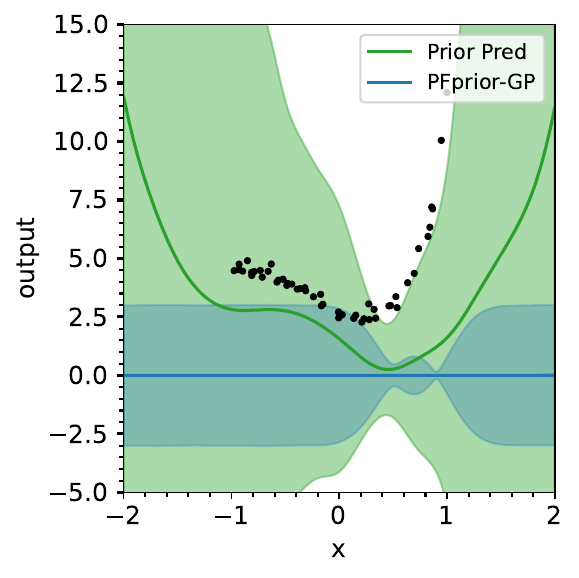}}
\subcaptionbox{LOGP posterior predictive \label{fig:LOGP_pp_ext}}{\includegraphics[width=0.33\linewidth]{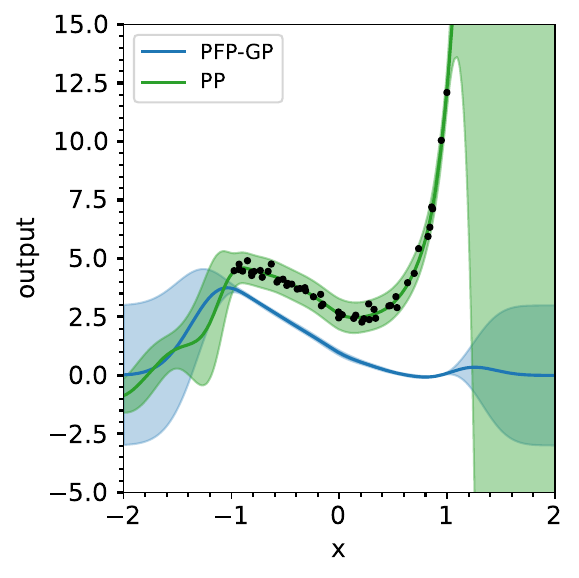}}
\subcaptionbox{ROGP prior predictive \label{fig:ROGP_prior_ext}}{\includegraphics[width=0.33\linewidth]{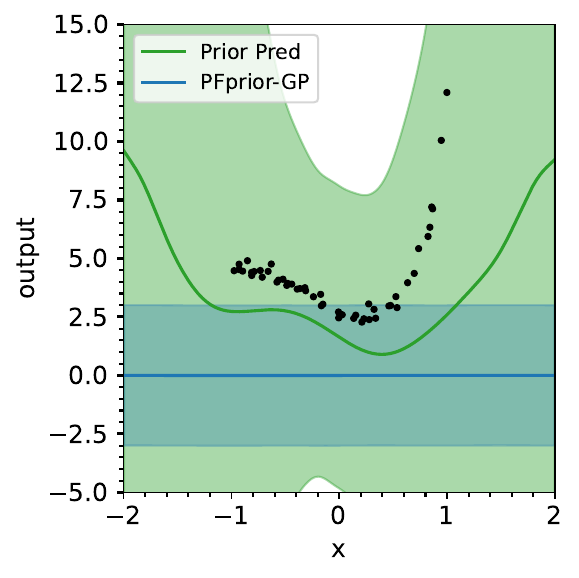}}
\subcaptionbox{ROGP posterior predictive \label{fig:ROGP_pp_ext}}{\includegraphics[width=0.33\linewidth]{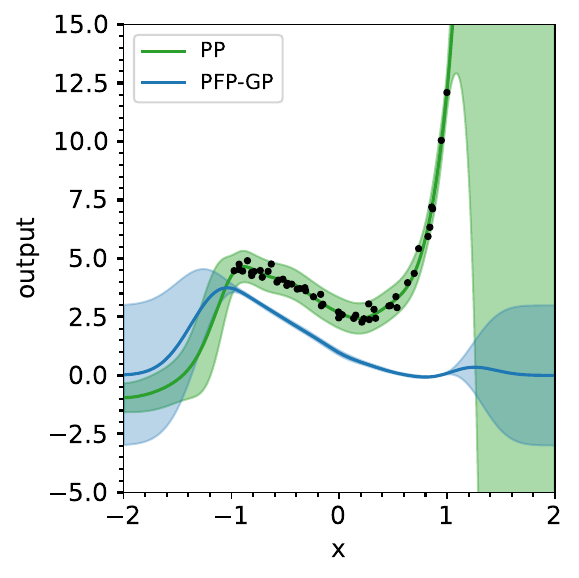}}
\caption{Panels (a) and (b) show the prior predictive and posterior predictive distributions obtained with LOGP and panels (c) and (d) show the same for ROGP. Solid lines correspond to mean predictions and bands show $\pm$ three standard deviations.}
\label{fig:ext_pp}
\end{figure}

Finally, the posterior predictive distributions along with push-forward posteriors through the GP (PFP-GP) when extrapolating outside the data regime using LOGP and ROGP are shown in Fig.~\ref{fig:ext_pp}. 
First, note that the prior predictive distributions differ between the two methods because LOGP employs basis functions derived from a modified covariance kernel, whereas ROGP uses basis functions corresponding to the unmodified kernel. 
As a result, the LOGP prior predictive distribution is narrower than that of ROGP, particularly over the interval 
[0, 1], reflecting the influence of the modified covariance. 
In contrast, the PFP-GP distributions for both LOGP and ROGP match closely and successfully recover the push-forward prior distributions in regions sufficiently far from the data. 

\FloatBarrier
\newcommand{\cL}{{\ensuremath{\mathcal{L}}}}
\subsection{Advection-Diffusion-Reaction PDE}\label{sec:appl_ADR}
Lastly, we consider a model linear non-homogeneous advection--diffusion--reaction PDE problem adapted from~\cite{raissi2017inferring}. 
We define the truth model as
\begin{align*}
\cL(u) :=\frac{\partial u}{\partial t} + \frac{\partial u}{\partial x} - \frac{\partial^2 u}{\partial x^2} - u = e^{-t}\left(2\pi \cos(2 \pi x) + 2(2\pi^2-1)\sin(2\pi x)\right)
\end{align*}
on the interval $x \in [0,1]\, , 0 \leq t \leq 1$ with initial and boundary conditions given by
$u(x,0) = \sin(2\pi x)$, $u(0,t) = 0$, and $u(1,t) = 0$. The fit PDE model considered is
\begin{align*}
\cL(u) = e^{-t}(2\pi^2-1)\sin(\lambda x)
\end{align*}
which is missing the cosine term and half of the sine term. 
Embedding a GP in the source term gives a GP augmented fit model of the form
\begin{align*}
\cL(u) = e^{-t}\left((2\pi^2-1)\sin(\lambda x) + \delta_{\boldsymbol{w}}(x)\right).
\end{align*}

Given data from the truth model, $
\mathcal{D} = \{u(x_i, t_i)\,|\, i = 1, \cdots, N\},$
our objective is to infer both the fit model parameter $\lambda$ and the GP weights $\boldsymbol{w}$.  
It's noteworthy that, being in the WS context, by inferring the GP weights we effectively learn the structural form of the missing source term in the fit model, thereby enabling model discovery. 

\begin{figure}[ht]
\centering
\subcaptionbox{\label{fig:adr_m_lam}}{\includegraphics[width=0.4\linewidth]{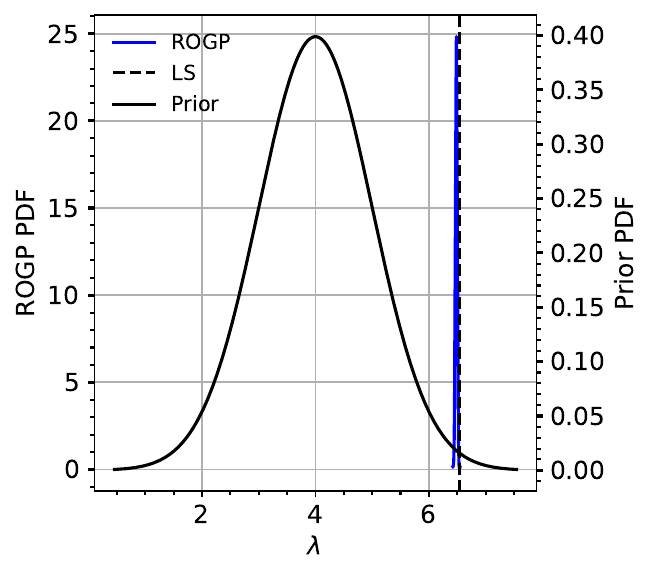}}
\subcaptionbox{\label{fig:adr_m_lam_zoom}}{\includegraphics[width=0.4\linewidth]{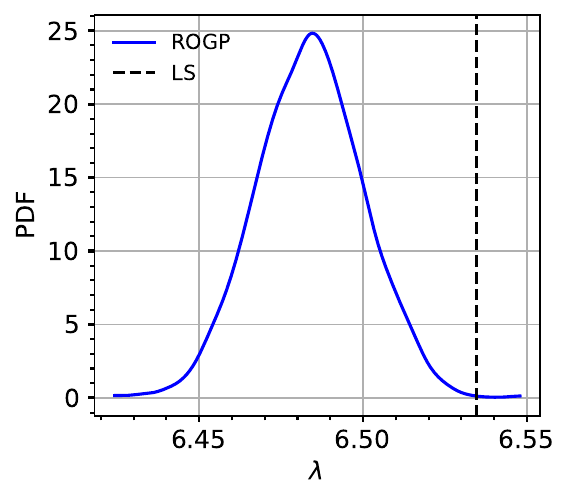}}
\caption{Panel (a) shows the marginal prior, posterior and least-squares estimate of $\lambda$. Panel (b) presents the zoomed-in view of the marginal posterior.}
\label{fig:adr_marg_lam}
\end{figure}

We consider randomly sampled data points from the spatio-temporal field of the truth model with measurement noise $\sigma_d = 0.02$.  
In this example, for the ease of demonstration, we only show results with ROGP as LOGP would require the evaluation of a 4-D integral in Eq.~\eqref{eq:em_LOGP_intH}. Nonetheless, with an appropriate quadrature rule and finite-difference based  gradient computations, LOGP can also be implemented.

For this problem, it was necessary to optimize the SQE kernel hyperparameters to accurately capture the missing source term with the GP. 
To this end, we optimized for the hyperparameter values by maximizing the log posterior in Eq.~\eqref{eq:rogp_log_post} while using 10 basis functions in the GP representation and 100 data points from the truth model. 
The optimization resulted in $\sigma_f = 100$ and $l = 0.6$ and we use these values for subsequent inference of GP weights and model parameters. 



\begin{figure}[ht]
    \centering
    \includegraphics[width=0.5\linewidth]{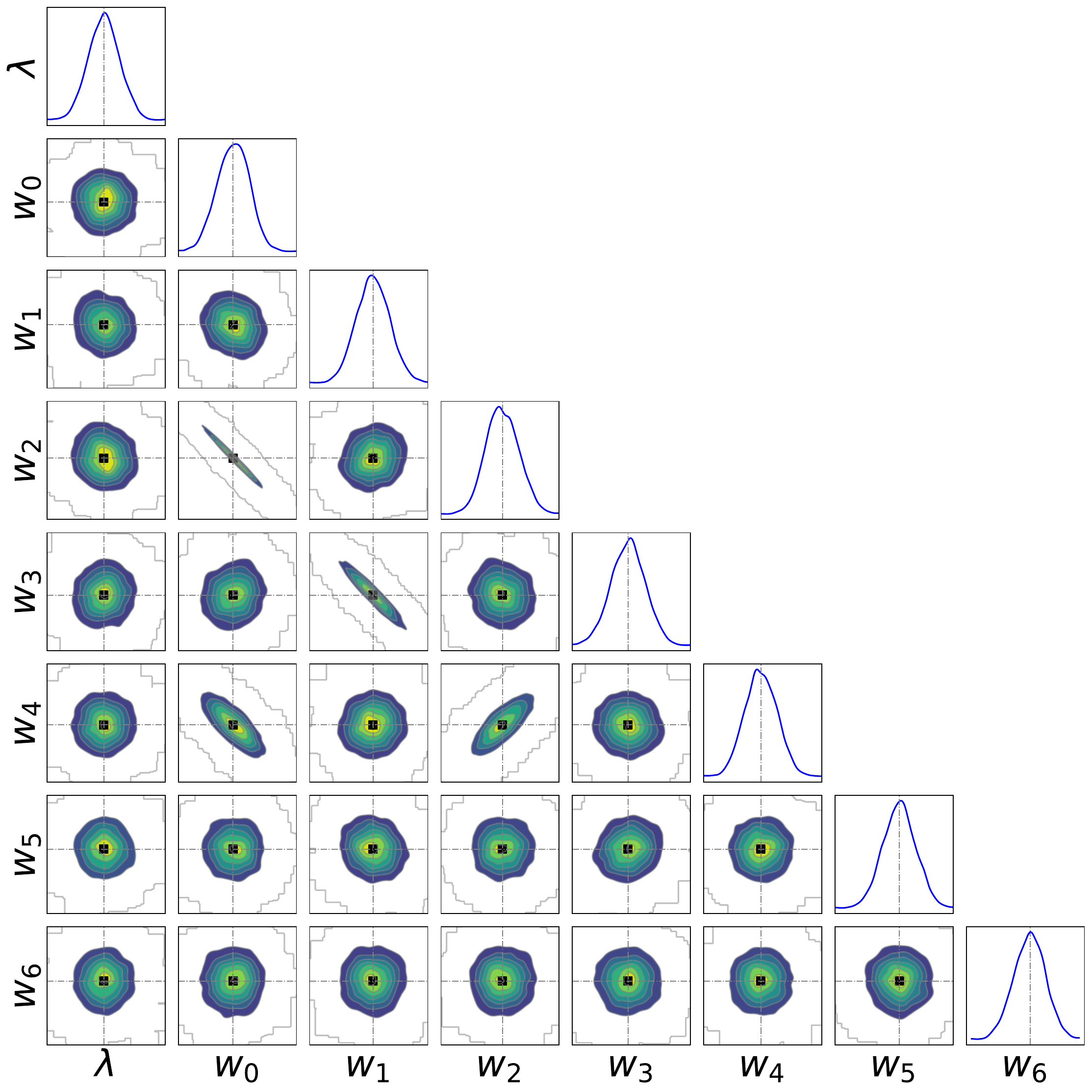}
    \caption{Joint posterior distributions of $\lambda$ and first few GP weights.}
    \label{fig:ADR_joint}
\end{figure}

Figure~\ref{fig:adr_marg_lam} shows the marginal posterior distribution of $\lambda$, compared to the least-squares estimate.  
The ROGP posterior aligns very closely with both the least-squares solution and the true value of~$2\pi$ used in the truth model. 
Joint distributions of the parameters are shown in Fig.~\ref{fig:ADR_joint}. 
While $\lambda$ remains uncorrelated with the GP weights, strong correlations are observed between some of the GP weights, specifically $w_0$-$w_2$ and $w_1$-$w_3$ which contribute to capturing the sine and cosine terms in the missing source. 

\begin{figure}[ht]
\centering
\subcaptionbox{\label{fig:adr_gp_src}}{\includegraphics[width=0.4\linewidth]{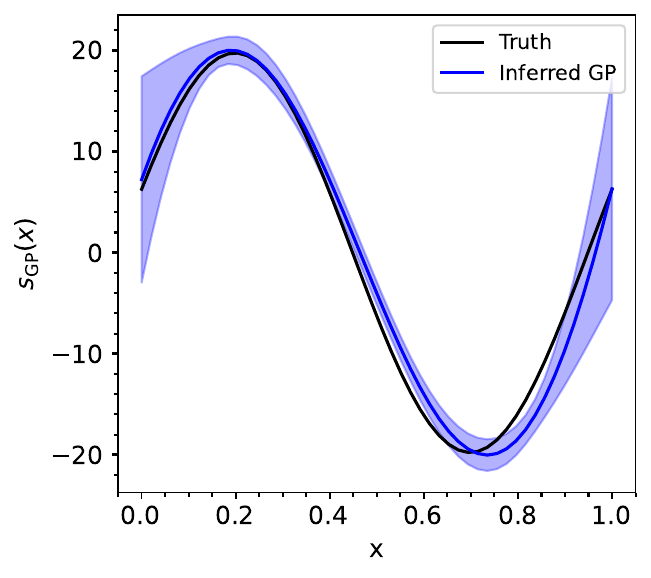}}
\subcaptionbox{\label{fig:adr_full_src}}{\includegraphics[width=0.4\linewidth]{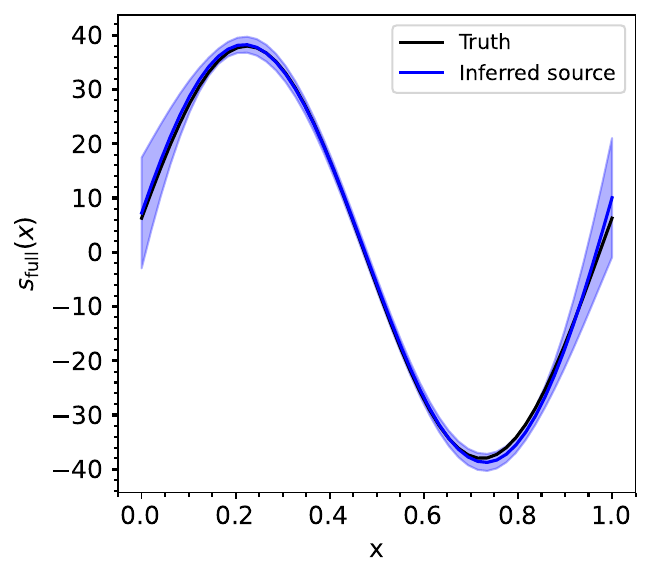}}
\caption{Panel (a) compares the inferred GP and the true missing source term, $s_{\text{GP}}(x) := 2\pi \cos({2\pi x}) + (2\pi^2 - 1)\sin{2 \pi x}$. Panel (b) compares the inferred source term and the spatial component of the true source term, $s_{\text{full}}(x) := 2\pi \cos({2\pi x}) + 2(2\pi^2 - 1)\sin{2 \pi x}$. Bands correspond to $\pm$ three standard deviations.}
\label{fig:adr_src}
\end{figure}

Figure~\ref{fig:adr_src} compares the true and recovered source terms.   
In particular, Fig.~\ref{fig:adr_gp_src} compares the mean and three standard deviation intervals of the learned GP with the true missing source term.
The slight difference between the learned GP and the true missing term is expected because the least-squares estimate of $\lambda$ is not exactly equal to $2\pi$.
The full source term, which also accounts for the contribution of $\lambda$, is in excellent agreement with the true source term as shown in Fig.~\ref{fig:adr_full_src}.   
Incorporating the learned source term into subsequent predictions reduced the maximum absolute error from 0.4 (in the absence of the GP source term) to 0.008.

\FloatBarrier
\section{Conclusions}\label{sec:concl}

We have developed a flexible framework for embedded model error by adopting the weight-space representation of GPs.
WS GPs allow us to work with a finite set of weights, associated with the eigenfunctions of the underlying covariance kernel, enabling inference when the GP is embedded within non-linear models. 
Predictions with WS and FS GPs were compared, showing that the two match very closely near the training data. 
However, inevitable basis truncation led to a collapse of the predictive distribution with the WS GP when predictions were made far enough from the data, highlighting the need for a judicious choice of the number of basis functions, based on the extent of extrapolation, for accurate extrapolation relative to the FS GP.

To disambiguate the contributions of model parameters and model bias, we adapted the OGP formulation originally developed for additive model error~\cite{plumlee2017bayesian} to the embedded model error setting in general nonlinear models. 
The key idea is to enforce orthogonality between the model bias and the model gradient evaluated at optimal parameter values defined by a chosen loss function. 
We proposed two approaches: one based on model linearization and the other based on regularization of the bias prior distribution without any linearization. 
In LOGP, linear constraints are used to modify the prior covariance of the embedded GP, and eigenfunctions of the modified covariance are computed numerically.
In contrast, with ROGP, the model evaluated at the optimal parameter values is considered directly in the loss function leading to non-linear constraints on the embedded GP. 
These non-linear constraints are introduced as regularization terms weighted by a penalty parameter along with the prior distribution of the embedded GP. 

The WS GP representation increases the dimensionality of the inference problem, which we addressed using the LIS method~\cite{spantini2015optimal,cui2014likelihood} to identify a low-dimensional subspace where the likelihood most strongly informs the prior. 
This substantially reduced the effective dimension in inference and enabled efficient MCMC sampling.

The proposed framework was demonstrated on three examples -- a linear algebraic model, a non-linear algebraic model with interacting sub-models, and a linear non-homogeneous advection–diffusion–reaction PDE.
For the linear case, both LOGP and ROGP reduce to the standard additive OGP formulation. Convergence studies were used to determine the number of WS basis functions, while the use of LIS allowed large GP representations with minimal additional computational cost.
The orthogonality constraints, with \emph{iid} Gaussian noise on the data, not only successfully captured the least-squares estimates in the supports of the parameter posteriors but also translated to zero off-diagonal terms in the posterior covariance resulting in uncorrelated model parameters and GP weights. 

In the non-linear example, model parameters remained weakly correlated with the GP weights, although correlations among the GP weights themselves were observed. 
For LOGP, LIS can be applied directly, yielding uncorrelated LIS parameters and efficient sampling. 
For ROGP, the penalty parameter controls the strength of orthogonality enforcement and influences posterior correlations.
Hence, it must be chosen to balance constraint enforcement and sampling efficiency. 
In all cases, extrapolation far from the training data recovered the prior predictive distribution when a sufficiently rich WS representation was used. 
In the ADR example, MAP-based hyperparameter optimization was necessary for the embedded GP to accurately capture the missing source term.
Subsequent predictions with the embedded GP led to a substantial reduction in the maximum absolute error. 

Future work can include several directions.
The proposed framework can be naturally extended to constructions including both internal and external GP corrections, for which orthogonality constraints can be derived for each GP. 
The internal GP would make the model more flexible while adhering to the physics of the problem while the external GP would capture residual model error. 
As noted earlier, the LIS framework must be modified for use with the ROGP approach, specifically by accounting for the regularization terms in the construction of the global LIS.
While the methodology in this work was demonstrated using a weighted $L^2$ loss, other loss functions with unique minimizers for the model parameters can also be considered, for instance the loss function based on inner products in the Sobolev space as in~\cite{plumlee2017bayesian} as well as user-defined losses such as those based on moment matching.
Orthogonality constraints for these cases can be derived following the general procedure proposed in this work.

\appendix

\section{Likelihood Informed Subspace Summary}\label{sec:LIS_sum}

We summarize in the following the LIS construction of \cite{cui2014likelihood,spantini2015optimal}. Recall that our model to be calibrated is given by 
\begin{align*}
    y = \tilde{f}(\mathbf{x}; \boldsymbol{\lambda}, \boldsymbol{\phi}(\mathbf{x})^\top\boldsymbol{w}) + \varepsilon
\end{align*}
with calibration parameters $\boldsymbol{\theta} = \{\boldsymbol{\lambda}, \boldsymbol{w}\} \in \mathbb{R}^{p+m}$ and prior $\pi(\boldsymbol{\theta}) = \mathcal{N}(\bm{\theta}|\mu_p, \Sigma_p)$. 
For notational convenience, let $\tilde{\mathbf{f}}(\boldsymbol{\theta}) = [\tilde{f}(x_1; \boldsymbol{\theta}), \cdots, \tilde{f}(x_N; \boldsymbol{\theta})]$. 
Further, let the covariance matrix of the zero-mean measurement noise $\varepsilon$ be $\Sigma_d$.   
For non-linear models, the global LIS basis is built by forming a Monte Carlo approximation of the expected prior-preconditioned Gauss-Newton Hessian (ppGNH), 
\begin{align}
    S = \frac{1}{n} \sum_{k=1}^{n} L^T H(\boldsymbol{\theta}^{(k)})L \label{eq:LIS_ppGNH}
\end{align}
where $\boldsymbol{\theta}^{(k)} \sim \pi(\boldsymbol{\theta}|\mathbf{y})$ and $L$ is obtained from the symmetric decomposition of the prior covariance $\Sigma_p = LL^\top$. 
The local Hessian, $H(\boldsymbol{\theta}^{(k)})$, is approximated by
\begin{align*}
    H(\boldsymbol{\theta}^{(k)}) = \nabla_{\boldsymbol{\theta}} \tilde{\mathbf{f}} \big|_{\boldsymbol{\theta}^{(k)}}^\top \Sigma_d^{-1} \nabla_{\boldsymbol{\theta}} \tilde{\mathbf{f}} \big|_{\boldsymbol{\theta}^{(k)}}.
\end{align*}
Let $(\delta_i^2, v_i)$ denote the eigenvalue–eigenvector pairs of the ppGNH, ordered such that $\delta_i^2 > \delta_{i+1}^2$. The LIS basis vectors are given by $u_i = L v_i$, and the rank-$r$ LIS basis is $U_r = [u_1, \dots, u_r]$. The associated rank-$r$ projector is
\begin{align*}
    P_r = U_r W_r^{\top},
\end{align*}
where $W_r = [w_1, \dots, w_r]$ with $w_i = L^{-\top} v_i$. Choosing $V_{\perp}$ such that $[V_r \; V_{\perp}]$ forms an orthonormal basis of $\mathbb{R}^{p+m}$, the projector onto the CS becomes
\begin{align*}
    P_{\perp} = I - P_r = U_{\perp} W_{\perp}^{\top},
\end{align*}
where $U_{\perp} = L V_{\perp}$ and $W_{\perp} = L^{-\top} V_{\perp}$. Further, since $P_r$ is self-adjoint with respect to the inner product induced by the prior precision, any parameter vector admits the decomposition
\begin{align*}
    \boldsymbol{\theta}
    = P_r \boldsymbol{\theta} + P_{\perp} \boldsymbol{\theta}
    = U_r \boldsymbol{\theta}_r + U_{\perp} \boldsymbol{\theta}_{\perp},
\end{align*}
with $\boldsymbol{\theta}_r = W_r^{\top} \boldsymbol{\theta} \in \mathbb{R}^{r}$ and 
$\boldsymbol{\theta}_{\perp} = W_{\perp}^{\top} \boldsymbol{\theta} \in \mathbb{R}^{p+m-r}$. The prior distribution factorizes accordingly:
\begin{align*}
    \pi(\boldsymbol{\theta})
    = \pi_r(\boldsymbol{\theta}_r)\,\pi_{\perp}(\boldsymbol{\theta}_{\perp}),
\end{align*}
where $\pi_r(\boldsymbol{\theta}_r) = \mathcal{N}(\boldsymbol{\theta}_r |W_r^{\top}\mu_p, I_r)$ and 
$\pi_{\perp}(\boldsymbol{\theta}_{\perp}) = \mathcal{N}(\boldsymbol{\theta}_{\perp} | W_{\perp}^{\top}\mu_p, I_{\perp})$. 
Finally, the posterior admits the approximation 
\begin{align}
    \pi(\boldsymbol{\theta} | \mathbf{y}) \approx \hat{\pi}(\boldsymbol{\theta} | \mathbf{y}) 
    &\propto \pi(\mathbf{y}|P_r\boldsymbol{\theta}) \pi(\boldsymbol{\theta}) \nonumber \\
    &= \pi(\mathbf{y} | U_r \boldsymbol{\theta}_r) \pi_r(\boldsymbol{\theta}_r) \pi_{\perp}(\boldsymbol{\theta}_{\perp}) \nonumber \\
    &= \pi(\boldsymbol{\theta}_r | \mathbf{y})\pi_{\perp}(\boldsymbol{\theta}_{\perp})
\end{align}
so that MCMC sampling is required only for the $r$ LIS coordinates. Thus, the effective dimension of the inverse problem is reduced from $p+m$ to $r$, and typically $r \ll p+m$. To retain all meaningful likelihood-informed directions, we choose $r$ such that the $r$-th eigenvalue is less than one, i.e. $\delta_r^2 < 1$, and, in our embedded GP applications (see Section~\ref{sec:Appl}), we select $r$ such that $\delta_r^2 \le 0.1$. Finally, the posterior samples needed to approximate the ppGNH in Eq.~\eqref{eq:LIS_ppGNH} are obtained using \emph{Algorithm 1} of~\cite{cui2014likelihood}.
This is an adaptive procedure to find the global LIS, where we start with the posterior mode and its corresponding local LIS. 
A \emph{subspace} MCMC chain in this local LIS, with the initial state at the posterior mode projected onto the LIS, is simulated until we obtain an uncorrelated sample. 
The initial LIS is modified using this new sample, and Eq.~\eqref{eq:LIS_ppGNH}, to start a new \emph{subspace} MCMC chain. 
This adaptation is terminated after a maximum allowable number of Hessian evaluations or using subspace distance metrics. 
Once the global LIS is discovered, a much longer MCMC chain can be initialized to obtain posterior samples from this global LIS.

\section*{Acknowledgments}
MK, KS, and HNN acknowledge support by the U.S. Department of Energy, Office of Science, Office of Advanced Scientific Computing Research (ASCR), Scientific Discovery through Advanced Computing (SciDAC) program, under the FASTMath Institute. MK and MP also acknowledge support by the U.S. Department of Defence, Office of Naval Research (ONR), under the MURI grant no. N00014-21-1-2475.
This article has been co-authored by employees of National Technology and Engineering Solutions of Sandia, LLC under Contract No. DE-NA0003525 with the U.S. Department of Energy (DOE). The employees co-own right, title and interest in and to the article and are responsible for its contents. The United States Government retains and the publisher, by accepting the article for publication, acknowledges that the United States Government retains a non-exclusive, paid-up, irrevocable, world-wide license to publish or reproduce the published form of this article or allow others to do so, for United States Government purposes. The DOE will provide public access to these results of federally sponsored research in accordance with the DOE Public Access Plan \url{https://www.energy.gov/downloads/doe-public-access-plan}. 
%
The views and conclusions contained herein are those of the authors and should not be interpreted as necessarily representing the official policies or endorsements, either expressed or implied, of the ONR or the US government.

\end{document}